\documentclass[hidelinks,twoside]{article}

\usepackage{PRIMEarxiv}

\usepackage[english]{babel}
\usepackage{latexsym}
\usepackage{amssymb}
\usepackage{amsmath}
\usepackage{amsfonts}
\usepackage{mathtools}
\usepackage{graphics}
\usepackage{enumitem}
\usepackage{afterpage}
\usepackage{fancybox}
\usepackage{scalerel}
\usepackage{xspace}
\usepackage[normalem]{ulem}
\usepackage{hyperref}
\usepackage[dvipsnames]{xcolor}
\usepackage{todonotes}

\newcommand{\M}[1]{\mbox{\boldmath$#1$}}

\newcommand{\mim}[2][]{%
  \todo[backgroundcolor=SkyBlue!20,#1]{\textbf{Mimmo:} \small #2}%
}

\usepackage{imakeidx}         
\usepackage{multicol}        
\makeindex
    \newcommand\gobbleone[1]{}
    \newcommand*{\seeonly}[2]{\ (\emph{\seename} #1)}

\DeclareMathOperator{\sgn}{sgn}

\newcommand{\red}[1]{{\textcolor{red}{#1}}}
\newcommand{\ind}{\mathit{ind}}

\newcommand{\al}{\alpha}
\newcommand{\be}{\beta}
\newcommand{\ga}{\gamma}

\newcommand{\ep}{\varepsilon}

\newcommand{\fhi}{\varphi}
\newcommand{\Kappa}{\mbox{\Large$\kappa$}}

\newtheorem{theorem}{Theorem}[section]
\newtheorem{proposition}{Proposition}[section]
\newtheorem{lemma}{Lemma}[section] 
\newtheorem{corollary}{Corollary}[section]
\newtheorem{definition}{Definition}[section]
\newtheorem{example}{Example}[section]
\newtheorem{remark}{Remark}[section]

\newcommand{\dimostraz}{\noindent\textit{Proof}:\ }
\newcommand{\qed}{{\hfill\ensuremath{\M{\dashv}}}}
\newcommand{\eod}{\qed}
\newcommand{\ita}{\textit}
\newcommand{\infy}{\infty}
\newcommand{\I}{\textit{I}}

\newcommand{\Nv}{\color{navy}}

\newcommand{\nor}{\mbox}

\newcommand{\mR}{\mathbb{R}}

\newcommand{\mN}{\mathbb{N}}

\newcommand{\sK}{\mathcal{K}}
\newcommand{\sI}{\mathcal{I}}
\newcommand{\sF}{\mathcal{F}}
\newcommand{\sM}{\mathcal{M}}

\newcommand{\imp}{\Rightarrow}
\newcommand{\frec}{\longrightarrow}

\newcommand{\x}{\cdot}
\newcommand{\sse}{\longleftrightarrow}

\newcommand{\et}{\wedge}
\newcommand{\vel}{\vee}

\renewcommand{\preceq}{\preccurlyeq}
\renewcommand{\leq}{\leqslant}
\renewcommand{\geq}{\geqslant}
\newcommand{\defFoAs}{\; \overset{\mbox{\tiny Def}}{\longleftrightarrow} \;}
\newcommand{\putAs}{\coloneqq}

\usepackage{color}

\newcommand{\pp}[1]{pp.{#1}}
\newcommand{\Lem}[1]{Lemma~\ref{#1}}
\newcommand{\Lems}[2]{Lemmas~\ref{#1} and \ref{#2}}
\newcommand{\Def}[1]{Def.~\ref{#1}}
\newcommand{\Defs}[2]{Defs~\ref{#1} and~\ref{#2}}
\newcommand{\Fig}[1]{Figure~\ref{#1}}
\newcommand{\Cor}[1]{Corollary~\ref{#1}}
\newcommand{\Prop}[1]{Proposition~\ref{#1}}
\newcommand{\Sec}[1]{Sec.~\ref{#1}}
\newcommand{\Subsec}[1]{Subsec.~\ref{#1}}

\newcommand{\Rem}[1]{Remark~\ref{#1}}
\newcommand{\Exa}[1]{Example~\ref{#1}}
\newcommand{\Exas}[2]{Examples~\ref{#1} and \ref{#2}}
\newcommand{\abs}[1]{\left|{#1}\right|}
\newcommand{\COMMENT}[1]{}
\newcommand{\CILC}[1]{}

\newcommand{\EAR}{\mbox{EAR}\xspace}
\newcommand{\RMCF}{\ensuremath{\textit{RMCF}}\xspace}
\newcommand{\RMCFp}{\ensuremath{\textit{RMCF}^{\hspace{0.05em}+}}\xspace}
\newcommand{\RDF}{\ensuremath{\textit{RDF}}\xspace}
\newcommand{\RDFp}{\ensuremath{\textit{RDF}^{\hspace{0.05em}+}}\xspace}
\newcommand{\RDFs}{\ensuremath{\textit{RDF}^{\hspace{0.05em}*}}\xspace}
\newcommand{\RDFn}{\ensuremath{\textit{RDF}^{\hspace{0.05em}n}}\xspace}
\newcommand{\RDFpFlat}{\ensuremath{\textit{RDF}^{\hspace{0.05em}+}_{\mathit{flat}}}\xspace}
\newcommand{\RDFpOrd}{\ensuremath{\textit{RDF}^{\hspace{0.05em}+}_{\mathit{ord}}}\xspace}

\def\sqdot{\mathbin{\scalerel*{\strut\rule{1.5ex}{1.5ex}}{\x}}}

\definecolor{navy}{rgb}{0,0,0.5}

\title{\large\bf Decision algorithms for fragments of real analysis.\ III:\\ A theory of differentiable functions with (semi-)open intervals
}

\author{Gabriele Buriola$^{[0000-0002-1612-0985]}$\thanks{Corresponding author.}\\
  Dept.\ of Computer Science\\
  University of Verona\\
  I-37129 Verona, Italy\\
  \texttt{gabriele.buriola@univr.it.} \\
  \And
  Domenico Cantone$^{[0000-0002-1306-1166]}$ \\
  Dept.\ of Mathematics and Computer Science \\
  University of Catania \\
  I-95125 Catania, Italy\\
  \texttt{domenico.cantone@unict.it} \\
   \And
  Gianluca Cincotti$^{[0000-0001-8460-1708]}$\\
  Dept.\ of Mathematics and Computer Science \\
  University of Catania \\
  I-95125 Catania, Italy\\
  \texttt{cincotti@dmi.unict.it} \\
  \And
    Eugenio G. Omodeo$^{[0000-0003-3917-1942]}$ \\
  Dept.\ of Mathematics, Informatics, and Earth Sciences\\
  University of Trieste \\
  I-34127 Trieste, Italy\\
  \texttt{eomodeo@units.it} \\
  \AND
    Gaetano T. Spart\`a$^{[0000-0002-8993-5851]}$\\ 
Pontificia Universit\`a Gregoriana,\\ Rome, Italy\\
\texttt{g.sparta@unigre.it}
}

\date{\today}

\usepackage{fancyhdr}
\begin{document}

\maketitle

\vspace{-0.7cm}

\begin{abstract}

This paper enriches preexisting satisfiability tests for unquantified languages, which in turn augment a fragment of Tarski's elementary algebra with unary real functions possessing a continuous first derivative. 

Two sorts of individual variables are available, one ranging over real numbers and
the other one ranging over the functions of interest. Numerical terms are built from real variables through constructs designating the four basic arithmetic operations and through the function-application constructs $f(t)$ and $D[\,f\,](t)$, where $f$ stands for a function variable, $t$ for a numerical term, and $D[\,\sqdot\,]$ designates the differentiation operator. Comparison relators can be placed between numerical terms. An array of predicate symbols are also available, designating various relationships between functions, as well as function properties, that may hold over intervals of the real line; those are: (pointwise)
function comparisons, strict and nonstrict monotonicity~/~convexity~/~concavity
properties, comparisons between the derivative of a function and a real term---here,  w.r.t.\ earlier research, they are extended to (semi)-open intervals.

The decision method we propose consists in preprocessing the given formula into an equisatisfiable quantifier-free
formula of the elementary algebra of real numbers, whose satisfiability can then be checked by means of Tarski's decision method. No direct reference to
functions will appear in the target formula, each function variable having been superseded by a collection of stub real variables; hence, in order to prove
that the proposed translation is satisfiability-preserving, we must figure out a sufficiently flexible family of interpolating $C^1$ functions that can accommodate a model for the source formula whenever the target formula turns out to be satisfiable.

\smallskip

\noindent{\bf Key words:}\ Decidable theories, Tarski's elementary algebra, Functions of a real variable.\\~
\noindent{\bf MS Classification {2020}:}\ 03B25, 26A06.
\end{abstract}

\newpage

\tableofcontents

\section*{Introduction}


\noindent The primary aim of this paper is to present the completeness proof for a satisfiability decision algorithm applicable to a fragment of analysis introduced in \cite{GC00}. The formal language under consideration, known as \RDFp, has been revised and improved on two occasions, in \cite{DC07} and \cite{BCCOS20}. However, neither of these works provided a detailed completeness proof, leaving a gap that this paper intends to fill. To ensure that the paper is self-contained, it includes a comprehensive description of the syntax and semantics of \RDFp, along with the corresponding satisfiability decision algorithm.

In the early 1930s, Alfred Tarski proposed an axiomatic system for elementary geometry based on first-order predicate calculus. Using a quantifier-elimination method, he proved the completeness of this system. This completeness revealed that the truth problem for sentences in the elementary algebra of real numbers is algorithmically solvable \cite{AT 67,AT 51}. Subsequent studies of Tarski's decision algorithm pursued two main goals: improving computational efficiency in specific cases and extending its applicability beyond algebra to encompass constructs relevant to real analysis. One notable extension was the \RMCF theory (Theory of Reals with Monotone and Convex Functions), initially studied by D. Cantone, A. Ferro, E.G. Omodeo, and J.T. Schwartz \cite{DC87}. This theory enriched the existential theory of real numbers by introducing various predicates about real functions. Other decidability results, inspired by techniques similar to those underlying \RMCF, include \RMCFp (Augmented Theory of Reals with Monotone and Convex Functions) by D. Cantone, G. Cincotti, and G. Gallo \cite{DC06}, \RDF (Theory of Reals with Differentiable Functions) by D. Cantone and G. Cincotti \cite{GC00,DC07}, and \RDFp (Augmented Theory of Reals with Differentiable Functions), which evolved from \cite{GC00}. The \RDFp theory, the subject of this paper, represents an enriched version of findings initially presented in an earlier proceedings paper \cite{BCCOS20}.

\COMMENT{\noindent
Around 1930, Alfred Tarski put forward an axiomatic system of elementary geometry
based on first-order predicate calculus and proved, through a quantifier-elimination
method, the completeness of his axiomatic theory. Completeness yielded the
algorithmic solvability of the truth problem as referred to sentences in the
elementary algebra of real numbers \cite{AT 67,AT 51}. Subsequent improvements of
the decision algorithm aimed, on the one hand, at making its complexity
affordable in special subcases; on the other hand, at broadening its applicability
beyond algebra, so as to handle entities and constructs relevant to the realm of
real analysis. One of the decidable extensions was the \ita{RMCF} theory (Theory of Reals with Monotone and Convex Functions), first investigated by D. Cantone, A. Ferro, E.G. Omodeo and J.T. Schwartz \cite{DC87}, whose language extends the existential theory of real numbers with various predicates about real functions. Other decidability results were achieved through ideas close to the ones which had led to the decidability of \ita{RMCF}; those results regard: $RMCF^+$ (Augmented theory of Reals with Monotone and Convex Functions) by D. Cantone, G. Cincotti and G. Gallo \cite{DC06}, \ita{RDF}  (Theory of Reals with Differentiable Functions) by D. Cantone and G. Cincotti \cite{GC00,DC07}, and finally the \RDFp  theory (Augmented Theory of Reals with Differentiable Functions) (stemmed from \cite{GC00}), the subject of this paper which is an enrichment of a previous proceedings paper \cite{BCCOS20}.}

Let us outline the basic idea behind proving the decidability of \RMCF, \RMCFp, \RDF, and \RDFp. Consider an existential sentence $\psi \putAs \exists x_1 \cdots \exists x_n \theta$ in the \emph{Elementary Algebra of Real numbers} (\EAR), where $\theta$ is a quantifier-free formula involving only the variables $x_1, \dots, x_n$. Such a sentence $\psi$ is \emph{true} if and only if $\theta$ is \emph{satisfiable}---that is, if there exist real numbers $a_1, \dots , a_n$ such that substituting them for the variables $x_1, \dots, x_n$ makes $\theta$ true. Thanks to Tarski's decision procedure for sentences (particularly  existential ones) of \EAR, we can also establish the satisfiability of quantifier-free formulas in \EAR. The four theories mentioned---\RMCF, \RMCFp, \RDF, and \RDFp---extend the existential fragment of \EAR by introducing a new sort of variable (representing functions) and new predicate symbols (expressing special properties of, or relations between, functions). To establish their decidability, formulas from these theories are translated into formulas within the existential theory of real numbers in a way that preserves satisfiability. This reduction allows us to rely on Tarski's original decision result for \EAR.

To stress the difference between the syntactic level and the intended model (consisting of real functions), we adopt the following convention: given a real function $f\colon A \subseteq \mR \frec \mR$, we denote by $f'$ and $f''$ the first and the second derivatives of $f$, respectively (when they exist), while reserving $D[\,\sqdot\,]$ as a syntactic symbol for differentiation within our language. Moreover, when topological concepts are considered, for example, bounded or open intervals, we refer to the standard topology over $\mR$ \cite[p.~81]{Munkres:book}. Finally, for standard logical notions, such as validity and satisfiability of a formula, we refer to any introductory text in mathematical logic (e.g.,~\cite{Enderton:book}).

\begin{center}------------\end{center}

The paper is structured as follows. \Sec{sec:RFDplus} introduces the syntax and semantics of the language of interest and demonstrates its expressive power---due in part to the ease of introducing useful derived constructs---through a series of illustrative examples. \Sec{sec:decisionAlgorithm} details the proposed decision algorithm, discussing an example illustrating its application through a step-by-step walkthrough of how it processes a specific valid formula. In \Sec{sec:correctness}, we provide a proof of the correctness of the proposed decision algorithm. Finally, \Sec{sec:final} compares our work with related approaches and offers concluding remarks.

As discussed in \Sec{sec:final}, the matter of this paper paved the way for decidable extensions of \RDFp including one, \cite{BCCOS23}, that incorporates the pointwise addition of functions.

\section{The \RDFp  Theory}\label{sec:RFDplus}

The theory \RDFp extends \RDF, the quantifier-free theory of \emph{Reals with Differentiable Functions}~\cite{GC00}, by increasing its expressive power while preserving its quantifier-free nature. It focuses on real numbers and $C^1$ functions---real-valued functions of a single real variable with a continuous first derivative. In \RDFp, some operators ($\,\cdot\,$, $+\,$, $-$) represent fundamental arithmetic operations, while another construct denotes differentiation of real functions. The theory’s predicate symbols capture key properties of $C^1$ functions, including strict and nonstrict monotonicity, convexity, and concavity over both bounded and unbounded intervals. Additionally,  \RDFp includes predicates for equality and strict comparison ($=$,\,$>$) between real numbers and (pointwise) between $C^1$  functions, as well as comparisons between first derivatives and real terms. This section introduces the syntax of \RDFp, clarifies its intended interpretation, and presents illustrative examples of its use.

\subsection{Syntax and semantics}\label{subsec:syntax}
\index{syntax of \RDFp|(}
The language \RDFp has two infinite supplies of individual variables: \emph{numerical variables},\index{variable!numerical v.} denoted by $x,y,z,\dots$, and \emph{function\index{variable!function v.} variables}, denoted by $f,g,h, \dots$.  Numerical and function variables are assumed to range, respectively, over the set $\mR$ of real numbers and over the collection of $C^1$ real functions. Four constants are also available:
\begin{itemize}
\item the symbols $\mathbf{0}$ and $\mathbf{1}$,  designating the numbers $0$ and $1$, respectively;\footnote{As will turn out, the constants $\mathbf{0},\mathbf{1}$ are eliminable from our language without loss of expressive power, since $z= 0 \et u= 1$ is the sole solution to $u=u\x u>z\x z=z$.}

\item the distinguished symbols $+\infy$ and $-\infy$, which occur only as ends of \emph{interval specifications} (see below).
\end{itemize}   

The syntax of {\em numerical terms}\index{term!numerical t.} is recursively specified as follows.
\begin{definition}
\textsc{Numerical terms} of \RDFp:
\begin{quote}
\begin{enumerate}[label=\arabic*)]
\item\label{lab:numterm1} numerical variables and the constants $\mathbf{0},\mathbf{1}$ are \emph{numerical terms};

\item\label{lab:numterm2} if $s$ and $t$ are numerical terms, then the following are also \emph{numerical terms}:
\[\begin{array}{llcl}s+t\:,&s-t\:,& \mbox{and} &s\x t\:,\end{array}\]
intended to represent the sum, difference, and product of $s$ and $t$;

\item\label{lab:numterm3} if $t$ is a numerical term and $f$ is a function variable, then 
\[
f(t) \quad \text{and} \quad D[f](t)
\]
are also \emph{numerical terms}, intended to represent the application of $f$ to $t$ and the derivative of $f$ evaluated at $t$, respectively;

\item numerical terms are all and only the expressions built up according to the above rules~\ref{lab:numterm1}, \ref{lab:numterm2}, and \ref{lab:numterm3}.
\end{enumerate}\end{quote}
\end{definition}
In the ongoing, we often use the term ``numerical variable" to refer either to numerical variable proper or to one of the constants $\mathbf{0}$ and $\mathbf{1}$.
Moreover, we will use the term ``\emph{extended} numerical variable" to refer to either a numerical variable or one of the symbols $-\infy$, $+\infy$; and the term ``\emph{extended} numerical term" to refer to either a numerical term or one of the symbols $-\infy$, $+\infy$.

\smallskip

In preparation for the syntax of {\em atoms} and {\em formulas} of \RDFp, we introduce notation for specifying intervals---bounded or unbounded, and open, half-open, or closed.

\begin{definition}[Specification of \RDFp intervals]\phantom{aa}
\begin{quote}
An \textsc{interval spec}\index{interval spec} $A$ is an expression of any of the four forms 
\[
[e_1, e_2], \quad [e_1, e_2[, \quad ]e_1, e_2], \quad \text{and} \quad ]e_1, e_2[,
\]
where $e_1$ denotes either a numerical term or $-\infty$, and $e_2$ denotes either a numerical \hbox{term or $+\infty$}.
We call the extended numerical terms $e_1$ and $e_2$ the \textsc{ends} of the interval spec $A$\,.
\end{quote}


\em (When it comes to semantics, we will call the quantities designated by the ends of a specified interval the {\em endpoints}\index{endpoints (of an interval)} of that interval.) \eod
\end{definition}

\begin{definition}[\RDFp atoms]\label{def:atom} An \textsc{atom}\index{atom (of \RDFp)}\index{relator!primitive r.\seeonly{atom}|gobbleone} of \RDFp  is an expression of one of the following forms:
\emph{\[
\begin{array}{rccrccrccrcc}
s=t    & , &\phantom{xs}& s>t &,&\phantom{xs}&
(f=g)_A    & , &\phantom{xs}& (f>g)_A &,\\
\nor{Up}(f)_A    & , &&
\nor{Strict}\_\nor{Up}(f)_A  &,&&
\nor{Down}(f)_A  & , &&
\nor{Strict}\_\nor{Down}(f)_A   &,\\
\nor{Convex}(f)_A & , && \nor{Strict}\_\nor{Convex}(f)_A  &,&&\nor{Concave}(f)_A & , && \nor{Strict}\_\nor{Concave}(f)_A &,\\(D[f] \bowtie t)_A &,
\end{array}
\]}
$\!\!$where ${\bowtie} \in \{ {}=,<,\ >,\ \leq ,\ \geq\}$, and $A$ is an interval spec. 
The expressions $s$ and $t$ stand for numerical terms, and $f$ and $g$ for function variables.

A \textsc{formula}\index{formula (of \RDFp)} of \RDFp is any truth-functional combination of \RDFp atoms.
\em (For definiteness, we build \RDFp formulas from \RDFp atoms using the usual propositional connectives $\neg$, $\et$, $\vel$, $\longrightarrow$, and  $\longleftrightarrow$.)  \eod
\end{definition}

\COMMENT{
For convenience, we enrich our language with some useful derived expressions, introduced through the following syntactic shorthands.
\begin{definition}[Some derived expressions]
\begin{align*}
 s \neq t      & \quad \defFoAs \quad  (s>t) \vel (t > s) \,,\\
 s < t      & \quad \defFoAs \quad   t > s\:,                   \\
 s \geq t   & \quad \defFoAs \quad  (s>t) \vel (s=t) \,,   \\
 s = t_1/t_2  & \quad \defFoAs \quad   (t_1=s \x t_2) \et 
                              (t_2\x t_2> \mathbf{0})\,, \\
 s > t_1/t_2  & \quad \defFoAs \quad   \big((t_1 < s \x t_2) \et (t_2 > \mathbf{0})\big) \vel \big((t_1 > s \x t_2) \et (t_2 < \mathbf{0})\big)\,,    \\[0.09cm]
 (D[f]\neq t)_A  & \quad \defFoAs \quad (D[f]< t)_A \vel (D[f]> t)_A\,,\footnotemark
 \end{align*}
 where $s$, $t$, $t_1$, and $t_2$ are numerical terms, $f$ is a function variable, and $A$ is an interval spec.\footnotetext{The rationale for the abbreviation $(D[f]\neq t)_A$ will become clearer in light of the intended semantics: it relies on the continuity of the derivative of $f$.} Similar syntactic definitions can also be given for the following expressions:\\ \centerline{\rule{0pt}{2.9ex} $s \leq t$, \quad $s \geq t_1/t_2$, \quad $s \leq t_1/t_2$, \quad and  \quad $s < t_1/t_2$.} 

\red{Also, expressions of the form $s = q$ are expressible in \RDFp, for any rational number $q$ ...} \eod
\end{definition}

Later on, we will show how terms of the form $t_1 / t_2$ can be introduced in arbitrary contexts.

\begin{remark}\label{remark:updownflex} 
Note that although we have chosen to regard them as primitive constructs, the relators
$\nor{\rm Up}$ and $\nor{\rm Down}$ are expressible in terms of differentiation, via the equivalences:
\[\begin{array}{rcl}
\nor{\rm Up}(f)_A   & \sse & (D[f]\geq \mathbf{0})_A \vel (e_1=e_2)\,,\\
\nor{\rm Down}(f)_A & \sse & (D[f]\leq \mathbf{0})_A \vel (e_1=e_2)\,,
\end{array}
\]
where $e_1$ and $e_2$ are the ends of $A\,$. {In each equivalence, if either $e_1=-\infty$ or $e_2=+\infty$, the second disjunct is omitted.}

Other constructs---easily introduced as derived relators---capture notions such as \emph{constancy}, \emph{linearity}, and 
\emph{pointwise upward and downward monotonicity}:

\begin{align*}
\nor{\rm Constant}(f)_{A} & \defFoAs~~  (D[f] = \mathbf{0})_{A} \,,\\[0.3cm]
\nor{\rm Linear}(f)_{A} & \defFoAs~~  \nor{\rm Concave}(f)_A \et \nor{\rm Convex}(f)_A\,,\\[0.3cm]
%
%
\nor{\rm Up}(f,s)   & \defFoAs~~  D[f](s) \geq \mathbf{0}\,, \\[0.3cm]
\nor{\rm Down}(f,s)   & \defFoAs~~  D[f](s) \leq \mathbf{0}\,, \\[0.3cm]
{\color{red}\nor{\rm Strict\_up}(f,c)_{[x,y]}} & {\color{red}\defFoAs~~ x < c \wedge c < y \wedge \nor{\rm Strict\_Up}(f)_{[x,y]}\,,} \\[0.3cm]
{\color{red}\nor{\rm Flex}(f,c)_{[x,y]}} & {\color{red}\defFoAs~~ }
  {\color{red}\big( x < c \wedge c < y \big) \wedge} \\[0.2cm]
  &\hspace{1.4cm} {\color{red}\Big(\big(
    \nor{\rm Strict\_Convex}(f)_{]x,c[} \wedge \nor{\rm Strict\_Concave}(f)_{]c,y[}\big)} \\
  &\hspace{2cm} {\color{red}\vel\ 
    \big(\nor{\rm Strict\_Concave}(f)_{]x,c[} \wedge \nor{\rm Strict\_Convex}(f)_{]c,y[}
  \big)\Big)\,.} \tag*{\eod}
\end{align*}
\end{remark}
\mim[inline]{ Ritengo che i due esempi evidenziati in rosso debbano essere rimossi, in quanto fanno riferimento a proprietà locali che non dovrebbero essere espresse in relazione a un intervallo $[x, y]$. In particolare, il secondo esempio---che dovrebbe caratterizzare la presenza di un flesso in $c$ di $f$---risulta scorretto, poiché la condizione proposta è solo sufficiente, ma non necessaria.\\
Tuttavia, nel contesto in cui vengono presentati---ovvero per illustrare l’espressività del linguaggio---sarebbe opportuno sostituirli con altre caratterizzazioni significative, ad esempio proprietà del tipo ${\mathrm{Linear}}(f)_A$, piuttosto che eliminarli del tutto. (Per il momento, ho aggiunto $\nor{\rm Constant}(f)_{A}$ e $\nor{\rm Down}(f,s)$ (duale di $\nor{\rm Up}(f,s)$).)}
}

\index{syntax of \RDFp|)}

\index{semantics of \RDFp|(}
\indent The semantics of \RDFp revolves around the
designation rules listed in our next definition, with which any truth-value \ita{assignment} for the formulas of \RDFp  must comply.

\begin{definition}\label{DEF:assignments}
An \textsc{assignment}\index{assignment (for \RDFp)} for \RDFp  is a mapping $M$ whose domain consists of all terms and formulas of \RDFp,   satisfying the following conditions:
\begin{enumerate}
\setcounter{enumi}{-1}
\item $M\mathbf{0}$ and $M\mathbf{1}$ denote the real numbers $0$ and $1$, respectively. 

Moreover, $M(-\infty) \putAs -\infty$ and $M(+\infty) \putAs +\infty$.

\item For each numerical variable $x$, $Mx$ is a real number. 

\item For every function variable $f$, the associated function $Mf$ is a real-valued function of a real variable, defined on the entire real line, differentiable  everywhere, and with a continuous derivative.

\item For each numerical term of the form $ t_1 \otimes t_2 $, with ${\otimes} \in \{+,-,\x \}$, define $M(t_1 \otimes t_2) \coloneqq Mt_1 \otimes Mt_2$.

\item For each numerical term of the form $f(t)$, define $M(f(t)) \coloneqq (Mf)(Mt)$.

\item For each numerical term of the form $D[f](t)$,  define $M(D[f](t)) \coloneqq (Mf)'(Mt)$.

\item For each interval specification $A$, $MA$ is an interval of $\mR $ of the appropriate kind, whose endpoints are the evaluations via $M$ of the ends of $A$.
\\
For example, when $A={]}s,t]$, then $MA={]}Ms, Mt ]$. Hence, in particular, if $Ms > Mt$, then $MA$ is the empty interval.

\item Truth values are assigned to formulas of \RDFp 
 according to the following rules (where $s,t$ stand for numerical terms, and $f,g$ for function variables):
 \begin{enumerate}[label={\bf\alph*}$)$]
 \item $s=t$ (resp., $s>t$) is true iff $Ms=Mt$ (resp., $Ms>Mt$) holds;
 \item\label{item:f=g} $(f=g)_A$ is true iff $(Mf)(x)=(Mg)(x)$ holds for all $x$ in $MA$;
 \item $(f>g)_A$ is true iff $(Mf)(x)>(Mg)(x)$ holds for all $x$ in $MA$;
 \item $(D[f]\bowtie t)_A$, with ${\bowtie} \in  \{ =, <, >, \leq, \geq \}$, is true iff $(Mf)'(x)\bowtie M t$ holds for all $x$ in $MA$;
 \item\label{item:monotF} $\nor{\rm Up}(f)_A$ (resp., $\nor{\rm Strict\_Up}(f)_A$) is true iff ($Mf$) is a monotone non-decreasing (resp., strictly increasing) function in $MA$;
 \item\label{item:convF} $\nor{\rm Convex}(f)_A$ (resp.,  $\nor{\rm Strict\_Convex}(f)_A$) is true iff ($Mf$) is a convex (resp., strictly convex) function in $MA$;
 \item\label{item:secondToLast} \begin{sloppypar}the truth values of $\nor{\rm Down}(f)_A\,$, $\nor{\rm Concave}(f)_A\,$, $\nor{\rm Strict\_Down}(f)_A\,$, and  $\nor{\rm Strict\_Concave}(f)_A$ are defined similarly to items \ref{item:monotF} and \ref{item:convF};\end{sloppypar}
 
 \item the truth value that $M$ assigns to any formula whose leading symbol is one of $\neg$, $\et$, $\vel$, $\longrightarrow$, and $\longleftrightarrow$ complies with the standard semantics of the corresponding propositional connectives. 
 \end{enumerate}
\end{enumerate}

An assignment $M$ is said to \textsc{model}\index{model (of a set of formulas)} a set $\Phi$ of formulas if $M\varphi$ is true for every $\varphi$ in $\Phi$.
\end{definition}

\begin{remark} 
Note that, for an interval spec $A$ and an assignment $M$, if $MA$ is the empty interval, then the atoms listed in items~\ref{item:f=g}--\ref{item:secondToLast} of  \Def{DEF:assignments} are vacuously true. Likewise, if $MA$ evaluates to a singleton, the atoms
\[
\nor{\rm Up}(f)_{[t_1,t_2]}\quad \text{and} \quad \nor{\rm Convex}(f)_{[t_1,t_2]},
\]
as well as their strict variants and the duals of all such atoms, are vacuously true. However, in this case, atoms of the form
\[
\begin{array}{llll}
(f = g)_{[t_1, t_2]},\quad (f > g)_{[t_1, t_2]},\quad \text{and} \quad (D[f] \bowtie t)_{[t_1, t_2]},
\end{array}
\]
with ${\bowtie} \in  \{ =, <, >, \leq, \geq \}$, \emph{can} evaluate to false. This is because these atoms are required to hold at every point in the interpreted interval, which, in the singleton case, reduces to $\{M(t_1)\}$; failure at that single point suffices to falsify the atom. \eod
\end{remark}

We now illustrate the expressive power of the \RDFp language by showing how it can formally capture various classical constructs and propositions from elementary real calculus.

\index{semantics of \RDFp|)}

\subsection{Expressing classical results in \RDFp} \label{subsec:examples}
To facilitate the expression of classical results in \RDFp, we begin by introducing a number of derived syntactic forms and relators that extend the basic language. These notational enhancements not only streamline the formulation of statements but also highlight the expressive power of \RDFp, making it suitable for capturing a wide range of elementary results from real calculus. Several such results are presented below.

\begin{definition}[Some derived expressions]\label{DEF:derivedExpressions}
For convenience, we enrich our language with some useful derived expressions, introduced through the following syntactic shorthands:
\begin{align*}
 \boldsymbol{n}  & \quad \defFoAs \quad \underbrace{\mathbf{1}+\cdots+\mathbf{1}}_{\mbox{\scriptsize $n$ times}} \qquad \text{(for every $n \geq 2$)}\,,\\
 s \neq t      & \quad \defFoAs \quad  (s>t) \vel (t > s) \,,\\
 s < t      & \quad \defFoAs \quad   t > s\:,                   \\
 s \geq t   & \quad \defFoAs \quad  (s>t) \vel (s=t) \,,   \\
 s = t_1/t_2  & \quad \defFoAs \quad   (t_1=s \x t_2) \et 
                              (t_2\x t_2> \mathbf{0})\,, \\
 s > t_1/t_2  & \quad \defFoAs \quad   \big((t_1 < s \x t_2) \et (t_2 > \mathbf{0})\big) \vel \big((t_1 > s \x t_2) \et (t_2 < \mathbf{0})\big)\,,    \\[0.09cm]
 (D[f]\neq t)_A  & \quad \defFoAs \quad (D[f]< t)_A \vel (D[f]> t)_A\,,\footnotemark
 \end{align*}
 where $s$, $t$, $t_1$, and $t_2$ are numerical terms, $f$ is a function variable, and $A$ is an interval spec.\footnotetext{The rationale for the abbreviation $(D[f]\neq t)_A$ reflects the intended semantics introduced before, particularly the assumption that the derivative of $f$ is continuous.} Similar syntactic definitions can also be given for the following expressions:
 \[
 s \leq t, \quad s \geq t_1/t_2, \quad s \leq t_1/t_2, \quad \text{and} \quad s < t_1/t_2.\tag*{\eod}
 \]
\end{definition}

\begin{remark}\label{remark:updownflex} 
Note that although we have chosen to regard them as primitive constructs, the relators
$\nor{\rm Up}$ and $\nor{\rm Down}$ are expressible in terms of differentiation, via the equivalences:
\[\begin{array}{rcl}
\nor{\rm Up}(f)_A   & \longleftrightarrow & (D[f]\geq \mathbf{0})_A \vel (e_1=e_2)\,,\\
\nor{\rm Down}(f)_A & \longleftrightarrow & (D[f]\leq \mathbf{0})_A \vel (e_1=e_2)\,,
\end{array}
\]
where $e_1$ and $e_2$ are the ends of $A\,$. {In each equivalence, if either $e_1=-\infty$ or $e_2=+\infty$, the second disjunct is omitted.}

Other constructs---easily introduced as derived relators---capture notions such as \emph{constancy}, \emph{linearity}, and 
\emph{pointwise upward and downward monotonicity}:
\begin{align*}
\nor{\rm Constant}(f)_{A} & \defFoAs~~  (D[f] = \mathbf{0})_{A} \,,\\[0.2cm]
\nor{\rm Linear}(f)_{A} & \defFoAs~~  \nor{\rm Concave}(f)_A \et \nor{\rm Convex}(f)_A\,,\\[0.2cm]
%
%
\nor{\rm Affine}(f)_{A} & \defFoAs~~ \nor{\rm Constant}(f)_{A} \ \vel \ \nor{\rm Linear}(f)_{A} \,,\\[0.2cm]
\nor{\rm Up}(f,s)_A   & \defFoAs~~  D[f](s) \geq \mathbf{0}\ \wedge \ e_1 < s \ \wedge \ s < e_2\,, \\[0.3cm]
\nor{\rm Down}(f,s)_A   & \defFoAs~~  D[f](s) \leq \mathbf{0}\ \wedge \ e_1 < s \ \wedge \ s < e_2\,, 
\COMMENT{
{\color{red}\nor{\rm Strict\_up}(f,c)_{[x,y]}} & {\color{red}\defFoAs~~ x < c \wedge c < y \wedge \nor{\rm Strict\_Up}(f)_{[x,y]}\,,} \\[0.2cm]
{\color{red}\nor{\rm Flex}(f,c)_{[x,y]}} & {\color{red}\defFoAs~~ }
  {\color{red}\big( x < c \wedge c < y \big) \wedge} \\[0.2cm]
  &\hspace{1.4cm} {\color{red}\Big(\big(
    \nor{\rm Strict\_Convex}(f)_{]x,c[} \wedge \nor{\rm Strict\_Concave}(f)_{]c,y[}\big)} \\
  &\hspace{2cm} {\color{red}\vel\ 
    \big(\nor{\rm Strict\_Concave}(f)_{]x,c[} \wedge \nor{\rm Strict\_Convex}(f)_{]c,y[}
  \big)\Big)\,.} \tag*{\eod}
}
\end{align*}
where $e_1$ and $e_2$ are the ends of $A$.\footnote{If $e_1$ is $-\infty$ (resp., $e_2$ is $+\infty$), then the conjunct $e_1 < s$ (resp., $s < e_2$) is dropped.}\eod
\end{remark}

\begin{remark} 
Note that any constant function $x \mapsto q$ over a given interval $A$, where $q$ is a rational constant, can be easily characterized by means of an \RDFp formula. For instance, when $q$ is a positive rational number $n/m$  and $A$ is the interval spec $]\!-\infty,+\infty[$, such a function can be described using the derived relator:
\[
\begin{array}{rcl}
(f=\frac{\boldsymbol{n}}{\boldsymbol{m}})_{]-\infty,+\infty[} &\putAs& (D[f]=\mathbf{0})_{]-\infty,+\infty[}\ \et\ f(\mathbf{0})= \frac{\boldsymbol{n}}{\boldsymbol{m}}\,.
\end{array}
\]
In analogous terms, one can express that $f$ is a first-degree polynomial with fixed rational coefficients. For example, the function $x\mapsto 2x-\frac{1}{3}$ can be defined as:
\[
\begin{array}{rcl}\big(f=2x-\frac{1}{3}\big)_{]-\infty,+\infty[}&\putAs&(D[f]= \mathbf{2})_{]-\infty,+\infty[}\ \et\ f(\mathbf{0})= \mathbf{0} - \frac{\mathbf{1}}{\mathbf{3}}\,.\end{array} \tag*{\eod}
\] 
\end{remark}

The expressive capabilities of the \RDFp language make it well suited to capture a wide variety of elementary results from real calculus. In particular, the combination of interval-based quantification and derived syntactic forms allows for succinct and rigorous formalizations of basic analytical facts. 

\subsubsection{Examples of expressible properties}\label{subsec:examples}
We next show that several basic facts of elementary real calculus can be expressed by \RDFp formulas. 
These translations not only capture the semantic content of classical theorems, but also allow their validity to be formally verified. 
Indeed, each \RDFp formula below is valid—that is, evaluates to true under every assignment $M$—and this can be mechanically checked using the decision algorithm described in \Sec{sec:decisionAlgorithm}.

\COMMENT{\begin{example}\label{exa:updownflex}
Note the following elementary equivalences:
\[\begin{array}{rcl}\nor{Up}(f)_A   & \leftrightarrow & (D[f]\geq 0)_A \vel (e_1=e_2)\,,\\
\nor{Down}(f)_A & \leftrightarrow & (D[f]\leq 0)_A \vel (e_1=e_2)\,,
\end{array}
\]
where $e_1,e_2$ are the ends of $A\,$.
\end{example}}

\begin{example}\label{exa:SCOp168} \emph{A strictly convex curve and a concave curve defined over the same interval $[a,b]$ can meet in at most two points.}

This can be formalized in the \RDFp language as:
\begin{multline*}
\Big[\,\nor{Strict\_Convex}(f)_{[a,b]} \et \nor{Concave}(g)_{[a,b]} \et
f(x_1) = g(x_1) \et f(x_2) = g(x_2) \et f(x_3) = g(x_3) \\
\et a \leq x_1 \leq b \et a \leq x_2 \leq b \et a \leq x_3 \leq b\,\Big]  
\longrightarrow (\, x_1 = x_2 \vee x_1 = x_3 \vee x_2 = x_3\,)\;.\tag*{\eod}
\end{multline*}
\end{example}

\begin{example}\label{exa:SCOp177} \emph{Let $f$ and $g$ be real functions that take the same values at the endpoints of a closed interval $[a,b]$. Assume also that $f$ is strictly convex on $[a,b]$ and that $g$ is linear on $[a,b]$. Then
$f(c) < g(c)$ holds at every point $c$ in the interior of the interval $[a,b]$.}

This is captured in \RDFp as:
\[
\Big[\ \nor{Strict\_Convex}(f)_{[a,b]} \et \nor{Linear}(g)_{[a,b]} \et f(a) = g(a) \et f(b) = g(b) \Big] ~\longrightarrow~ (f < g)_{]a,b[}\;.
\]

Note that the formula remains valid even if the premise $\nor{Linear}(g)_{[a,b]}$ is  weakened to $\nor{Concave}(g)_{[a,b]}$.
\eod
\end{example}

\begin{example}\label{exa:two_one} \emph{Let $f$ be a real function on $[a,b]$, with continuous derivative $f'$. If $f'(x)>0$ for all $x \in ]a,b[$, then $f$ is strictly increasing on $[a,b]$.}

This is captured in \RDFp as:
\[
\left(D[f]> 0\right)_{]a,b[}\ ~\longrightarrow~ \nor{Strict\_Up}(f)_{[a,b]}\;.
\] 
In \Subsec{subsec:decAlgorithmAtWork}, we illustrate how the decision algorithm can be used to establish the validity of this formula. \eod
\end{example}

\begin{example} 
\emph{If a differentiable function $f$ has constant derivative on $[a,b]$, then it is linear on $[a,b]$:}
\[
(D[f] = t)_{[a,b]} ~\longrightarrow~ \nor{Linear}(f)_{[a,b]}\,.
\tag*{\eod}
\]
\end{example}

\begin{example}
\emph{Let $f$ be a real function defined on the interval $[a,b]$, with continuous derivative $f'$, and suppose that $a < b$. If $f(a) = f(b)$, then there exists a point $c \in\, ]a,b[$ such that $f'(c) = 0$.}

This result, a weak version of \emph{Rolle's theorem}, can be expressed in the \RDFp\ language as follows:
\[
\left[a < b \et f(a) = f(b)\right] ~\longrightarrow~ \neg \left( D[f] \neq 0 \right)_{]a,b[} \;.
\tag*{\eod}
\]
\end{example}

\begin{example}
\emph{If $f$ is differentiable with continuous $f'$ on $[a,b]$, then for some $c \in ]a,b[$ we have $f'(c) = \frac{f(b) - f(a)}{b - a}$.}

This result, a weak version of \emph{Lagrange's mean value theorem}, can be expressed in the \RDFp\ language as follows:
\[
\left[a < b \et t = \dfrac{f(b) - f(a)}{b - a} \right] ~\longrightarrow~ \neg \left( D[f] \neq t \right)_{]a,b[} \;.
\tag*{\eod}
\]
\end{example}

\begin{example}
\emph{Let $f$ and $g$ be real functions defined on the interval $[a,b]$, differentiable with continuous derivatives. If $f'(x) > 0$ and $g'(x) < 0$ for all $x \in\, ]a,b[\,$, and $f(a) = g(a)$, then $f(x) > g(x)$ for all $x \in\, ]a,b]$:}
\[
\left[\, (D[f] > 0)_{]a,b[} \et (D[g] < 0)_{]a,b[} \et f(a) = g(a) \,\right] ~\longrightarrow~ \left( f > g \right)_{]a,b]} \;.
\tag*{\eod}
\]
\end{example}

\begin{example}
\emph{If $f$ and $g$ are $C^1$ on $[a,b]$ and satisfy $f(a) = g(a)$, $f'(a) = g'(a)$, with $f$ strictly convex and $g$ concave, then $f(x) > g(x)$ for all $x \in ]a,b]$:}
\[
\Big[\, f(a) = g(a) \et D[f](a) = D[g](a) \et 
\nor{Strict\_Convex}(f)_{[a,b]} \et \nor{Concave}(g)_{[a,b]}\, \Big] 
~\longrightarrow~ (f > g)_{]a,b]} \;.
\tag*{\eod}
\]
\end{example}

\begin{example}
\emph{If $f$ and $g$ are linear and agree at two distinct points $a \neq b$, then they are equal everywhere:}
\[
\Big[\ \nor{Linear}(f)_{]-\infty,+\infty[} \et \nor{Linear}(g)_{]-\infty,+\infty[} \et f(a) = g(a) \et f(b) = g(b) \et a \neq b \ \Big] 
~\longrightarrow~ (f = g)_{]-\infty,+\infty[} \;.
\tag*{\eod}
\]
\end{example}

\begin{example}
\emph{If $f$ is strictly increasing and $g$ is strictly decreasing on $\mathbb{R}$, then $f(x) = g(x)$ has at most one solution:}
\[
\Big[\ \nor{Strict\_Up}(f)_{]-\infty,+\infty[} \et \nor{Strict\_Down}(g)_{]-\infty,+\infty[} \et 
f(x) = g(x) \et f(y) = g(y) \ \Big] ~\longrightarrow~ x = y\;.
\tag*{\eod}
\]
\end{example}

\begin{example}
\emph{Let $f$ be a $C^1$ function defined on $[a,b]$ such that $f'$ is non-negative on $[a,c]$ and non-positive on $[c,b]$, for some point $c \in {]}a , b[$. Then $f(c)$ is a maximum of $f$ on $[a,b]$:}
\[
\Big[\ a < c < b \et (D[f] \geq 0)_{[a,c]} \et (D[f] \leq 0)_{[c,b]} \et a \leq x \leq b \ \Big] ~\longrightarrow~ f(c) \geq f(x)\;.
\tag*{\eod}
\]
\end{example}

\begin{example}
\emph{Let $f$ be a $C^1$ convex function on $[a,b]$ such that $f'(c) = 0$ for some $c \in [a,b]$. Then $f(c)$ is a minimum of $f$ on $[a,b]$:}
\[
\Big[\, a \leq c \leq b \et D[f](c) = 0 \et \nor{Convex}(f)_{[a,b]} \et a \leq x \leq b \,\Big] 
~\longrightarrow~ f(c) \leq f(x)\;.
\tag*{\eod}
\]
\end{example}

Two of the above examples of valid \RDFp formulas—namely, \Exas{exa:SCOp168}{exa:SCOp177}—which do not involve the differentiation operator $D[\,\sqdot\,]$, appear in \cite[\pp{168,177}]{SCO11}. In that context, they are expressed in a richer language, \RMCFp, whose intended universe of functions is broader than that of \RDFp. In particular, functions in \RMCFp are assumed to be continuous, but not necessarily differentiable. Accordingly, any formula deemed valid by the decision procedure for \RMCFp remains valid in \RDFp, and its negation will be judged unsatisfiable by the decision algorithm described next (see~\cite[Footnote~1]{BCCOS23}).

\section{The Decision Algorithm}\label{sec:decisionAlgorithm}

When dealing with a quantifier-free language such as \RDFp, which is closed under propositional connectives, determining the validity of a formula $\psi$ amounts to establishing the unsatisfiability of its negation $\neg\psi$. Moreover, checking the satisfiability of $\neg\psi$ reduces to checking whether at least one disjunct in a disjunctive normal form of $\neg\psi$ is satisfiable. Thus, the central issue concerning the decidability of \RDFp becomes: how can we determine whether a given conjunction of \RDFp \emph{literals}---that is, \RDFp atoms or their negations---is satisfiable? 

Via routinary flattening techniques and exploiting the eliminability of the $\nor{Up}/\nor{Down}$ predicates (see \Rem{remark:updownflex}), we can reduce each instance of this problem to checking the satisfiability of arbitrary conjunctions $\fhi_0$ of literals, each of which is either an atom of the following forms or its negation:
\newcommand{\mypm}{}
\[\label{array:FlatForm}
\begin{array}{rccrccrccrc}
z = x+y & , & \phantom{xx} & \mypm (f=g)_A \COMMENT{(h=f+g)_A} &, & \phantom{xx} & \mypm \nor{Strict}\_\nor{Up}(f)_A & , & \phantom{xx} & \mypm \nor{Convex}(f)_A &,\\
z = x \cdot y  &,  & & \mypm (f>g)_A &,&& \mypm \nor{Strict}\_\nor{Down}(f)_A &,  & & \mypm \nor{Concave}(f)_A  &,\\
x > y&,&& z=f(x) &,&&\mypm (D[f] \bowtie z)_A &,&& \mypm \nor{Strict}\_\nor{Convex}(f)_A &,\\
& & &              &      &               &z = D[f](x)&,&& \mypm \nor{Strict}\_\nor{Concave}(f)_A  &,\end{array}\tag{\boldmath${\ddagger}$}
\]
where ${\bowtie} \in  \{ =, <, >, \leq, \geq \}$.
Here, 
$A$ denotes an interval spec whose ends may be numerical variables or extended reals $\{-\infty$, $+\infty\}$, and, as usual, $x,y,z$ represent numerical variables, and $f,g\COMMENT{,h}$ represent function variables.\footnote{Note that disequalities of the form $z \neq x+y$, $z \neq x\cdot y$, and so on, can be rewritten, as shown in \Def{DEF:derivedExpressions}, using inequalities. Furthermore, negated inequalites such as $\neg (x > y)$ can be expressed as the disjunction $x < y \vel x = y$.}

The family $\RDFpFlat$ of all conjunctions of \RDFp literals of the form~$(\ddagger)$ constitutes a significant subtheory of \RDFp, as captured by the following result:
\begin{lemma}\label{lem:stdForm}
The theory $\RDFp$ is decidable if and only if its fragment $\RDFpFlat$ is decidable.
\end{lemma}

Suppose, then, that we are given a formula $\fhi_0$ belonging to $\RDFpFlat\,$. Through a process that may branch at various points, $\varphi_0$ will undergo a sequence  of transformations
\[
\fhi_0 \leadsto {\fhi}_1 \leadsto {\fhi}_2 \leadsto {\fhi}_3 \leadsto {\fhi}_4 =\widehat{\fhi},
\]
ultimately yielding a formula $\widehat{\fhi}$ in which no function variables appear. This final formula $\widehat{\fhi}$ can be tested for satisfiability by means of Tarski's celebrated decision procedure~\cite{GC 75,AT 51}. The proper functioning of this method rests on certain assumptions about the detailed structure of $\fhi_0$, easy to ensure, which we clarify through the following definitions:
\begin{definition}\label{def:domainVars} 
Let $\varphi$ be a formula in $\RDFpFlat$, that is, a conjunction of \RDFp literals of the forms shown in~\eqref{array:FlatForm}.

A \textsc{domain variable} in $\fhi$ is a numerical variable $x$
that occurs in $\fhi$ either as
\begin{itemize}
  \item the argument of a term of the form $f(x)$ or $D[f](x)$, where $f$ is a function variable, or
  \item an endpoint of an interval appearing in $\varphi$ (\/for instance, as in $\nor{Convex}(f)_{[x, +\infty[}$\/). \eod
\end{itemize}
\end{definition}

\begin{definition}\label{def:orderedform} 
An $\RDFpFlat$ \ formula $\fhi_0$ is said to be in \textsc{ordered form} if it has the shape:
\[
\fhi\ \ \et\ \  \bigwedge_{i=1}^{n-1} x_i<x_{i+1},
\]
where $\{x_1,\dots,x_n\}$ is the set of all distinct domain variables in $\fhi$. 
\eod\end{definition}

The family $\RDFpOrd$ \ of all ordered formulas of $\RDFpFlat$ constitutes a significant subtheory of \RDFp, as shown by the following result:
\begin{lemma}\label{lem:ordForm}
$\RDFpFlat$ \ is decidable if and only if $\RDFpOrd$ \ is decidable.
\end{lemma}

Note that the (omitted) proofs of \Lems{lem:stdForm}{lem:ordForm} rely on effective syntactic rewritings.

\begin{remark}\label{rem:open}
Excluding literals of the types $(f>g)_A$, $(D[f]>z)_A$, and $(D[f]<z)_A$ (cf. step~\ref{itemAlgoOne}.\ref{alg1:actionTwo}) in the algorithm below), it will suffice to consider only  \emph{closed} intervals $A$. Indeed, by continuity, the literals not listed above hold on an open or half-open interval if and only if they hold on its closure; for example, $(f = g)_{]w_1, w_2[}$ holds if and only if $(f = g)_{[w_1, w_2]}$ holds. \eod
\end{remark}

\COMMENT{
When one deals with an unquantified language such as \RDFp, which is closed with respect to propositional connectives, being able to
determine algorithmically whether or not a formula is valid amounts to establishing whether the negation thereof
is \emph{satisfiable} or \emph{unsatisfiable}. We prefer to address the satisfiability problem for \RDFp in what follows, and our algorithm
will produce a yes/no answer, where `yes' means that $\fhi$ admits a model. Hence, indirectly, if we were to test a formula for validity, 
a negative response would mean that a counterexample exists.



   The idea is to transform, through a finite number of steps, the given \RDFp  formula $\fhi$ to be tested for satisfiability into a finite collection of formulas $\psi_i$, still devoid of quantifiers, each belonging to the elementary algebra of real numbers; this will be done so that $\fhi$ is satisfiable
   if and only if at least one of the resulting $\psi_i$s is satisfiable. Each resulting $\psi_i$ can be tested via Tarski's decision algorithm. 
   
   First we discuss how to reduce our formula $\fhi$ to a particular format, called \ita{ordered form}; second we explain the algorithm.

\subsubsection*{\centerline{Normalization}} 
Let $T$ be an unquantified, possibly multi-sorted, first-order theory, endowed with: equality $=$,
a denumerable infinity of individual variables $x_1,\ x_2,\dots $, function symbols $F_1,\ F_2,\dots $, and predicate symbols $P_1,\ P_2,\dots$.

\begin{definition}\label{def:flattenedForm} A formula $\fhi$ of $T$ is said to be in \textsc{normal form} if:
\begin{enumerate}[label=(\arabic*)] 
\item\label{flattening_one} every term occurring in $\fhi$ either is an individual variable or has the form $F(x_1,\ x_2,\dots ,\ x_n)$, where $x_1,\ x_2,\dots ,\ x_n$ are individual variables and $F$ is a function symbol; 
\item\label{flattening_two} every atom in $\fhi$ either has the form $x=t$, where $x$ and $t$ are an individual variable and a term, or has the form  $P(x_1,\ x_2,\dots ,\ x_n)$, where $x_1,\ x_2,\dots ,\ x_n$ are individual variables and $P$ is a predicate symbol.
\end{enumerate}
\end{definition}

Via routinary flattening techniques, one proves the following:
\begin{lemma} There is an effective procedure to transform any formula $\fhi$ of $T$ into a formula $\psi$ in normal form so that $\fhi$ and $\psi$ are equisatisfiable.
\end{lemma}
\dimostraz{To translate $\fhi$ into $\psi$, we proceed in two stages as follows. 

\begin{sloppypar}Repeatedly, as long as an atom of either the form $t_s \! =\! t$ or the form $P(t_1 , \dots , t_{s-1}, t_s , t_{s+1}, \dots ,t_n )$ occurs in $\fhi$, where $t_1 , \dots ,t_n , t$ are terms and $t_s$ is not a variable, we pick a variable $x^{t_s}$ not occurring in $\fhi$, of the same sort as $t_s$. Then we rewrite $\fhi$ as the conjunction $x^{t_s} = t_s\et\fhi'$ where $\fhi'$ is the formula resulting from the earlier $\fhi$
  through replacement of the variable $x^{t_s}$ to all occurrences of the term $t_s$. When $\fhi$ 
  is no longer rewritable, it will satisfy condition \ref{flattening_two} of \Def{def:flattenedForm}.\end{sloppypar}

Next, as long as a term $t_s$ which is not a variable occurs in a term \\\centerline{$F(t_1, \dots ,t_{s-1}, t_s ,t_{s+1} ,\dots ,t_n)$}
 inside $\fhi$, we pick a new variable $x^{t_s}$ of the same sort as $t_s$. Much as in the preceding stage, we rewrite $\fhi$ as the conjunction $x^{t_s} = t_s\et\fhi'$ where $\fhi'$ results from the earlier $\fhi$ when all occurrences of the term $t_s$ are replaced by $x^{t_s}$. At the end $\varphi$ will also satisfy condition \ref{flattening_one} of \Def{def:flattenedForm}, hence it can serve as the sought $\psi$. \qed}

A more restrained format than just ``normal form'' is defined here:
\begin{definition} A formula $\fhi$ of $T$ is said to be in \textsc{standard normal form} if it is a conjunction of literals of the forms:
\[\begin{array}{ccc}
x=y\,, & x=F(x_1,\dots ,x_n)\,, & x\neq y\,,\\[0.09cm]
\multicolumn{3}{c}{\begin{array}{ccc}
P(x_1,\dots ,x_n)\,, &&\neg P(x_1,\dots ,x_n)\,,
\end{array}}\end{array}\]
where $x,y,x_1,\dots ,x_n,$ stand for individual variables, $F$ for a function symbol, and $P$ for a predicate symbol.
\end{definition}

Let $S$ be the class of all formulas of $T$ in standard normal form; the following holds:
\begin{lemma}\label{lem:was3.4}
$T$ is decidable if and only if $S$ is decidable.
\end{lemma}
\dimostraz Clearly any algorithm for formulas in $T$ is also an algorithm for formulas in $S$. For the converse, suppose that an algorithmic satisfiability test for $S$ is available, and let $\fhi$ be any formula of $T$. By applying the normalization process to $\fhi$, we get a formula $\psi$, in normal form such that $\fhi$ and $\psi$ are  equisatisfiable. We can now bring $\psi$ to disjunctive normal form, thus obtaining a formula ${\psi}_1 \vel \cdots \vel {\psi}_\kappa$ where all ${\psi}_i$s are conjunctions, and we may assume that each ${\psi}_i$ is in standard normal form, because any literal of type $\neg \, x \! =\! F(y_1 , \dots ,y_n )$ within it  can be replaced by the conjunction $ \neg \, x=z \et z=F(y_1, \dots , y_n )$, where $z$ is a brand new variable. Since
\[
\fhi \ \ \mbox{is satisfiable} \sse \ \psi \ \mbox{is satisfiable} \ \sse \ {\psi}_i \ \mbox{is satisfiable for some}\ i
\]
and since all transformations used to build conjunctions ${\psi}_1 ,\dots , {\psi}_\kappa $ are effective, our lemma follows. \qed   

We are now almost ready to define an \emph{ordered form} for \RDFp  formulas.
\begin{definition} A \textsc{domain variable} in a formula $\fhi$ of \RDFp  is a numerical variable $x$
which occurs in $\fhi$ either as the argument of a term of one of the forms $f(x)$, $D[f](x)$, with
$f$ a function variable, or as an end of some interval mentioned in $\fhi$ (as exemplified by
$\nor{Convex}(f)_{[x,+\infy ]}$).
\end{definition}
\begin{definition}\label{def:orderedform} An \RDFp  formula is said to be in \textsc{ordered form} if it is in standard normal form and
has the form:
\[
\fhi\ \ \et\ \  \bigwedge_{i=1}^{n-1} (x_i<x_{i+1}),
\]
where $\{x_1,\dots,x_n\}$ is the set of all distinct domain variables in $\varphi$. 
\end{definition}
The family $\RDFpOrd$ of all ordered formulas of \RDFp  is a proper subset of \RDFp,   nevertheless the following holds:
\begin{lemma}\label{lem:ordForm}
\RDFp  is decidable if and only if $\RDFpOrd$ is decidable.
\end{lemma}
\dimostraz omitted. \qed
} 

\subsection{The algorithm}\label{subsec:algorithm}
We next describe the satisfiability decision algorithm for formulas of \RDFp.  
In view of \Lem{lem:ordForm}, we may assume, without loss of generality, that the \RDFp formula $\fhi_0$ to be tested for satisfiability is given in ordered form. As already announced, the algorithm reduces $\fhi_0$ through a sequence of transformations,
\[
\fhi_0 ~~\leadsto~~ \fhi_1 ~~\leadsto~~ \fhi_2 ~~\leadsto~~ \fhi_3 ~~\leadsto~~ \fhi_4 = \widehat{\fhi},
\]
ultimately producing a formula $\widehat{\fhi}$ such that:
\begin{enumerate}
\item $\fhi_0$ and $\widehat{\fhi}$ are equisatisfiable;\vspace{0.15cm}
\item $\widehat{\fhi}$ is a 
Tarskian formula, that is, a formula involving only numerical variables, the arithmetic operations $+$ and $\x$, and the predicates $=$ and $<$.
\end{enumerate}
A decision algorithm for \RDFp thus results from integrating Tarski's decision algorithm with the reduction $\fhi_0\leadsto\widehat{\fhi}$ described below.\footnote{By abuse of notation, we also use $\fhi_0 \leadsto \widehat{\fhi}$ to refer to the entire sequence of transformations from $\fhi_0$ to $\widehat{\fhi}$.} 

\medskip

Roughly speaking, the transformations $\fhi_{i-1}\leadsto\fhi_i$ ($i=1,2,3,4$) serve the following purposes:
\begin{enumerate}
\item Subdivide into cases each literal of the form $(f>g)_A$\,, $(D[f]>t)_{A}$\,, or $(D[f]<t)_{A}$ where $A$ is \emph{not} a closed interval of the form $[v,w]$. For example, the literal $(f>g)_{]v,w]}$ requires a case distinction on the left end, considering either $f(v)>g(v)$ or $f(v)=g(v)$, with each case requiring a separate treatment.

\item Replace each negative literal with an equivalent implicit existential formulation. For example, the negation $\neg \nor{Strict}\_\nor{Up}(f)_{[v,w]}$ introduces new variables
$x,y,x',y'$ subject to constraints 
\[
v\leq x<x'\leq w, \qquad y=f(x), \qquad  y'=f(x'), \qquad y\geq y'.
\]
This substitution captures the failure of strict monotonicity by asserting the existence of two points in $[v, w]$ where $f$ does not increase.

\item For each domain variable $v_j$ in $\varphi_0$, introduce new variables $y_j^f$ and $t_j^f$ for each function variable $f$, together with the constraints:
\[
y_j^f = f(v_j), \quad t_j^f = D[f](v_j)\,.
\]
These variables explicitly capture the value and the derivative of $f$ at $v_j$, enabling the elimination of function terms in subsequent transformations.

\item Eliminate all literals involving function variables whose graphs have already been encoded by the variables $y_j^f$ and $t_j^f$ introduced above. This elimination phase involves introducing additional variables, each subject to appropriate constraints, so as to replace the functional expressions by purely numerical conditions.
\end{enumerate}

\begin{center}------------\end{center}

In the following, $w_i$ denotes a numerical variable, whereas $z_i$ denotes an extended numerical variable.

The series of transformations we need goes as follows:
\begin{enumerate}[label=\arabic*., ref=\arabic*]

\item\label{itemAlgoOne} $\fhi_0 \leadsto {\fhi}_1$~~(\textsc{behavior at the endpoints})

Certain information implicit in literals of the forms $(f > g)_A$ and $(D[f] > t)_A$ is made explicit through the enrichments described below.

 \begin{enumerate}[label=\alph*), ref=\alph*]

 \item\label{alg1:actionOne}  We  rewrite each atom of the form 
 \[
  (f>g)_{]-\infy , w_2[}\;,
 \]
 where $f,\, g$ are function variables and $w_2$ is a numerical variable, as the conjunction
 \[
 (f>g)_{]-\infy, w_1]}\ ~\et~ \ (f>g)_{[w_1,w_2[}\  ~\et~ \ w_1\! <\! w_2\,,
 \]
 where $w_1$ is either the domain variable that immediately precedes $w_2$ in the ordering of domain variables (if such a variable exists), or a newly introduced domain variable otherwise.\footnote{Note that, here as well as in steps~\ref{itemAlgoOne}.\ref{alg1:actionTwo}\textbf{)}, \ref{itemAlgoOne}.\ref{alg1:actionThree}\textbf{)}, and \ref{itemAlgoOne}.\ref{alg1:actionFour}\textbf{)} below, whenever a new domain variable is introduced, its position with respect to the ordering of domain variables (see \Def{def:orderedform}) is uniquely determined.}
 
 We also perform the symmetric rewriting:
 {\[
 (f>g)_{]w_1,+\infy [} ~\leadsto~ (f>g)_{]w_1, w_2]}\ \et \ (f>g)_{[w_2, +\infy [}\ \et \ w_1<w_2\,.
 \]}
 
 Thanks to the rewritings just performed, every comparison between functions now refers either to a closed interval or to a bounded interval. (The purpose of the rewritings at step~\ref{alg1:actionThree}{\bf)} will be analogous.)
 
 \item\label{alg1:actionTwo} Let $a, b$ be real numbers such that $a\! <\! b$, and let $f$ and $g$ be real continuous functions on the closed interval $[a,b]$. Then  the inequality $f\! >\! g$ holds throughout the open interval $]a,b[$ if and only if one of the followings holds:\\
  \begin{tabular}{rl}
  i. & $f>g$ holds over $[a,b]$; \\
  ii. & $f>g$ holds over $[a,b[$, and $f(b)=g(b)$; \\
  iii. & $f>g$ holds over $]a,b]$, and $f(a)=g(a)$; \\
  iv. & $f>g$ holds over $]a,b[$, and $f(a)=g(a)$ and $f(b)=g(b)$.
  \end{tabular}
  
  \begin{enumerate}[label=\ref{alg1:actionTwo}$_\arabic*$), ref={\ref{itemAlgoOne}.{\ref{alg1:actionTwo}$_\arabic*$})}]
  \item\label{alg1:action2:case1} We can therefore rewrite any conjunct of the following forms
 \[
 (f>g)_{]w_1, w_2[}\,, \quad (f>g)_{[w_1, w_2[}\,, \quad \text{and} \quad (f>g)_{]w_1, w_2]}
 \]
 as an equivalent disjunction comprising four or two alternatives. In particular:
 {\footnotesize\[\begin{aligned}
  (f>g)_{]w_1, w_2[}&\leadsto (f>g)_{[w_1, w_2]}\ \vel\ \big[(f>g)_{[w_1, w_2[} \et f(w_2)=g(w_2) \big]\\\
 &  \phantom{\leadsto (f>g)_{[w_1, w_2]}~ }\ \vel \big[(f>g)_{]w_1, w_2]} \et f(w_1)=g(w_1) \big]\ \\
 & \phantom{\leadsto (f>g)_{[w_1, w_2]}~ }\ \vel \big[(f>g)_{]w_1, w_2[} \et f(w_1)=g(w_1) \et f(w_2)=g(w_2)\big]\,,\\[.1cm]
 (f>g)_{[w_1, w_2[}&\leadsto(f>g)_{[w_1, w_2]}\vel \big[(f>g)_{[w_1, w_2[} \et f(w_2)=g(w_2) \big]\,, \\[0.1cm]
 (f>g)_{]w_1, w_2]}&\leadsto(f>g)_{[w_1, w_2]}\vel \big[(f>g)_{]w_1, w_2]} \et f(w_1)=g(w_1) \big]\,.
    \end{aligned} \]}
 
  \item\label{alg1:action2:case2} Each such rewriting disrupts the structure of the overall formula. To restore it, we bring the formula again to disjunctive normal form by applying the distributive law:
  \[
  (\al \vel \be ) \et \ga \ \sse \ (\al \et \ga ) \vel ( \be \et \ga )\,.
  \]
  Subsequently, the algorithm proceeds by treating each disjunct separately.

  \item\label{alg1:action2:case3} Let $w_1, w_2$ be numerical variables and let $f,g$ be function variables. Consider any disjunct $\delta_i$ in which the literals $(f > g)_{]w_1, w_2[}$, $f(w_1) = g(w_1)$, and $f(w_2) = g(w_2)$ occur together. Assume that $w_1 < w_2$ in the ordering of domain variables and that there are no domain variables strictly between $w_1$ and $w_2$. In such a case, we enrich the disjunct by adding the literals
  \[
  w_1 < w,\quad w < w_2,\quad \text{and} \quad f(w) = z,
  \]
  where $w$ and $z$ are new numerical variables (with $w$ treated as a domain variable).
  Clearly, the formula thus obtained is equisatisfiable with the original one.
  \end{enumerate}

 \item\label{alg1:actionThree} We next rewrite each atom of the form
 \[
  (D[f]>t)_{]-\infy , w_2[}\;,
 \]
 where $f$ is a function variable and $t$, $w_2$ are numerical variables, as the conjunction
 \[
 (D[f]>t)_{]-\infy, w_1]}\ \et \ (D[f]>t)_{[w_1,w_2[}\  \et \ w_1\! <\! w_2\,,
 \]
 where $w_1$ is the first domain variable preceding $w_2$ (according to the ordering of domain variables), if such a variable exists; otherwise, $w_1$ is a newly introduced domain variable.
\\ We also perform the symmetric rewriting:
 {\footnotesize\[
 (D[f]>t)_{]w_1,+\infy [} \quad \leadsto \quad (D[f]>t)_{]w_1, w_2]}\ \et \ (D[f]>t)_{[w_2, +\infy [}\ \et \ w_1<w_2\,,
 \]}
 and proceed analogously in the two corresponding cases:
 \[
 (D[f]<t)_{]-\infy , w_2[} \quad \text{and} \quad (D[f]<t)_{]w_1, +\infy [}\,.
 \]
 Each of these rewritings yields a formula that is equisatisfiable with the original one.
 
 \item\label{alg1:actionFour} Let $a$, $b$, and $t$ be real numbers, and let $f\! \in C^1 ([a,b])$. Then $f'>t$ on $]a,b[$ if and only if one of the following conditions holds:\\
 \begin{tabular}{rl}
 i. & $f'>t$ in $[a,b]$,\\
 ii. & $f'>t$ in $[a,b[$ and $f'(b)=t$, \\
 iii. & $f'>t$ in $]a,b]$ and $f'(a)=t$, \\
 iv. & $f'>t$ in $]a,b[$, $f'(a)=t$ and $f'(b)=t$. 
 \end{tabular}
 
 The rewriting to be performed in this case is closely analogous to the one described in step~\ref{itemAlgoOne}.\ref{alg1:actionTwo}), i.e.:
  \begin{enumerate}[label=\ref{alg1:actionFour}$_\arabic*$), ref={\ref{itemAlgoOne}.{\ref{alg1:actionFour}$_\arabic*$})}]
  
  \item\label{alg1:action4:case1} We rewrite conjuncts of the forms
  \[
  (D[f]>t)_{]w_1,w_2[}\,,\quad (D[f]>t)_{[w_1,w_2[} \,, \quad \text{and} \quad (D[f]>t)_{]w_1,w_2]}
  \]
  as equivalent disjunctions. For instance: 
  \[
  \begin{array}{rcl}
  (D[f]>t)_{]w_1,w_2]}&\leadsto& (D[f]>t)_{[w_1,w_2]} \ \vel\ \ \big((D[f]>t)_{]w_1,w_2]} \ \et \ D[f](w_1)=t \big)\,.
  \end{array}
  \]
  
  \item\label{alg1:action4:case2} We bring the overall formula  back into disjunctive normal form, taking into account the distributive law
  \hbox{$(\al \vel \be ) \et \ga \ \sse \ (\al \et \ga ) \vel ( \be \et \ga )$} into account.
  
  \item\label{alg1:action4:case3} If three literals $(D[f]>t)_{]w_1, w_2[}$, $D[f](w_1)=t$, and $D[f](w_2)=t$ occur together in a formula, and if $w_1\! <\! w_2$ as ordered domain variables with no domain variables in between, then we add the literals $w_1 \! < w$, $w\! < w_2$, and $f(w)=z$, where $w$ and $z$ are new numeric variables (with $w$ treated as a domain variable).

  The treatment of literals of the form $(D[f]<t)_{]w_1, w_2[}$ proceeds analogously.\footnote{Literals of the forms $(D[f]\leq t)_{A}$\/, $(D[f]\geq t)_{A}$, and $(D[f]= t)_{A}$ require no similar treatment.} 
  \end{enumerate}
 \end{enumerate}
 By applying the rewriting steps~\ref{alg1:actionOne}) through~\ref{alg1:actionFour}) to a formula $\psi$ in ordered form, we obtain a finite disjunction of formulas ${\psi}_i$ such that $\psi$ is satisfiable if and only if at least one of the ${\psi}_i$ is satisfiable.
 
Each formula $\psi_i$ is then processed independently by the remainder of the algorithm.

\smallskip
  
\item\label{itemAlgoTwo} ${\fhi}_1 \leadsto {\fhi}_2$~~(\textsc{negative-clause removal})

From ${\fhi}_1$ we construct an equisatisfiable formula ${\fhi}_2$ that contains only positive predicates. The general idea of this step is to substitute each negative clause involving a function symbol with an implicit existential assertion.

For simplicity:
 \begin{itemize}
  \item  $x,x_1,x_2,x_3,y_1,y_2,y_3$ will denote numerical variables, distinct from those in the original formula;
  
  \item  we use the notation $x\! \preceq \! y$ as shorthand for $x\! \leq \! y$ when $x,\ y$ are numerical variables; otherwise
  (i.e., when $x = -\infty$ or $y = +\infy$), $x\! \preceq \! y$ stands for $0=0$.
  \end{itemize}
  \begin{enumerate}[label=\alph*), ref=\ref{itemAlgoTwo}.\alph*)]

 \item\label{alg2:actionA}  Replace any literal $\neg (f=g)_{[z_1,z_2]}$ occurring in ${\fhi}_1$ with:
 \[
 (z_1 \preceq x \preceq z_2) \et y_1=f(x) \et y_2=g(x) \et \neg (y_1=y_2).
 \]
 
 \item\label{alg2:actionB} Replace any literal $\neg (f>g)_{ [z_1,z_2]}$ occurring in ${\fhi}_1$ with:
 \[
 (z_1 \preceq x \preceq z_2) \et y_1=f(x) \et y_2=g(x) \et (y_1\leq y_2).
 \]
 \item\label{alg2:actionC} Replace any literal $\neg (D[f] \bowtie t)_{[z_1,z_2]}$, with ${\bowtie}  \in \{ <, \leq , =, >, \geq \}$, occurring in ${\fhi}_1$ with:
 \[
 (z_1 \preceq x \preceq z_2) \et y_1=D[f](x) \et  \neg (y_1 \bowtie t). 
 \]
 \item\label{alg2:actionD} Replace any literal $\neg$Strict\_Up $(f)_{[z_1,z_2]}$ (resp. $\neg$ Strict\_Down $(f)_{[z_1,z_2]}$) occurring in ${\fhi}_1$ with:
 \\\centerline{$
 \Gamma \et y_1 \geq y_2\ \ \ \ (\mbox{resp.}\  \leq )\,,
$}

\noindent where
 \[
 \Gamma \putAs (z_1 \preceq x_1 < x_2 \preceq z_2) \et y_1=f(x_1) \et y_2=f(x_2).
 \]
 \item\label{alg2:actionE} Replace any literal $\neg $ Convex$(f)_{[z_1,z_2]}$ (resp. $\neg $ Strict\_Convex $(f)_{[z_1,z_2]}$) occurring in ${\fhi}_1$ with:
 \[
 \Gamma \et (y_2-y_1)(x_3-x_1)>(x_2-x_1)(y_3-y_1)
 \hspace{0.5cm}(\mbox{resp.}\ \geq )\,,
 \]
 where
 \[
 \Gamma \putAs (z_1 \preceq x_1 < x_2 < x_3 \preceq z_2) \et y_1=f(x_1) \et y_2=f(x_2) \et y_3=f(x_3).
 \]
 Literals of the forms $\neg\, \nor{Concave}(f)_{[z_1,z_2]}$, $\neg\, \nor{Strict\_Concave}(f)_{[z_1,z_2]}$ are handled similarly.
 \end{enumerate}
With only slight modifications, the same treatment applies to literals over open and semi-open intervals. For example:
\[
\neg(f > g)_{]w_1, w_2]} \leadsto (w_1 < x \leq w_2) \et y_1 = f(x) \et y_2 = g(x) \et (y_1 \leq y_2).
\]
 
Equisatisfiability of the ${\fhi}_1$ and ${\fhi}_2$ is straightforward to establish.
Invoking \Lems{lem:stdForm}{lem:ordForm}, we can normalize ${\fhi}_2$ to obtain an equivalent formula in ordered form with domain variables $v_1,v_2,\dots ,v_r$.

\smallskip

\item\label{itemAlgoThree} ${\fhi}_2 \leadsto {\fhi}_3$~~(\textsc{explicit evaluation of function variables})

This step prepares for the elimination of functional clauses by explicitly evaluatiing function variables at domain variables.
For each such variable $v_j$ and for every function variable $f$ occurring in ${\fhi}_2$, introduce two new numerical variables $y^f_j$ and $t^f_j$, and add the literals $y^f_j \! =\! f(v_j)$ and $t^f_j \! =\! D[f](v_j)$ to ${\fhi}_2$. Moreover, for each literal of the form $x\! =\! f(v_j)$ already in ${\fhi}_2$, add the literal $x\! =\! y^f_j$; and for each literal $x\! =\! D[f](v_j)$ already in ${\fhi}_2$, insert the literal $x\! =\! t^f_j$ into $\varphi_3$.

The formula ${\fhi}_3$ resulting from these insertions is clearly equisatisfiable with the original ${\fhi}_2$.

\smallskip

\item\label{itemAlgoFour} ${\fhi}_3 \leadsto {\fhi}_4$~~(\textsc{elimination of function variables})

As a final step, we eliminate all literals that contain function variables.

Let $V \putAs \{ v_1,v_2, \dots ,v_r \}$ be the set of distinct domain variables occurring in ${\fhi}_3$. Define the index function $ind \colon  V \cup \{ -\infy , +\infy \} \frec \{1,2,\dots ,r\}$ by:
\[
\ind(x)\putAs
\left\{
 \begin{array}{cl}
 1 & \mbox{if}\ x=-\infy ,\\
 l & \mbox{if}\ x=v_l\ \mbox{for some}\ l \in \{1,2,\dots, r\} , \\
 r & \mbox{if}\ x=+\infy .
 \end{array}
\right.
\]

Then, for each function symbol $f$ occurring in ${\fhi}_3$, we proceed as follows (if needed, we introduce new numerical variables ${\ga}^f_0$ ,${\ga}^f_r$, $k^f_0$, and $k^f_r$ to facilitate the elimination of the functional literals):
\begin{enumerate}[label=\alph*), ref=\alph*]

 \item\label{alg4:actA}  For each literal $ (f\! =\! g)_{[z_1,z_2]}$ occurring in ${\fhi}_3$, add 
 all literals
 $y^f_i \! = \! y^g_i,\ t^f_i \! = \! t^g_i$
 for every subscript $i$ such that $\ind(z_1) \! \leq \! i \!  \leq \! \ind(z_2)$; moreover, if $z_1 =-\infy$, introduce the literal 
$ {\ga}^f_0 \! = \! {\ga}^g_0$,
 and if $z_2 = +\infy$ introduce the literal 
$ {\ga}^f_r \! = \! {\ga}^g_r$\,. 

\item\label{alg4:actB}  For literals of type $(f>g)_A$, we consider separately bounded and unbounded interval specifications:

  \begin{enumerate}[label=\ref{alg4:actB}$_\arabic*$), ref={\ref{itemAlgoFour}.{\ref{alg4:actB}$_\arabic*$})}]
     
     \item\label{alg4:actionB:case1}  For each literal $(f \! > \! g)_{[w_1,w_2]}$ (resp. $(f \! > \! g)_{]w_1,w_2[}$, $(f>g)_{[w_1,w_2[}$, or $(f \! > \! g)_{]w_1,w_2]}$) occurring in ${\fhi}_3$, add the literal
 \[
 y^f_i > y^g_i\,,
 \]
 for each $ \ind(w_1) \! \leq \! i\! \leq \! \ind(w_2)$ (resp. $ \ind(w_1) \! <\! i\! <\! \ind(w_2)$, $ \ind(w_1) \! \leq \! i\! <\! \ind(w_2)$, or $ \ind(w_1) \! <\! i\! \leq \! \ind(w_2)$). Moreover, if $w_1 \! <\! w_2$ as domain variables, in the case $(f>g)_{]w_1,w_2[}$ (resp. $(f>g)_{[w_1,w_2[}$ or $(f>g)_{]w_1,w_2]}$) add the following literals

 \[
 t^f_{\mathit{ind}(w_1)} \geq t^g_{\mathit{ind}(w_1)} ,\ \ t^f_{\mathit{ind}(w_2)} \leq t^g_{\mathit{ind}(w_2)}\ \ \ 
 \big(\mbox{resp.}\ t^f_{\mathit{ind}(w_2)} \leq t^g_{\mathit{ind}(w_2)}\ \ \mbox{or} \ \ t^f_{\mathit{ind}(w_1)} \geq t^g_{\mathit{ind}(w_1)} \big).
 \]
 
 \item\label{alg4:actionB:case2} For each literal $(f>g)_{]-\infty , + \infty[}$ (resp. $(f>g)_{]-\infty , w_1]}$ or $ (f>g)_{[w_1 , + \infty[}$) occurring in $\fhi_3$, add the literal
\[
 y^f_i > y^g_i\,,
 \]
 for each $1 \! \leq \! i \! \leq \! r$ (resp. $ 1 \! \leq i \! \leq \! \ind(w_1)$ or $\ind(w_1) \! \leq \! i \! \leq \! r$), and the literals
 \[
  k_0^f  \geq  k_0^g, \, \ k_r^f  \geq  k_r^g \ \ \ (\mbox{resp.} \ k_0^f \geq k_0^g\ \ \mbox{or}\ \ k_r^f \geq k_r^g ).
 \]
   \end{enumerate}
 
 \item\label{alg4:actC} For each literal $(D[f]\bowtie y)_{[z_1,z_2]}$ occurring in ${\fhi}_3$, where \hbox{${\bowtie}  \in  \{=, <, >, \leq ,\geq \}$,} add the following formulas:
 \[
 t^f_i \bowtie y,\ \ \ %
 \textstyle \frac{y^f_{j+1}-y^f_j}{v_{j+1}-v_j} \bowtie y ,
 \]
 for each $\ind(z_1) \leq i \leq \ind(z_2)$ and each $\ind(z_1) \leq j < \ind(z_2)$; moreover, if ${\bowtie} \in \! \{ \leq ,\geq \}$, then for each $\ind(z_1) \leq j < \ind(z_2)$ add the formulas:
 \[
 \textstyle\left( \frac{y^f_{j+1}-y^f_j}{v_{j+1}-v_j} = y  \right) \frec (t^f_j =y \et t^f_{j+1} =y);
 \]
 finally, if $z_1 = -\infy $, introduce the literal 
 ${\ga}^f_0 \bowtie y,$
 whereas, if $z_2 =+\infy $, introduce the literal 
 $ {\ga}^f_r \bowtie y. $
 
 \item\label{alg4:actD} For each literal $(D[f]\bowtie y)_{]w_1,w_2[}$ (resp. $(D[f]\bowtie y)_{]w_1,w_2]}$ or $(D[f]\bowtie y)_{[w_1,w_2[}$) occurring in ${\fhi}_3$, where ${\bowtie} \in \! \{=, <, >, \leq ,\geq \}$, add the formulas:
\[
 t^f_i \bowtie y,\ \ \ %
\textstyle \frac{y^f_{j+1}-y^f_j}{v_{j+1}-v_j} \bowtie y
 \]
 for each $\ind(w_1) \! \leq \! j \! < \! \ind(w_2)$ and for each $\ind(w_1) \! <\! i\! <\! \ind(w_2)$ (resp. $\ind(w_1) \! < \! i \! \leq \! \ind(w_2)$ or $\ind(w_1) \! \leq \! i \! < \! \ind(w_2)$); moreover, if ${\bowtie} \in \! \{ \leq ,\geq \}$, then for each $\ind(z_1) \leq j < \ind(z_2)$ add the formulas:
 \[
 \textstyle\left( \frac{y^f_{j+1}-y^f_j}{v_{j+1}-v_j} = y  \right) \frec (t^f_j =y \et t^f_{j+1} =y).
 \] 
 
 \item\label{alg4:actE} For each literal Strict$\_$Up$(f)_{[z_1,z_2]}$ (resp. Strict$\_$Down$(f)_{[z_1,z_2]}$) occurring in ${\fhi}_3$, add the literals
 \[
 t^f_i \geq 0\ \ (\mbox{resp.}\ \leq ),\ \ \ %
 y^f_{j+1} > y^f_j \ \ (\mbox{resp.}\ <),
 \]
 for each $\ind(z_1) \leq  i,j \leq \ind(z_2),\ j\not= \ind(z_2)$; moreover, if $z_1 \! =\! -\infy $, introduce the literal
 ${\ga}^f_0 \! > \! 0$  $(\mbox{resp.}\ <)$\ %
 and, if $z_2 = +\infy $, introduce the formula 
 ${\ga}^f_r \! > \! 0\ \ (\mbox{resp.}\ <).$
 
\item\label{alg4:actF} For each literal Convex$(f)_{[z_1,z_2]}$ (resp. Concave$(f)_{[z_1,z_2]}$) occurring in ${\fhi}_3$, add the following formulas:
 \[
\textstyle t^f_i \leq \frac{y^f_{i+1}-y^f_i}{v_{i+1}-v_i}  \leq t^f_{i+1}\ \ (\mbox{resp.}\ \geq ),
 \]
 \[
 \textstyle \left( \frac{y^f_{i+1}-y^f_i}{v_{i+1}-v_i} =t^f_i\ \vel \ \frac{y^f_{i+1}-y^f_i}{v_{i+1}-v_i} = t^f_{i+1} \right) \frec ( t^f_i =t^f_{i+1} ),
 \]
 for each $\ind(z_1) \! \leq \! i\! <\! \ind(z_2)$; moreover, if $z_1 =-\infy $ introduce the literal 
 ${\ga}^f_0 \leq t^f_1 \ \ (\mbox{resp.}\ \geq ),$
 and, if $z_2 =+\infy $, introduce the literal 
 ${\ga}^f_r \geq t^f_r \ \ (\mbox{resp.} \ \leq ).$
 
 \item\label{alg4:actG} For each literal Strict$\_$Convex$(f)_{[z_1,z_2]}$ (resp. Strict$\_$Concave$(f)_{[z_1,z_2]}$) occurring in ${\fhi}_3$, add the following formulas:
 \[
 \textstyle t^f_i < \frac{y^f_{i+1}-y^f_i}{v_{i+1}-v_i} < t^f_{i+1}\ \ (\mbox{resp.}\ >),
 \]
 for each $\ind(z_1)\! \leq \! i\! <\! \ind(z_2)$; moreover, if $z_1 =-\infy $, introduce the 
  literal 
 ${\ga}^f_0 < t^f_1 \ \ (\mbox{resp.}\ >),$
 and, if $z_2 =+\infy $, introduce the literal 
 ${\ga}^f_r > t^f_r \ \ (\mbox{resp.}\ <).$

 \item\label{alg4:actH}  If there are literals involving variables of type $k$, i.e., literals of the form $k_i^f \! \geq k_i^g$ with $i\! \in \! \{0,r\}$ and $f,g$ function variables, modify the current formula according the following steps:

 \begin{enumerate}
 
    \item for each variable $k_i^f$,  add the formula $-1  \leq  k_i^f  \leq  +1$, with $i\! \in \! \{0,r\}$;
    
    \item if both literals $k_i^f \geq k_i^g$ and $k_i^g \geq k_i^h$ occur, add literals $k_i^f \geq k_i^h$ and $y_i^f > y_i^h$, with $i\! \in \! \{0,r\}$;
 
    \item if literals $k_0^f \geq k_0^g$, $\ga_0^f \trianglerighteq m$ and $\ga_0^g \trianglelefteq n$ occur together, with $\trianglerighteq  \, \in  \{ \geq, >, = \}$ and $\trianglelefteq \, \in  \{ \leq, <, = \}$, add literal $m \leq n$; specularly, in the case $k_r^f \geq k_r^g$, $\ga_r^f \trianglelefteq m$ and $\ga_r^g \trianglerighteq n$, add the literal $m\geq n$.
 \end{enumerate}

 \item\label{alg4:actI} Drop all literals involving any function variable.
 
 \end{enumerate}
 
\end{enumerate} 

We have thus eliminated all function variables originally occurring in $\fhi_0$; the resulting formula ${\fhi}_4$ can now be tested for satisfiability using Tarski's method.


\subsection{The decision algorithm at work}\label{subsec:decAlgorithmAtWork} 

To illustrate the decision algorithm for \RDFp in action, we examine a paradigmatic case: the formula
\[
(D[f]>0)_{]a,b[} \ \frec\ \nor{Strict\_Up}(f)_{[a,b]},
\]
in which $a$ and $b$ are numerical variables. This example, introduced as \Exa{exa:two_one} in \Sec{sec:RFDplus}, asserts that if the derivative of $f$ is strictly positive on the open interval $]a,b[$, then $f$ is strictly increasing on the closed interval $[a,b]$.

To establish that this formula is valid---that is, true under any assignment of values to $a$, $b$, and $f$---we proceed by verifying that its negation is unsatisfiable by means of our algorithm and Tarski's decision method. This yields the conjunction:
\[
(D[f]>0)_{]a,b[}\ \ \et\ \ \neg \nor{Strict\_Up} (f)_{[a,b]}\,.
\]
which we denote by $\varphi$. If no assignment satisfies $\varphi$, then the original implication holds universally. We now apply the decision algorithm to $\varphi$ step by step.

\paragraph{$\fhi \leadsto {\fhi}_1$}~~(\textsc{behavior at ends}):

Following step~\ref{itemAlgoOne}.\ref{alg1:actionFour}), we expand the atom $(D[f]>0)_{]a,b[}$ into a disjunction:
\[
\begin{array}{lcl}
(D[f]>0)_{]a,b[} &\leadsto & (D[f]>0)_{[a,b]} ~\vel~ \left[(D[f]>0)_{[a,b[} ~\et~ D[f](b)=0 \right]  \\[0.12cm]
 && \hspace{2.25cm} ~\vel~ \left[ (D[f]>0)_{]a,b]} ~\et~ D[f](a)=0 \right] \\[0.12cm]
 && \hspace{2.25cm}~\vel~ \left[ (D[f]>0)_{]a,b[} ~\et~ D[f](a)=0 ~\et~ D[f](b)=0 \right].                 
\end{array}
\]
We now distribute this disjunction over the conjunction with $\neg \nor{Strict\_Up}(f)_{[a,b]}$ to obtain the disjunctive normal form:
\begin{equation}\label{array:temporaryDNF}\tag{$\dagger$}
\begin{aligned}
 &\big[ (D[f]>0)_{[a,b]} \et \neg \nor{Strict\_Up}(f)_{[a,b]} \big] \\
 &  ~~~~~~         \vel\ \ \big[ (D[f]>0)_{[a,b[} \et D[f](b)=0   \et \neg \nor{Strict\_Up}(f)_{[a,b]} \big]\\
 &  ~~~~~~         \vel\ \ \big[ (D[f]>0)_{]a,b]} \et D[f](a)=0 \et \neg \nor{Strict\_Up}(f)_{[a,b]} \big]\\
 &  ~~~~~~         \vel\ \ \big[ (D[f]>0)_{]a,b[} \et D[f](a)=0 \et D[f](b)=0 \et \neg \nor{Strict\_Up}(f)_{[a,b]} \big].
\end{aligned}
\end{equation}
       
We now apply the rest of the algorithm to each disjunct. For brevity, we focus on the first one.

\smallskip

\paragraph{${\fhi}_1 \leadsto {\fhi}_2$}~~(\textsc{negative-clause removal}):

To eliminate the literal $\neg \nor{Strict\_Up}(f)_{[a,b]}$, we apply its definitional expansion:
\[
\begin{array}{lcl}
(D[f]>0)_{[a,b]} \et \neg \nor{Strict\_Up}(f)_{[a,b]} &\leadsto & (D[f]>0)_{[a,b]} \et a\leq x_1 < x_2 \leq b \\
 && \et f(x_1)=y_1 \et f(x_2)=y_2 \et y_2 \leq y_1 .
\end{array}
\]         
This yields our formula ${\fhi}_2$.

\smallskip

\paragraph{${\fhi}_2 \leadsto {\fhi}_3$}~~(\textsc{explicit evaluation of function variables}):

To streamline notation, we rename variables as follows:
\[
a \leadsto v_1 ,\ \ x_1 \leadsto v_2 ,\ \ x_2 \leadsto v_3 ,\ \ b \leadsto v_4 ,\ \ y_1 \leadsto y_2 ,\ \ y_2 \leadsto y_3\,.
\] 
The formula now becomes:
\[
(D[f]>0)_{[v_1 ,v_4]}\ \et \ v_1 \leq v_2 < v_3\ \leq v_4 \ \et \ f(v_2)=y_2 \et \ f(v_3)=y_3 \ \et \ y_3 \leq y_2 .
\]
We evaluate $f$ at all domain variables in $V = \{v_1, v_2, v_3, v_4\}$ and introduce corresponding literals:
{\small
\begin{align*}
{\fhi}_3 &\putAs  (D[f]>0)_{[v_1 ,v_4]} \et v_1 \leq v_2 < v_3 \leq v_4 \et f(v_2)=y_2 \et f(v_3)=y_3 \et y_3 \leq y_2 \\
          &\phantom{\putAs  (D[f]>0)_{[v_1 ,v_4]}~} \et f(v_1)=y_1^f \et f(v_2)=y_2^f \et f(v_3)=y_3^f \et f(v_4)=y_4^f \et y_2 = y_2^f \et y_3 = y_3^f \\
          &\phantom{\putAs  (D[f]>0)_{[v_1 ,v_4]}~} \et D[f](v_1)=t_1^f \et D[f](v_2)=t_2^f \et D[f](v_3)=t_3^f \et D[f](v_4)=t_4^f . 
\end{align*}
}

\paragraph{${\fhi}_3 \leadsto {\fhi}_4$}~~(\textsc{elimination of function variables}):

We now replace all functional literals with purely arithmetic constraints, obtaining:
\begin{align*}
{\fhi}_4 &\putAs  v_1 \leq v_2 < v_3 \leq v_4 \et y_3 \leq y_2 \et y_2 =y_2^f \et y_3 =y_3^f \\
   &\phantom{\putAs  v_1 \leq v_2 < v_3 \leq v_4~ } \et \, t_1^f >0 \et t_2^f >0 \et t_3^f >0 \et t_4^f >0 \\
   &\phantom{\putAs  v_1 \leq v_2 < v_3 \leq v_4~ } \textstyle \et \, \frac{y_2^f -y_1^f}{v_2 - v_1}>0 \et \frac{y_3^f -y_2^f}{v_3 - v_2}>0 \et  \frac{y_4^f -y_3^f}{v_4 - v_3}>0.
\end{align*}
This formula contains only numerical variables and is thus suitable for Tarski’s decision method. 
 To demonstrate unsatisfiability of ${\fhi}_4$, consider the sub-conjunction:
\[
\textstyle v_2 < v_3 \et y_2 =y_2^f \et y_3 =y_3^f \et y_3 \leq y_2 \et \frac{y_3^f -y_2^f}{v_3 -v_2} >0\,.
\]
Since $y_3^f = y_3 \leq y_2 = y_2^f$ and $v_2 < v_3$, the last inequality leads to a contradiction.\\
A similar analysis of the remaining disjuncts in~\eqref{array:temporaryDNF} establishes the overall unsatisfiability of $\fhi$, and hence the validity of the original implication.

\section{Correctness of the Algorithm} \label{sec:correctness}

To establish the correctness of our algorithm, it suffices to show that each transformation $\fhi_i \leadsto {\fhi}_{i+1}$ (for $i=0,\dots,3$) preserves satisfiability. 

The correctness of the first three steps—--addressing behavior at endpoints, eliminating negative clauses, and evaluating function variables explicitly---follows directly from the explanations given in the algorithm's description. 

This section focuses on the final transformation ${\fhi}_3 \leadsto {\fhi}_4$, which is less immediate than the preceding ones. 

Recall that ${\fhi}_4$ is obtained from ${\fhi}_3$ by eliminating all literals involving function variables and replacing them with formulas over numerical variables. The key challenge, therefore, lies in relating a formula that encodes properties of functions to one that refers solely to numerical quantities.

To prove that ${\fhi}_3$ and ${\fhi}_4$ are \emph{equisatisfiable}, we proceed as usual by proving both \textbf{soundness} and \textbf{completeness}: 
\begin{itemize}

   \item \textbf{Soundness}:  Any model of ${\fhi}_3$ can be extended to satisfy the additional constraints introduced in ${\fhi}_4$. These added formulas correctly reflect the pointwise and differential properties of the functions referenced in ${\fhi}_3$, evaluated at specific arguments.

   \item \textbf{Completeness}: Conversely, given a model for ${\fhi}_4$, one can construct function interpretations that satisfy ${\fhi}_3$ by piecing together appropriate interpolating functions consistent with the numerical data.   
\end{itemize}
In our construction, these interpolating functions are defined piecewise, combining linear segments with controlled ``elastic'' perturbations. These perturbations are generated using quadratic and exponential expressions that ensure continuity and differentiability while respecting the numerical constraints.

\begin{remark}
The family of interpolating functions introduced below is tailored to the current decision algorithm. In future extensions of the \RDFp language, we may enrich the syntax with operations such as function addition and scalar multiplication. To support such constructs, it will be necessary to ensure that the underlying class of admissible functions forms a \emph{vector space} (e.g., of class $C^1$ or $C^n$). This requirement would prompt a revision of the function construction method. \eod
\end{remark}

\subsection{Mathematical tools}\label{subsec:tools}

Let ${\mR}^+$ denote the set of all positive real numbers, and let ${\mR}^+_0$ denote the set of non-negative reals. For each $k \in {\mR}^+_0$ and each pair $d=(d_1,d_2)\in {\mR }^{+}_{0}\!\! \times \! {\mR }^{+}_{0}$, we define the functions $Exp_{[k]} \colon [0, \frac{1}{2}] \frec [0, \frac{1}{2} ]$ and $\textit{Alter}_{[d]} \colon [0,1] \frec [0,1]$ as follows:
\begin{align*}
\textstyle Exp_{[k]} (x) &~\putAs~ \frac{1}{2} + \left( x-\frac{1}{2}\right) \x e^{-kx},\ \ \forall x \in \left[ 0,\frac{1}{2}\right];\\
\textit{Alter}_{[d]}(x) & ~\putAs~  \left\{ 
    \begin{array}{ll}
    Exp_{[d_1]}(x) &  \mbox{if}\ x\in \left[0,\frac{1}{2}\right],\\[0.15cm]
    1-Exp_{[d_2]}(1-x) & \mbox{if}\ x\in \ \left]\frac{1}{2},1\right].
    \end{array}
 \right.  
\end{align*}
Then, for any $\al \in \mR$, let ${\I}_{\al}\putAs [\min(0,\al),\max(0,\al)]$, so that
${\I}_{\al}$ is a  bounded interval contained in $\mR$.  We then define the  quadratic function
\[\begin{array}{rcl}
\textit{Parabola}_{[\al]}(x) &\putAs& 4\al x(1-x),\ \ \forall x\in [0,1],
\end{array}\]
which maps $[0,1]$ into ${\I}_{\al}$.

\begin{definition}\label{def:1.2} We dub \textsc{$[\al]$-elastic family}\index{elastic family}, for any $\al\in\mR$, the collection of all real-valued functions of the form
\[\begin{array}{rcl}
\textit{Elastic}_{[\al ,d]}(x)&\putAs&\textit{Parabola}_{[\al]}(\textit{Alter}_{[d]}(x))\ \ \forall x\in [0,1],
\end{array}
\]
with $d$ ranging over ${\mR }^{+}_{0}\!\! \times \! {\mR }^{+}_{0}$. Each such function $\textit{Elastic}_{[\al,d]}$ maps $[0,1]$ into ${\I}_{\al}$.
\end{definition}

Under this definition, the function $C(x)\putAs \textit{Elastic}_{[\al,d]}(x)$ associated with a specific pair $d=(d_1,d_2)$ satisfies the following properties:
\begin{enumerate}
\item\label{itemC:one} An explicit expression for $C(x)$ is:
\[
C(x)= \left\{
     \begin{array}{ll}
     \al [1-(2x-1)^2 e^{-2d_1 x} ] & \mbox{if}\ x\in \left[ 0,\frac{1}{2}\right] , \\[0.12cm]
     \al [1-(2x-1)^2 e^{-2d_2 x} ] & \mbox{if}\ x\in \left] \frac{1}{2},1 \right] .
     \end{array}
     \right.
\] 
This function satisfies: $C(0)\! =\! C(1)\! =\! 0$ and $C(\frac{1}{2})\! =\! \al$. In particular, $\abs{C(x)} \! \leq  \! \abs{\al}$ for all $x \! \in \! [0,1]$, and if $\al \! =\! 0$, then $C(x)$ is identically zero on $[0,1]$.

\item\label{itemC:two} The function $C(x)$ is continuous and differentiable on $[0,1]$. Its derivative is given by:
\[
C'(x)=\left\{
      \begin{array}{ll}
         2\al (2x-1)[d_1 (2x-1)-2]e^{-2d_1x} & \mbox{if}\ x\in \left[ 0,\frac{1}{2}\right] , \\[0.1cm] 
        -2\al (2x-1)[d_2 (2x-1)+2]e^{-2d_2 (1-x)} & \mbox{if}\ x\in \left] \frac{1}{2},1 \right].
      \end{array}
      \right.
\]
This function is continuous on $[0,1]$, and satisfies $C'\big(\frac{1}{2}\big) = 0$. Moreover, if $\al \! > \! 0$ (resp. $\al \! < \! 0$), then $C(x)$ is strictly increasing (resp. strictly decreasing) on $[0,\frac{1}{2}[$ and strictly decreasing (resp. strictly increasing) on $]\frac{1}{2},1]$. \vspace{0.12cm}

\item\label{itemC:three} The function $C'(x)$ is differentiable in $[0,1]$ except possibly at $x=\frac{1}{2} $. The second derivative is given by:
\[
C''(x)=\left\{
      \begin{array}{ll}
         4\al [-4d_1^2 x^2 + 4d_1(d_1+2)x -d_1^2 -4d_1 -2] e^{-2d_1 x} & \mbox{if}\ x\ \mbox{in}\ \left[ 0,\frac{1}{2}\right] , \\[0.1cm]
        -4\al [4d_2^2 x^2 + 4d_2(2-d_2)x + d_2^2 -4d_2 +2] e^{-2d_2 (1-x)} & \mbox{if}\ x\in \left] \frac{1}{2},1 \right] .
      \end{array}
      \right.
\]

In particular, if $\al \! > \! 0$ (resp. $\al \! < \! 0$), then the $C(x)$ is strictly concave (resp. strictly convex) on $[0,1]$.
\end{enumerate}

We now use the elastic families just defined to introduce the functions employed in the correctness proof. Let $\al$, ${\theta}_1$, and ${\theta}_2$ be real numbers.

\begin{definition}\label{def:1.3} A function $C \colon [0,1] \frec \mR $ is said to be \textsc{single-$[\al ,{\theta}_1, {\theta}_2 ]$-defined}\index{single-defined} if and only if:\footnote{This definition---as well as the preceding one,  along with related properties)---can be readily adapted to arbitrary bounded and closed intervals $[a, b]\subset \mR$.}
\begin{enumerate}
 \item $C$ belongs to the $[\al]$-elastic family;

 \item the derivative of $C$ satisfies $C'(0)={\theta}_1$ and $C'(1)={\theta}_2$. 
\end{enumerate}
\end{definition}

\begin{figure}[!htb]\begin{center}
\doublebox{\resizebox{10cm}{!}{
  \includegraphics{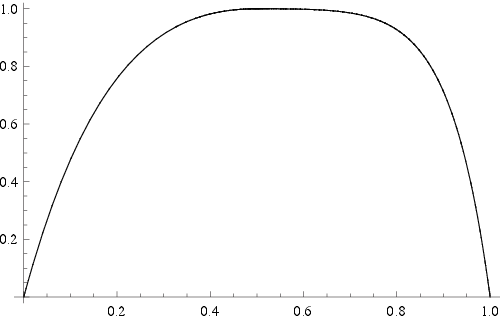}
}}\end{center}
\caption{\label{fig:one}\Nv\footnotesize Graph of the single-$[\al ,{\theta}_1 ,{\theta}_2]$-defined function with $\al \! =\! 1,\ {\theta}_1 \! =\! 6 ,\ {\theta}_2 \! =\! -12$.}
\end{figure}

A single-$[\al, {\theta}_1, {\theta}_2]$-defined function $C(x)$ is continuous and differentiable on $[0,1]$, with a continuous derivative. By \Def{def:1.3}, the straight lines tangent to $C(x)$ at $(0,0)$ and $(1,0)$ have slopes ${\theta}_1$ and ${\theta}_2$, respectively. Moreover, $C(x)$ is identically zero on $[0,1]$ if and only if ${\theta}_1 = {\theta}_2 = 0$.

The following proposition provides sufficent conditions for the existence of such functions:
\begin{lemma}\label{lem:1.1} A single-$[\al ,{\theta}_1 ,{\theta}_2]$-defined function exists if and only if one of the following conditions holds:
\begin{itemize}

 \item $\al={\theta}_1 = {\theta}_2 =0$,
 
 \item ${\theta}_1 \x {\theta}_2 < 0$, \quad $0<|4\al | \leq \min(|{\theta}_1|, |{\theta}_2|)$ \quad and \quad $\sgn(\al)=\sgn(\theta_1)$.\footnote{According to the usual definition of the sign function: 
$\sgn(x) \ \putAs \ \begin{cases}
    -1&\text{if }x<0,\\
    0&\text{if }x=0,\\
    1&\text{if }x>0.
\end{cases}
$
}
\end{itemize}
When such a function exists, it is unique.
\end{lemma}
\dimostraz (`$\Leftarrow$')\ If $\al={\theta}_1 = {\theta}_2 =0$, the sought function must be null in all $[0,1]$.

In the second case we have ${\theta}_1 \x {\theta}_2 \! < \! 0$, $0 \! < \!  |4\al | \! \leq  \! \min(| {\theta}_1 |, |{\theta}_2 | )$ and $\sgn(\al)=\sgn(\theta_1) $. We will determine non-negative reals $d_1,d_2$ such that the function $C(x)\putAs \textit{Elastic}_{[\al,d]}(x)$ with $d=(d_1,d_2)$ is a single-$[\al ,{\theta}_1 ,{\theta}_2 ]$-defined function; let us fix:
\[
\textstyle (d_1 ,d_2 )= \left( \frac{{\theta}_1}{2\al}-2, -\frac{{\theta}_2}{2\al}-2 \right).
\]
It remains to verify that $d_1,d_2$ are nonnegative. By hypothesis, we have $|4\al |\leq |{\theta}_1|$, hence $|\frac{{\theta}_1}{\al} | \geq 4$ and also $\frac{{\theta}_1}{\al}\geq 4$ because $\sgn(\al)=\sgn(\theta_1)$; thus $d_1 \geq 0$ and, analogously, $d_2 \geq 0$. Moreover, it is easy to check that $C'(0)=2\al (d_1 +2)={\theta}_1$ and $C'(1) \! = \! -2\al (d_2 +2) \! = \! {\theta}_2$.

(`$\Rightarrow$')\ The converse implication is easily obtained through \Defs{def:1.2}{def:1.3}.    \qed 

\medskip

Some properties of single-$[\al ,{\theta}_1 ,{\theta}_2]$-defined functions are elicited below.
\begin{lemma}\label{lem:1.2} Any single-$[\al ,{\theta}_1 ,{\theta}_2]$-defined function $C(x)$  is either convex or concave in $[0,1]$.
\end{lemma}
\dimostraz The claim is trivial if $C(x)$ is the null function.

Otherwise, the function $C(x)$ belongs to the $[\al]$-elastic family, for some $\al \! \in \! \mR $. Consider the second-order derivative of $C(x)$ in the intervals $[0,\frac{1}{2} [$, $]\frac{1}{2}, 1]$; we have $\sgn(C''(x))=-\sgn(\al)$ for all $x$.
Furthermore, $C$ has the same concavity or convexity for all $x$ in $[0,1]$. \qed 


\begin{corollary}\label{cor:1.1} Let $C(x)$ be a single-$[\al ,{\theta}_1 ,{\theta}_2]$-defined function. The function $C(x)$ is convex (resp. concave) in $[0,1]$ if and only if ${\theta}_1 \! \leq \! 0$ (resp. $\theta_1 \geq 0 $). Furthermore, the function $C(x)$ is strictly convex (resp. strictly concave) in $[0,1]$ if and only if ${\theta}_1 \! < \! 0$ (resp. $\theta_1 > 0$).
\end{corollary}
\dimostraz Straightforward from \Lems{lem:1.1}{lem:1.2}; in fact, $\sgn(\al)=\sgn({\theta}_1)$ holds, unless $C(x)$ is null all over $[0,1]$. $\phantom{xxx}$\hfill\qed

\begin{corollary}\label{cor:1.2} Let $k\in \mR $ and $C(x)$ be a single-$[\al ,{\theta}_1 ,{\theta}_2]$-defined function. If ${\theta}_1,{\theta}_2 \bowtie k$, where ${\bowtie} \in \{\leq, <, >, \geq \}$, then $C'(x) \bowtie k$ for all $x\in [0,1]$.
\end{corollary}
\dimostraz The claim follows from the monotonicity of the function $C'(x)$ in $[0,1]$, due to \Lem{lem:1.2}. \qed 

\medskip

The treatment of elastic families developed so far could be extended and generalized to arbitrary bounded and closed intervals $[a, b]\subset \mR$, but this seems unnecessary. For example, in the following we need to scale down the single-$[\al ,{\theta}_1 ,{\theta}_2]$-defined properties to the intervals $[0,\frac{1}{2}]$ and $[ \frac{1}{2} ,1]$; to do so, we can simply rely on the projection functions $p \colon [0,\frac{1}{2}] \frec [0,1]$ and $q \colon [\frac{1}{2} ,1] \frec [0,1]$, respectively, defined as:
\[
\textstyle p(x)=2x,\ \forall x\in \left[0,\frac{1}{2} \right]\ \ \mbox{and}\ \ q(x)=2x-1,\ \forall x\in \left[\frac{1}{2},1\right].
\]

Let $\al ,{\theta}_1 ,{\theta}_2$ be real numbers.

\begin{figure}[!htb]
\begin{center}
\doublebox{\resizebox{10cm}{!}{
  \includegraphics{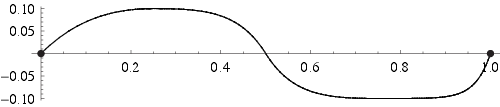}
}}
\end{center}
\caption{\label{fig:two}\Nv\footnotesize Graph of the double-$[\frac{1}{10},1,3 ]$-defined function.}
\end{figure}

\begin{definition}\label{def:1.4} A function $C \colon  [0,1]\frec \mR$ is called \textsc{double-$[\al ,{\theta}_1 ,{\theta}_2]$-defined}\index{double-defined}, if and only if it can be specified as
\[
C(x)=
  \left\{
     \begin{array}{ll}
     C_1(p(x)) & \text{if~ } x \in   [0,\frac{1}{2}],\\[.01cm]
     C_2(q(x)) & \text{if~ } x \in \ ]\frac{1}{2},1],
     \end{array}
  \right.
\]   
where:\\
\begin{enumerate}
\item $C_1(x)$ is a single-$[\frac{1}{2}\al ,\frac{1}{2}{\theta}_1 ,\frac{1}{2}\theta]$-defined function,

\vspace{0.1cm}

\item $C_2(x)$ is a single-$[-\frac{1}{2}\al ,\frac{1}{2}\theta ,\frac{1}{2}{\theta}_2]$-defined function,

\vspace{0.1cm}

\item $\theta \in \mR$ is such that $\theta=-4\al$.
\end{enumerate}
\end{definition}
As an example, see \Fig{fig:two} that charts the double-$[\frac{1}{10},1,3 ]$-defined function.

The real number $\theta $ of the above definition is called the \ita{\textsc{middle slope}} of $C(x)$. A  double-$[\al, {\theta}_1, {\theta}_2]$-defined function $C(x)$ is continuous and differentiable (with a continuous derivative) in $[0,1]$ and we also have $C'(0)=C_1'(p(0))\x p'(0)={\theta}_1$, $C'(1)={\theta}_2$, and $C'(\frac{1}{2})=\theta$. Moreover, $C(x)$ is the null function in $[0,1]$ if and only if $\al ={\theta}_1 = {\theta}_2 =0$.

As for single-defined functions, we have the following existence conditions:
\begin{lemma}\label{lem:1.3}
Let $\al ,{\theta}_1, {\theta}_2$ be real numbers. A double-$[\al ,{\theta}_1, {\theta}_2]$-defined function  exists if and only if either of the following conditions holds:
\begin{itemize}
\item $\al = {\theta}_1 ={\theta}_2 =0$,

\item ${\theta}_1 \x {\theta}_2 >0$, $0<|4\al|\leq min(|{\theta}_1|,|{\theta}_2|)$, and $\sgn(\al)=\sgn(\theta_1)$.
\end{itemize}
Either of these existence conditions uniquely defines the function in question.
\end{lemma}
\dimostraz Let us first check the backward implication, which, when $\al = {\theta}_1 ={\theta}_2 =0$, is trivial. On the other hand, if ${\theta}_1 \x {\theta}_2 >0$, $|4\al | \leq \mbox{min} (|{\theta}_1|, |{\theta}_2| )$ and $\sgn(\al)=\sgn(\theta_1)$, we put $\theta=-4\al$. By \Lem{lem:1.1}, a single-$[\frac{1}{2}\al , \frac{1}{2}{\theta}_1 ,\frac{1}{2} \theta ]$-defined function $C_1(x)$ exists because $\frac{1}{2} {\theta}_1 \x \frac{1}{2} \theta <0$ and $0 \! < \!  |2\al | \! \leq \! \mbox{min}(|\frac{1}{2}{\theta}_1 |,| \frac{1}{2} \theta |)$. Analogously, the existence of the single-$[-\frac{1}{2}\al , \frac{1}{2}\theta , \frac{1}{2} {\theta}_2]$-defined function $C_2(x)$ can be shown; therefore, the function $C(x)$ resulting from $C_1$ and $C_2$ as stated in \Def{def:1.4} is double-$[\al ,{\theta}_1, {\theta}_2]$-defined. 

The converse implication is obtained by coming back through the proof.  \qed 

We also have the following statement about the derivative of a double-$[\al, {\theta}_1, {\theta}_2]$-defined function:
\begin{lemma}\label{lem:1.4}
Let $k\in \mR$ and $C(x)$ be a double-$[\al, {\theta}_1, {\theta}_2]$-defined function with $\theta \! \in \! \mR$ its middle slope. If ${\theta}_1,{\theta}_2, \theta \bowtie k$, where ${\bowtie} \in \{\leq, <, >, \geq \}$, then $C'(x) \bowtie k$ for all $x\in [0,1]$.
\end{lemma}
\dimostraz Let $C_1, C_2,$ be the component of the function $C(x)$ as defined in \Def{def:1.4}. We have:
\[
C'(x)= \begin{cases}
C_1'(p(x))\x p'(x) & \mbox{if}\ x\in [ 0,\frac{1}{2}] ,\\[0.1cm]
C_2'(q(x))\x q'(x) & \mbox{if}\ x\in \, ] \frac{1}{2}, 1] .
\end{cases}
\]
By hypothesis, $\frac{1}{2}{\theta}_1 >\frac{1}{2}k,\ \frac{1}{2}\theta  >\frac{1}{2}k$ hold, and, by \Cor{cor:1.2}, $C_1'(\ep)>\frac{1}{2}k$ also holds for all $\ep \in [0,1]$. Thus, for all $x\in [0, \frac{1}{2}]$ we have $C'(x)=2\x C_1'(p(x))>k$; and, analogously, $C'(x)>k$ for all $x\in ]\frac{1}{2}, 1]$. \qed 

Let $\al ,{\theta}_1, {\theta}_2 $ be real numbers.

\begin{definition}\label{def:1.5} A function $C \colon  [0,1]\frec \mR $ is called \textsc{special-$[\al ,{\theta}_1, {\theta}_2]$-defined}\index{special-defined} if and only if $C(x)$ can be expressed in the following manner:
\[
C(x)=
  \begin{cases}
  C_1(p(x)) & \text{if }\ x\in [0,\frac{1}{4}], \\[.01cm]
  C_2(x)+\frac{\al}{2} & \text{if }\ x\in \ ]\frac{1}{4}, \frac{3}{4}], \\[.01cm]
  C_3(q(x)) & \text{if }\ x\in \ ]\frac{3}{4}, 1],
  \end{cases}
\]
where:
\begin{enumerate}
\item $C_1(x)$ is a single-$[\al ,{\theta}_1, -{\theta}_1]$-defined function,

\item $C_2(x)$ is a double-$[\frac{\al}{2}, {\theta}_1+{\theta}_2, {\theta}_1+{\theta}_2]$-defined function,

\item $C_3(x)$ is a single-$[\al ,-{\theta}_2, {\theta}_2]$-defined function.
\end{enumerate}
\end{definition}

\begin{figure}[!htb]
\begin{center}
\doublebox{\resizebox{10cm}{!}{
  \includegraphics{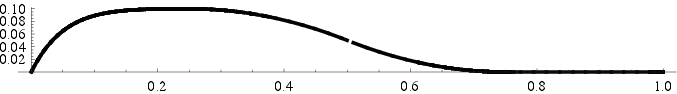}
}}
\end{center}
\caption{\label{fig:three}\Nv\footnotesize Graph of the special-$[\frac{1}{10},2,0 ]$-defined function.}
\end{figure}

\Fig{fig:three} charts the special-$[\frac{1}{10},2,0 ]$-defined function.

\medskip

The previous definition of middle slope naturally generalizes to a special-$[\al ,{\theta}_1, {\theta}_2]$-defined function: referring to the definition of such just made, this will be the middle slope of $C_2(x)$, i.e., $\theta \in \mR $ with $\abs{\theta}=\abs{4\al}$.\\
A special-$[\al , {\theta}_1, {\theta}_2]$-defined function is continuous and differentiable (with a continuous derivative) in $[0,1]$; we also have $C'(0)={\theta}_1,\ C'(1)={\theta}_2$ and $C'(\frac{1}{2})\! =\! \theta$. Moreover, $C(x)$ is the null function if and only if $\al ={\theta}_1 ={\theta}_2 =0$. 

As before, we have the following existence conditions:
\begin{lemma}\label{lem:1.5}
Let $\al ,{\theta}_1 ,{\theta}_2
$ be real numbers, with ${\theta}_1 \x {\theta}_2=0$. A special-$[\al ,{\theta}_1, {\theta}_2]$-defined function exists if and only if either of the following holds:
\begin{itemize}
\item $\al ={\theta}_1 = {\theta}_2 =0$,

\item $0<|4\al|\leq |{\theta}_1+{\theta}_2|$.
\end{itemize}
Moreover, when the function at issue exists it is unique.
\end{lemma}
\dimostraz  If $\al ={\theta}_1 ={\theta}_2 =0$, the claim is trivial.\\
Otherwise, let $0 \! < \! |4\al | \! \leq \! |{\theta}_1 +{\theta}_2 |$ and, without loss of generality, let us suppose that ${\theta}_1 =0\not= {\theta}_2$ and that $C_1(x)$ is the everywhere null function on $[0,1]$.
By \Lem{lem:1.3}, a double-$[\al ,{\theta}_1, {\theta}_2 ]$-defined function $C_2(x)$ exists, as ${\theta}_2 \x {\theta}_2 >0$ and $0<|4\al |\leq |{\theta}_2|$ hold by hypothesis.
Moreover, by \Lem{lem:1.1}, a single-$[\al ,-{\theta}_2 , {\theta}_2 ]$-defined function $C_3(x)$ exists, as $-{\theta}_2 \x {\theta}_2 <0 $ and $0<|4\al |\leq |{\theta}_2|$ hold.
Thus, the function $C(x)$ obtained as by \Def{def:1.5} from $C_1 ,C_2 ,C_3$ is a special-$[\al ,{\theta}_1 ,{\theta}_2 ]$-defined.
The converse implication can be obtained coming back through the proof. \qed 

As for single- and double-defined function, we have the following result about the derivative of a special-$[\al ,{\theta}_1, {\theta}_2]$-defined function:
\begin{lemma} 
Let $k\in \mR$ and let $C(x)$ be a special-$[\al, {\theta}_1, {\theta}_2]$-defined function, with $\theta \in \mR$ its middle slope. If ${\theta}_1,{\theta}_2, \theta \bowtie k$ (where ${\bowtie} \in \{\leq, <, >, \geq \}$), then $C'(x) \bowtie k$ for all $x\in [0,1]$. 
\end{lemma}
\dimostraz Let $C_1, C_2 ,C_3$ be the components of the function $C(x)$ as defined in \Def{def:1.5}. We have:
\[
C'(x)= 
\begin{cases}
C_1'(p(x))\x p'(x) & \text{if }\ x\in [ 0, \frac{1}{4} ] , \\[0.1cm]
C_2'(x) & \text{if}\ x\in \, ] \frac{1}{4}, \frac{3}{4} ] , \\[0.1cm]
C_3'(q(x))\x q'(x) & \text{if }\ x\in \, ] \frac{3}{4}, 1 ] .
\end{cases}
\]
It is easy to check that $C'(x)>k$ is true for all $x\in [0,1]$. \qed

We can now combine the definitions given above to obtain:
\begin{definition} A real function $C \colon   [0,1]\frec \mR $ is said to be \textsc{$[\al, {\theta}_1, {\theta}_2]$-defined}  if and only if it is either single-$[\al, {\theta}_1, {\theta}_2]$-defined, double-$[\al, {\theta}_1, {\theta}_2]$-defined or special-$[\al, {\theta}_1, {\theta}_2]$-defined.
\end{definition}

An $[\al, {\theta}_1, {\theta}_2]$-defined  function $C(x)$ is continuous and differentiable (with a continuous derivative) in $[0,1]$. We also have $C'(0)={\theta}_1$, $C'(1)={\theta}_2$ and
\[
\textstyle \abs{C'\left (\frac{1}{2}\right)} =
    \left\{
    \begin{array}{ll}
    0 & \mbox{if}\ C(x)\ \mbox{is}\ \mbox{single-}[\al ,{\theta}_1, {\theta}_2]\mbox{-defined}, \\[0.1cm]
    |4\al| & \mbox{otherwise}.
    \end{array}
    \right.
\]
Moreover, $C(x)$ is the null function in $[0,1]$ if and only if ${\theta}_1 \! = \! {\theta}_2 \! =0$.

\begin{lemma}\label{lem:1.7}
Let $\al, {\theta}_1, {\theta}_2$ be real numbers and let $C(x)$ be an $[\al, {\theta}_1, {\theta}_2]$-defined function; then,$|C(x)|\leq |\al|$ holds for all $x\in [0,1]$.
\end{lemma}
\dimostraz Straightforward from the definitions given before. \qed 

Even for an $[\al, {\theta}_1, {\theta}_2]$-defined function there are existence conditions:
\begin{lemma}\label{lem:1.8} Let $\al ,{\theta}_1, {\theta}_2$ be real numbers. An $[\al, {\theta}_1, {\theta}_2]$-defined function exists if and only if either of the following holds:
\begin{itemize}
\item $\al={\theta}_1 \! =\! {\theta}_2 \! =0$,

\item $0<|4\al|\leq \min \{ |x| \,|\, x\not= 0\ \mbox{and}\ x \in \{ {\theta}_1,{\theta}_2 \} \}$.
\end{itemize}
Moreover, when the function at issue exists it is unique.
\end{lemma}
\dimostraz Straightforward by Lemmas \ref{lem:1.1}, \ref{lem:1.3} and \ref{lem:1.5}. \qed 

\begin{corollary}\label{cor:1.3} Let $\al ,{\theta}_1, {\theta}_2, {\theta}_3, {\theta}_4$ be real numbers, and let $C_1(x),\ C_2(x)$ be, respectively, an $[\al ,{\theta}_1 ,{\theta}_2]$-defined function and an $[\al ,{\theta}_3, {\theta}_4]$-defined function. Then $C_1(x)\! =\! C_2(x)$ for all $x\! \in \! [0,1]$ if and only if ${\theta}_1 ={\theta}_3$ and ${\theta}_2 = {\theta}_4$.
\end{corollary}
\dimostraz Straightforward by the above lemma. \qed 

We now prove some other properties for $[\al ,{\theta}_1, {\theta}_2]$-defined functions that will be used in the next chapter.

\begin{lemma}\label{lem:1.9} Let $f$ be a single-$[\al ,{\theta}_1, {\theta}_2 ] $-defined function and $\be \! \in \! \mR $ such that $|\al | \! > \! |\be |$, then we have $|f| \! \geq \! |g|$ in $[0,1]$ with $g$  single-$[\be, {\theta}_1, {\theta}_2 ]$-defined function.
\end{lemma}
\dimostraz First we suppose $\al \! > \!  \be \! > \! 0$. Since $f$ and $g$ are single-defined, we can analyze separately the two intervals $[0, \frac{1}{2}]$ and $[\frac{1}{2},1]$. Let us consider the first one. The fact that $f$ and $g$ are single-defined yields:
\[
\textstyle f(x)=\al [1-(2x-1)^2 e^{-2d_{\al}x}]\ \nor{with}\ d_{\al}=\frac{{\theta}_1}{2 \al}-2\ \nor{in}\  \left[ 0,\frac{1}{2} \right]
\]
and
\[
\textstyle g(x)=\be [1-(2x-1)^2 e^{-2d_{\be}x}]\ \nor{with}\ d_{\be}=\frac{{\theta}_1}{2 \be}-2\ \nor{in}\ \left[ 0,\frac{1}{2} \right].
\]
Since $\al >\be >0$, we can consider $\be = \frac{\al}{c}$ with $c>1$, so:
\[
\textstyle f(x)=\al \left[1-(2x-1)^2 e^{-2\left( \frac{{\theta}_1}{2 \al}-2 \right) x} \right]\  \nor{in}\  \left[ 0,\frac{1}{2} \right]
\]
and
\[
\textstyle g(x)=\frac{\al}{c} \left[ 1-(2x-1)^2 e^{-2 \left( \frac{{\theta}_1}{2 \left( \frac{\al}{c} \right)}-2 \right) x} \right]\  \nor{in}\ \left[ 0,\frac{1}{2} \right].
\]
For the sake of simplicity, we replace $\frac{{\theta}_1}{\al}$ by $A$ (with $A>0$), and since $\al < \frac{{\theta}_1}{4}$ (by the definition of a single-defined function), we also have $A>4$. \\ In order to prove that:
\[
\textstyle \al \left[1-(2x-1)^2 e^{-2\left( \frac{A}{2}-2 \right) x} \right] \geq \frac{\al}{c} \left[ 1-(2x-1)^2 e^{-2 \left( \frac{cA}{2}-2 \right) x} \right]\  \nor{in}\ \left[ 0,\frac{1}{2} \right] ,
\]
we argue as follows:
\begin{align*}
~~~~~~~~~~~~~~~~~~ & \textstyle \hspace{-2cm}\al \left[1-(2x-1)^2 e^{-2\left( \frac{A}{2}-2 \right) x}  \right] \geq \frac{\al}{c} \left[ 1-(2x-1)^2 e^{-2 \left( \frac{cA}{2}-2 \right) x} \right]\  \nor{in}\ \left[ 0,\frac{1}{2} \right] \\ 
&\textstyle ~~\Longleftrightarrow ~~
c \left[1-(2x-1)^2 e^{(-A+4)x} \right] \geq  \left[ 1-(2x-1)^2 e^{(-cA+4)x} \right]\  \nor{in}\ \left[ 0,\frac{1}{2} \right] \\
&\textstyle ~~\Longleftrightarrow ~~c-c(2x-1)^2e^{-Ax+4x}\ \geq 1-(2x-1)^2 e^{-cAx+4x}\ \nor{in}\ \left[ 0,\frac{1}{2} \right] \\
&\textstyle ~~\Longleftrightarrow ~~c-1\ \geq c(2x-1)^2e^{-Ax+4x}-(2x-1)^2 e^{-cAx+4x}\ \nor{in}\ \left[ 0,\frac{1}{2} \right] \\
&\textstyle ~~\Longleftrightarrow ~~ c-1\ \geq c(2x-1)^2e^{-Ax}e^{4x}-(2x-1)^2 e^{-cAx}e^{4x}\ \nor{in}\ \left[ 0,\frac{1}{2} \right] \\
&\textstyle ~~\Longleftrightarrow ~~ c-1\ \geq (2x-1)^2e^{4x}[ce^{-Ax}-e^{-cAx}]\ \nor{in}\ \left[ 0,\frac{1}{2} \right].
\end{align*}
So, let us consider the functions $B(x)=(2x-1)^2e^{4x}$, $A(x)=ce^{-Ax}-e^{-cAx}$, and $C(x)=B(x)A(x)$. We have:
\[
B(0)=1,\ B\left(\frac{1}{2} \right )=0\ \nor{and}\ B'(x)= 8x(2x-1)e^{4x}.
\]
Thus, $B'(x)\leq 0$ in $\left[0, \frac{1}{2} \right]$, therefore $B(x)$ is a decreasing and positive function in $\left[0, \frac{1}{2} \right]$.

For  $A(x)$, we have:
\[
A(0)=c-1,\quad A\left(\frac{1}{2}\right) = e^{-\frac{A}{2}}\left( c - e^{-(c-1)\frac{A}{2}} \right) > 0,\quad \nor{and}\quad A'(x)=cA(-e^{-Ax}+e^{-cAx}).
\]
Thus, $A'(x)\leq 0$ in $\left[0, \frac{1}{2} \right]$ (since $A>1$), so $A(x)$ is a decreasing and positive function in $\left[0, \frac{1}{2} \right]$ too.

Let us consider $C(x)=B(x)A(x)$. We have:
\[
C(0)=c-1, \quad  C\left(\frac{1}{2}\right) =0, \quad \nor{and} \quad C'(x)=B'(x)A(x)+A'(x)B(x).
\]
Therefore, thanks to the properties of $A(x),B(x)$ and of their derivatives, also $C(x)$ is a decreasing and positive function in $\left[0, \frac{1}{2} \right]$ and in particular $C(0)=c-1$. Hence:
\[
c-1\ \geq (2x-1)^2e^{4x}[ce^{-Ax}-e^{-cAx}]\ \nor{in}\ \left[ 0,\textstyle{\frac{1}{2}} \right].
\]
For the second interval, $\left[\frac{1}{2}, 1 \right]$, the proof is similar.

The proof for the case $\al < \be < 0$ proceeds likewise. \qed 

This result can be extended to two $[\al, {\theta}_1, {\theta}_2]$-defined functions.

\begin{lemma} Let $f$ be an $[\al, {\theta}_1, {\theta}_2]$-defined function and $g$ a $[\be, {\theta}_1, {\theta}_2]$-defined function of the same type (both single-, double- or special-defined), with $| \al | > |\be |$; then $ |f|\geq |g|$ in $[0,1]$.
\end{lemma}
\dimostraz The single-single case is \Lem{lem:1.9}. Let us consider the double-double case. By \Def{def:1.4}, both $f$ and $g$ are built up from two single-defined functions rescaled in the interval $[0, \frac{1}{2}]$ and $[\frac{1}{2}, 1]$. The claim easily follows by applying the previous lemma to these single-defined functions. 

The special-special case is similar, with the interval $[0,1]$ split into three parts, instead of two.  \qed 

In a specific step of the algorithm-correctness proof, we will need the following lemma, together with its subsequent generalization.

\begin{lemma}\label{lem:opens} Let $\sI=\{I_k \, | \, k \! \in  \sK \}$ be a finite set of open, bounded non-empty real intervals indexed by a finite set $\sK$, with $I_k={]\al_k, \be_k[}$, equipped with a partial order $\leq$. If the condition
\[
\forall i,j \in \sK \ \big( I_j \leq I_i \Rightarrow \al_j < \be_i \big)
\]
holds, then there exists a collection  $\{ \phi_k \, | \, k \! \in \! \sK \}$ of elements such that $\phi_k \! \in \! I_k$ and if $I_j \leq I_i$, then $\phi_j \leq \phi_i$.     
\end{lemma}

\dimostraz Given $I_k \in \sI$, let us define:
\[
\bar{\be}_k \putAs \min \{ \be_j \, | \, I_j > I_k \} \ \ \mbox{and} \ \ \bar{\phi}_k \putAs \max \{ \phi_j \, | \, I_k > I_j \}.
\]
$\phi_k$ is defined by induction over $\leq$ as follows:
\begin{itemize}

    \item if $I_k$ is a minimal element with respect to $\leq$, we set:
    \[
       \phi_k \putAs \frac{\al_k + \min\{\be_k, \bar{\be}_k\}}{2};
    \]
    \item assume that, for all $I_j \leq I_k$, $\phi_j$ is defined, then we set:
    \[
       \phi_k \putAs \frac{\max\{\al_k,\bar{\phi}_k\} + \min\{\be_k, \bar{\be}_k\}}{2}.
    \]
\end{itemize}
The following two properties are now proved by induction on $\leq$:
\begin{itemize} 

    \item $\phi_k \! \in \! I_k$, i.e., $\al_k < \phi_k < \be_k$;

    \item if $I_j \leq I_k$, then $\phi_j \leq \phi_k$.
    
\end{itemize}

\noindent \ita{Base case:} If $I_k$ is minimal with respect to $\leq$, then $\al_k < \phi_k = \frac{\al_k + \min\{\be_k, \bar{\be}_k\}}{2} < \be_k$ and the second property is vacuously true.

\noindent \ita{Inductive step:} In this case $\phi_k = \frac{\max\{\al_k,\bar{\phi}_k\} + \min\{\be_k, \bar{\be}_k\}}{2}$. Assuming that $\bar{\phi}_k = \phi_h$ , with $I_h < I_k$, and that the two properties are valid for all $I_j < I_k$, we prove that they also apply to $I_k$. The following properties hold:
\[
\begin{array}{rl} \vspace{0.1 cm}
{\al}_k \! < \! {\be}_k : & \mbox{by construction}, \\ \vspace{0.1 cm}
{\al}_k \! < \! {\bar{\be}}_k: & \mbox{because if}\  I_g \! \geq \! I_k \ \mbox{then}\ {\be}_g \! > {\al}_k , \\ \vspace{0.1 cm}
{\phi}_h \! < \! {\be}_k: & \mbox{since}\ {\phi}_h=\frac{\max \{ {\al}_h , {\bar{\phi}}_h \} + \min \{ {\be}_h , {\bar{\be}}_h \} }{2}\ \mbox{and thus}\ {\phi}_h \! < \! {\bar{\be}}_h \! \leq \! {\be}_k ,\\ \vspace{0.1 cm}
{\phi}_h \! < \! {\bar{\be}}_k : & \mbox{because}\ I_k \! \geq \! I_h \ \mbox{and thus}\ {\phi}_h \! < \! {\bar{\be}}_h \! \leq \! {\bar{\be}}_k ,\\
\max \{ {\al}_k, {\bar{\phi}}_k \}\! < \! \min \{ {\be}_k , {\bar{\be}}_k \} : & \mbox{by the previous properties}.
\end{array}
\]
From the last point we obtain that:
\[
\al_k \leq \max \{ {\al}_k, {\bar{\phi}}_k \} <  \phi_k  <  \min \{ {\be}_k , {\bar{\be}}_k \}  \leq \be_k,
\]
concluding the inductive step. \eod

The previous lemma can be generalized as follows.

\begin{lemma}\label{lem:intervals} Let $\sI=\{I_k \, | \, k \! \in \! \sK \}$ be a finite set of bounded non-empty real intervals indexed by $\sK$, with $\al_k \putAs \inf I_k$ and $ \be_k \putAs \sup I_k$, furnished with a partial order $\leq$. If the condition
\[
\forall i,j \in \sK \ \big( I_j \leq I_i \, \Rightarrow \, \exists x,y \,( x \! \in \! I_j \et y \! \in \! I_i \et x \leq y ) \big)
\]
holds, then there exists a collection of elements $\{ \phi_k \, | \, k \! \in \! \sK \}$, with $\phi_k \! \in \! I_k$, such that if $I_j \leq I_i$, then $\phi_j \leq \phi_i$. 
    
\end{lemma}
\dimostraz The proof of \Lem{lem:opens} can be easily extended to the general case. \eod

At the end of this part, we recall some useful properties about continuous and differentiable real functions and their derivatives.
\begin{lemma}\label{lem:1.11} Let $f \colon[a,b]\frec \mR$ be a continuous and differentiable function defined in $[a,b]$. Let $k\in \mR$ and suppose $f'(x)\geq k$ (resp. $\leq $) for all $x\in [a,b]$. If $\frac{f(b)-f(a)}{b-a}=k$ then $f$ is linear in $[a,b]$, and $f'(x)=k$ holds for all $x\in [a,b]$.
\end{lemma}
\dimostraz The proof is a straightforward application of the mean value theorem and is omitted here. \qed 

\begin{lemma}\label{lem:1.12}
Let $f \colon[a,b]\frec \mR$ be a continuous and differentiable function defined in $[a,b]$. Suppose $f$ be convex (resp. concave) in $[a,b]$. If $\frac{f(b)-f(a)}{b-a}$ is equal to $f'(a)$ or $f'(b)$, then $f$ is linear in $[a,b]$, and $f'(x)=f'(a)$ holds for all $x\in [a,b]$.
\end{lemma}
\dimostraz We omit the proof, which is again a straightforward application of the mean value theorem. \qed 

\begin{lemma}\label{lem:1.13}
Let $f,\ g$ be real functions defined and differentiable in the closed and bounded interval $[ a ,b ]$, endowed with the second derivative in $a$. If $f(a) \! = \! g(a)$, $f'(a)=g'(a)$ and $f''(a) > g''(a)$, then there exists $c \! \in ]a , b[$ such that $f(x) \! > \! g(x)$ for all $x \! \in ]a, c[$.
\end{lemma}
\dimostraz We have:
\[
f''(a) > g ''(a) ~~\Longleftrightarrow~~ f''(a) - g''(a) >0 ~~\Longleftrightarrow~~ (f'-g')'(a) >0.
\]
Therefore, $(f'-g')'$ is an increasing function in $a$, then there exists $c \! \in ]a , b[$ such that:
\[
f'(x)-g'(x) > f'(a)-g'(a) = 0,
\]
for all $x \! \in ]a , c[$. Let $\bar{x} \! \in ]a ,c]$. Because of the Mean Value theorem there exists $ \ep \! \in ]a, \bar{x} [$:
\[
(f(\bar{x})-g(\bar{x}))-(f(a)-g(a)) = (f'(\ep) - g'(\ep )) \x (\bar{x} - a).
\]
We have $f'(\ep ) - g'( \ep ) > 0$ and $f(a) \! = \! g(a)$. Then $f(\bar{x}) - g(\bar{x}) \! > \! 0$, thus $ f(\bar{x}) \! > \! g(\bar{x})$. \qed 

\begin{lemma}\label{lem:1.14} Let $f,\ g$ be two real functions defined and derivable in the closed and bounded interval $[a , b]$, with $f(a) \! = \! g(a)$ (resp. $f(b) \! = \! g(b)$). If $f(x) \! > \! g(x)$ for all $x \! \in ]a , b[$, then $f'(a) \! \geq \! g'(a)$ (resp. $f'(b) \! \leq \! g'(b)$).
\end{lemma} 

\dimostraz We consider the case $f(a) \! = \! g(a)$; the case $f(b) \! = \! g(b)$ can be treated likewise.\\
From the hypotheses we get:
\[
\scalebox{1.2}{$\frac{ (g(x)-f(x))-(g(a)-f(a))}{x-a}$} <0
\]
for every $x \in {]a ,b[}$; then
\[
\textstyle g'(a)-f'(a)=(g-f)'(a) = \lim_{x\frec a^+} \scalebox{1.2}{$\frac{(g(x)-f(x))-(g(a)-f(a))}{x-a}$}  = \lim_{x\frec a^+} \scalebox{1.2}{$\frac{(g(x)-f(x))-(g(a)-f(a))}{x-a}$} \leq 0.
\]

Thus, $f'(a) \! \geq \! g'(a)$. \qed

\subsection{Correctness}\label{subsec:correctness}

In this section, we prove the correctness of the decision procedure previously described.

\begin{theorem} The two last formulas ${\fhi}_3$ and ${\fhi}_4$ of the decision algorithm for \RDFp  theory are equisatisfiable.
\end{theorem}
\dimostraz The idea behind the proof is that if a model, however complex, for ${\fhi}_3$ exists, then it is possible to build a simplified model that involves only functions defined as piecewise exponential functions. 

Let $F$ and $V \! =\! \{v_1, v_2, \dots , v_r\} $ be, respectively, the set of the function variables and the set of the ordered domain variables of ${\fhi}_3$.

Given an arbitrary model $M$, let $MV = \{{\eta}_i \in \mR \, |\,  {\eta}_i = Mv_i,\ \mbox{with}\ v_i \in V \}$ be the set of all real values associated with the domain variables in $V$ by $M$.
Moreover, let $\bar{v}$ denote the interpretation $Mv$ of the numerical variable $v$ under the model $M$; this notation also holds for numeric variables that are not domain variables.

The proof consists of two parts.

\noindent {\boldmath\large${\fhi}_4 \imp {\fhi}_3 )$} Let $M$ be a model for $\fhi_4$; our goal is to construct a model that satisfies $\fhi_3$. Specifically, we aim at extending $M$ to include the function variables appearing in $\fhi_3$ in a way ensuring that all clauses in ${\fhi}_3$ are satisfied.

The approach involves considering, for each function variable $f$, an appropriate perturbation of the piecewise function that passes through all points $(\bar{v}_i,\bar{y}_i^f)$ where $v_i \! \in \! V$. The perturbation we apply will take the form of an exponential adjustment built up using $[\al ,{\theta}_1, {\theta}_2]$-defined functions; the proof will amount to showing that for $\al \! >\! 0$  small enough all clauses in ${\fhi}_3$ will be satisfied.

For simplicity, we define the real value
\[
{\theta}_i^f (u) \putAs ({\eta}_{i+1}-{\eta}_i )\x u -({\bar{y}}_{i+1}^f -{\bar{y}}_i^f ),
\] 
for each function variable $f \! \in \! F$ and for each $i \! \in \! \{ 1,2,\dots , r-1\}$, where $u\in \mR $ is a suitable parameter.

Consider the following sets:
\begin{align*}
A_1 & \putAs    \left\{{\theta}_j^f (u)\! \in \! \mR |\, j \! \in \! \{ 1,2,\dots ,r-1 \} \ \mbox{and} \ u \! \in \! \{ {\bar{t}}_j^f ,{\bar{t}}_{j+1}^f \} \right\} ,\\
A_2 & \putAs    \left\{ \textstyle{\frac{1}{2}}({\bar{y}}_j^f -{\bar{y}}_j^g ) \! \in \! \mR |\, (f>g)_{[w_1, w_2 ]}\ \mbox{occurs in}\ {\fhi}_3 \ \mbox{and} \ j \! \in \! \{ \ind(w_1), \dots , \ind(w_2) \}   \right\}, \\
A_3 & \putAs    \left\{ {\theta}_j^f (\bar{y}) \! \in \! \mR |\, (D[f]\bowtie y)_{[z_1 ,z_2]}\ \mbox{occurs in} \ {\fhi}_3 \ \mbox{and} \ j\! \in \! \{ \ind(z_1), \dots, \ind(z_2)-1 \}    \right\}, \\
A_4 & \putAs   \left\{ {\theta}_j^f (0) \! \in \! \mR |\, \mbox{Strict}\_ \mbox{Up} (f)_{[z_1, z_2]}\ \mbox{or}\ \mbox{Strict}\_ \mbox{Down}(f)_{[z_1 z_2]}\ \mbox{occur in}\ {\fhi}_3    \mbox{ and }  j \! \in \! \{ \ind(z_1), \dots , \ind(z_2)-1 \}   \right\},
\end{align*}
and, let
\[
A \putAs \left\{ |x| \in {\mR}^+  \Bigg|  \ x\in \bigcup_{i=1}^4 A_i,\, x\not= 0 \right\} .
\]

In the trivial case $A  =  \emptyset $, put $m  =  0$ and $\I =\{ 0\}$, the following proof is then straightforward; otherwise, pick $m \! \in \! {\mR}^+$ so that it satisfies $0<m<\mbox{min}(A)$ and let $\I$ be the set $]0, \frac{1}{8}m[$.

For each function variable $f \in  F$, we introduce the following function family $\{ (Mf)_{\al}{\}}_{\al \in \I }$, with $(Mf)_{\al} \colon \mR \frec \mR $ defined as:\footnote{To properly treat literals of the form $(f>g)_A$ over unbounded intervals, a refinement of this definition is needed; to avoid cluttering, this further enrichment is presented only in the part of the proof where it is required.}
\begin{align*}
(Mf)_{\al}(\eta ) &\putAs 
\begin{cases}
{\bar{y}}_1^f +({\bar{\ga}}_0^f -{\bar{t}}_1^f )(1-e^{\eta -{\eta}_1})+{\bar{\ga}}_0^f ({\eta}_1 -\eta ) & \text{if }\ \eta \! \in ]-\infty , {\eta}_1[,\\[.1cm]
s_i^f (p_i (\eta ))+ c_i^f (p_i (\eta )) & \text{if }\ \eta \! \in \! [{\eta}_i, {\eta}_{i+1} [,\\[.1cm]
{\bar{y}}_r^f +( {\bar{t}}_r^f - {\bar{\ga}}_r^f)(1-e^{{\eta}_r -\eta})+{\bar{\ga}}_r^f (\eta -{\eta}_r ) & \text{if }\ \eta \! \in \! [ {\eta}_r, +\infty[,
\end{cases}
\intertext{with:}
s_i^f (x) &\putAs \bar{y}_i^f +(\bar{y}_{i+1}^f -\bar{y}_i^f)\x x, & \forall x \! \in \! [0,1] \\
p_i (\eta ) &\putAs \frac{\eta -{\eta}_i}{{\eta}_{i+1} -{\eta}_i}, & \forall \eta \! \in \! [{\eta}_i ,{\eta}_{i+1}]
\end{align*}
and $c_i^f  \colon  [0,1]\frec \mR$ is the $[\al ,{\theta}_i^f (\bar{t}_i^f ), {\theta}_i^f (\bar{t}_{i+1}^f )]$-defined function.

Here, $p_i$ denotes the projection from $[{\eta}_i ,{\eta}_{i+1}]$ to $[0,1]$; thus, $s_i^f (p_i (\eta ))$ is simply the linear interpolant between $({\eta}_i ,\bar{y}_i^f )$ and $({\eta}_{i+1}, \bar{y}_{i+1}^f )$, whereas $c_i^f (p_i (\eta ))$ is the exponential perturbation that adjust the function to satisfy the conditions on the derivative.

The following result is the basis of our theorem:
\begin{lemma}\label{lem:2.2} Let $f\! \in \! F$ be a function variable occurring in ${\fhi}_3$ and let $\al \! \in \! \I$. Then $(Mf)_{\al}$ is a continuous and differentiable function defined on $\mR $, with a continuous derivative,  and we have, for each $i\! \in \! \{ 1, 2, \dots ,r\}$:
\[
(Mf)({\eta}_i)=\bar{y}_i^f ,\ \ and\ \ (Mf)'({\eta}_i)=\bar{t}_i^f\,.
\]
\end{lemma}
\dimostraz Let $\al \! \in \! \I$. Initially we have to verify that the function $(Mf)_{\al}(\eta)$ is well-defined for each $\eta \! \in \! \mR$. For each $i \! \in \! \{ 1, 2, \dots , r-1 \}$, we have ${\theta}_i^f ({\bar{t}}_i^f )= {\theta}_i^f ({\bar{t}}_{i+1}^f)=0$ or
\[
0<|4\al |<m<\mbox{min}\{ |x| \in T|\ x\not= 0\},\ \mbox{where}\ T=\{ {\theta}_i^f ({\bar{t}}_i^f),  {\theta}_i^f ({\bar{t}}_{i+1}^f) \},
\]
because of the minimality of $m$ with respect to $A$; the existence of the $[\al , {\theta}_i^f ({\bar{t}}_i^f), {\theta}_i^f ({\bar{t}}_{i+1}^f)]$-defined functions $c_i^f$ is then given by \Lem{lem:1.8}.\\
We have $(Mf)({\eta}_r)={\bar{y}}_r^f$, and for each $i \! \in \! \{ 1, 2, \dots , r-1 \}$ it holds that $ (Mf)({\eta}_i)=s_i^f (p_i ({\eta}_i)) + c_i^f (p_i ({\eta}_i )) ={\bar{y}}_i^f $ because $c_i^f (0)=0$.\\
The function is continuous and differentiable, with a continuous derivative, in each interval $]-\infty , {\eta}_1 [,\ ]{\eta}_i , {\eta}_{i+1}[$ with $i \! \in \! \{1, 2, \dots , r-1 \}$, $]{\eta}_r, +\infty [$. Furthermore, for each $j \! \in \! \{ 1, 2, \dots , r\}$:
\[
\lim_{\eta \, \frec {\eta}_j^{-}} (Mf)_{\al}(\eta) = (Mf)_{\al}({\eta}_j )= \lim_{\eta \, \frec {\eta}_j^{+}} (Mf)_{\al}(\eta ),
\] 
thus, the function is continuous in $\mR$.

About differentiability, we have for each $j\in \{ 1, 2, \dots , r-1 \}$:
\begin{align*}
\lim_{\eta \frec {\eta}_j^{+}} (Mf)_{\al}'(\eta ) & = \lim_{\eta \frec {\eta}_j^{+}} \left[ (s_j^f (p_j))'(\eta ) + (c_j^f(p_j))'(\eta) \right] \\
 & = \lim_{\eta \frec {\eta}_j^{+} }   p_j'(\eta) \x { \left[ (s_j^f)'(x) + \x (c_j^f)'(x)\right] }_{ x=p_j (\eta) } \\
 & = \lim_{x \frec 0^{+}}   p_j'({\eta}_j ) \x \left[ (s_j^f)'(x) + (c_j^f)'(x) \right] \\
&  = \left[ ({\bar{y}}_{j+1}^f - {\bar{y}}_j^f )+ {\theta}_j^f ({\bar{t}}_j^f )\right] \x \frac{1}{{\eta}_{j+1}- {\eta}_j} \\
& = {\bar{t}}_j^f =\lim_{\eta \frec {\eta}_j^{-}} (Mf)_{\al}'(\eta ),
\end{align*}
because $(c_j^f)'(0)= {\theta}_j^f ({\bar{t}}_j^f)$.
 
Moreover, the following holds:
\[
\lim_{\eta \, \frec {\eta}_1^-}(Mf)_{\al}'(\eta) = \lim_{\eta \, \frec {\eta}_1^-} \left[ {\bar{\ga}}_0^f + ({\bar{t}}_1^f - {\bar{\ga}}_0^f )e^{\eta - {\eta}_1 } \right] = {\bar{t}}_1^f = \lim_{\eta \, \frec {\eta}_1^+}(Mf)_{\al}'(\eta );
\]
and the point ${\eta}_r$ can be treated in the same way.

Thus, $(Mf)_{\al}$ is differentiable with a continuous derivative on $\mR$, and we have $(Mf)_{\al}'({\eta}_j )= {\bar{t}}_j^f$ for each $j \in \{ 1, 2, \dots ,r \}$. \qed 

The remaining nontrivial task is to demonstrate that, for a sufficiently small $\al$ (since $\al$ is the only free parameter in the definition of $(Mf)_{\al}$), $(Mf)_{\al}$ satisfies all clauses of $\fhi_3$, such as $(f>g)_{[w_1 ,w_2 ]}$ or $(f \not= g)_{[z_1 ,z_2 ]}$. To establish this, we examine each type of clause and prove that, for some appropriate $\al>0$, these clauses hold true in our extended real model $M$. Given that there are a finite number of clause types and intervals, this suffices to prove the claim.
In the discussion that follows, we generally omit the subscript $\al$ for simplicity.

\paragraph{Case \boldmath($f$):}
Let $x\! =\! f(v_j)$ be a literal occurring in ${\fhi}_3$ for some $j\! \in \! \{ 1, 2, \dots ,r\}$.

 We have:
\[
M(f(v_j))=(Mf)(Mv_j)=(Mf)({\eta}_j)=\bar{y}_j^f =My_j^f =Mx
\]
where the last equality is true because the literal $x=y_j^f$, introduced in step \ref{itemAlgoThree}), occurs in ${\fhi}_4$.

\paragraph{Case \boldmath($D$):}
Let $x\! = \! D[f](v_j)$ be a literal occurring in ${\fhi}_3$ for some $j\! \in \! \{ 1, 2, \dots ,r\}$. Analogously to the previous case, in step \ref{itemAlgoThree}) the literal $x\! =\! t_j^f$ was added, and we have:
\[
M(D[f](v_j))=(Mf)'(Mv_j)=(Mf)'({\eta}_j)=\bar{t}_j^f =Mt_j^f =Mx.
\]

\paragraph{Case \boldmath$(f\mathord{=})$:}
Let $(f=g)_{[z_1 z_2]}$ be a literal occurring in ${\fhi}_3$ and let $Mz_1 \! \leq \! Mz_2$, otherwise the literal is vacuously true. We have to verify that for each $\eta \! \in \! [Mz_1, Mz_2]\subset \mR$ the following holds:
\[
(Mf)(\eta )=(Mg)(\eta ).
\]
Such equality holds for each ${\eta}_i \! \in \! MV$ with $i\! \in \! \{ \ind(z_1), \dots ,\ind(z_2)\}$ since, from clauses in step \ref{itemAlgoFour}.\ref{alg4:actA}), we have:
\[
(Mf)({\eta}_i )=\bar{y}_i^f =\bar{y}_i^g =(Mg)({\eta}_i ).
\] 
The equality also holds for each $\eta \! \in  ]{\eta}_j, {\eta}_{j+1}[$ with $j\! \in \! \{ \ind(z_1) ,\dots , \ind(z_2)-1 \}$:
\[
\begin{array}{lll} \vspace{0.05 cm}
(Mf)(\eta ) & = & s_j^f (p_j (\eta )) +c_j^f (p_j (\eta ))= \\
            & = & s_j^g (p_j (\eta )) + c_j^g (p_j (\eta ))=(Mg)(\eta ),
\end{array}            
\]
where $c_j^f (p_j (\eta ))\! =\! c_j^g ( p_j (\eta ))$ for each $\eta \! \in \! [{\eta}_j ,{\eta}_{j+1}]$ derives from \Cor{cor:1.3} and the conditions ${\theta}_j^f (\bar{t}_j^f )\! =\! {\theta}_j^g (\bar{t}_j^g )$ and ${\theta}_j^f (\bar{t}_{j+1}^f ) \! =\! {\theta}_j^f (\bar{t}_{j+1}^f )$, which hold because of the literals $t^f_i=t^g_i$ introduced in step \ref{itemAlgoFour}.\ref{alg4:actA}).

The proof for the extremal intervals, $]\! -\! \infty , {\eta}_1 ]$ and $[{\eta}_r, +\infty [ $, derives directly from the definition of $(Mf)$ in them, since ${\bar{\ga}}_0^f = {\bar{\ga}}_0^g$ and  ${\bar{\ga}}_r^f = {\bar{\ga}}_r^g$, again because of the literals added in step \ref{itemAlgoFour}.\ref{alg4:actA}).

\paragraph{Case \boldmath $(f\mathord{>})$:} Since the treatment of atoms of the form $(f>g)_A$ required different cases and sub-cases, we start with a presentation of the general structure of the completeness proof for this type of literals. 

\emph{First, we consider three subcases: bounded and closed intervals; bounded and open or semi-open intervals; unbounded and closed intervals. Moreover, the treatment for open or semi-open intervals requires itself some case distinctions depending on the type of the interpolating functions involved, i.e. single-, double- or special-defined.}

\paragraph{Subcase \boldmath $(f\mathord{>})$, bounded and closed intervals:}

We start analyzing this case by treating bounded and closed intervals.

Let $(f>g)_{[w_1, w_2]}$ be a literal occurring in ${\fhi}_3$ and let $Mw_1 \leq Mw_2 $, otherwise the literal is vacuously true. We have to verify that for each $\eta \in [Mw_1 ,Mw_2 ] \subset \mR$ the following holds:
\[
(Mf)(\eta) > (Mg)(\eta) .
\] 
This is true for each ${\eta}_i \! \in  \! MV$ with $i \! \in \! \{ \ind(w_1), \dots, \ind(w_2) \}$, since $(Mf)({\eta}_i) = {\bar{y}}_i^f > {\bar{y}}_i^g = (Mg)({\eta}_i)$ from clauses introduced in step \ref{alg4:actionB:case1}.

Moreover, the above inequality also holds for each $\eta \! \in \! [{\eta}_j , {\eta}_{j+1} ]$ with $j\in \{ \ind(w_1), \dots, \ind(w_2)-1 \}$. In fact, consider the points $\left( 0, \frac{1}{2} ({\bar{y}}_j^f +{\bar{y}}_j^g) \right)$ and $\left( 1, \frac{1}{2} ({\bar{y}}_{j+1}^f +{\bar{y}}_{j+1}^g) \right)$ in ${\mR}^2 $, and let $r_j(x)$ be the straight line interpolating such points:
\[
r_j(x)= \frac{1}{2} \left({\bar{y}}_{j+1}^f +{\bar{y}}_{j+1}^g - {\bar{y}}_j^f -{\bar{y}}_j^g \right)\x x + \frac{1}{2}\left({\bar{y}}_j^f + {\bar{y}}_j^g\right),\ \ \forall x\in [0,1].
\]
The sought inequality $(Mf)(\eta) > (Mg)(\eta)$ follows from the fact that, for $\al$ sufficiently small, the functions $(Mf)(\eta)$ and $(Mg)(\eta)$ lie, respectively, above and below the straight line $r_j (p_j (\eta))$ when $\eta$ ranges over the interval $[{\eta}_j ,{\eta}_{j+1}]$. In particular, let
\[
l \putAs \mbox{max}\left\{ \frac{1}{2} ({\bar{y}}_j^g - {\bar{y}}_j^f), \frac{1}{2} ({\bar{y}}_{j+1}^g - {\bar{y}}_{j+1}^f) \right\};
\]
we have:
\[
l=\underset{x\in [0,1]}{\mbox{max}} [r_j (x) - s_j^f (x) ]\geq r_j(x) - s_j^f (x),\ \ \forall x\in [0,1]. 
\]
By the minimality of $\al$ with respect to $A$,  $0<\! \al \! <\! \frac{1}{2} ({\bar{y}}_j^f - {\bar{y}}_j^g )$ and $0<\! \al \! <\!  \frac{1}{2}({\bar{y}}_{j+1}^f - {\bar{y}}_{j+1}^g)$ hold, and by \Lem{lem:1.7}, we have $| c_j^f(x)| \leq \al$ for each $x\in [0,1]$; hence, we conclude:
\[
c_j^f(x) \geq -\al > l\geq r_j(x)-s_j^f(x),\ \ \forall x\in [0,1].
\]
Analogously, we have $c_j^f (x) < r_j(x) - s_j^g(x)$ for each $x\in [0,1]$.

Thus, for each $\eta \in  [{\eta}_j, {\eta}_{j+1}]$:
\[
(Mf)(\eta)= { \left[ s_j^f(x) + c_j^f(x) \right] }_{x=p_j (\eta)} > {[ r_j (x)]}_{x=p_j (\eta)} > (Mg)(\eta) .
\]
Now we consider the case of open or semi-open bounded intervals that is the most involved one.

\smallskip

\paragraph{Subcase \boldmath $(f\mathord{>})$, bounded and open or semi-open intervals:}

Let $(f>g)_{]w_1, w_2[}$ be a literal that occurs in ${\fhi}_3$ and $w_1 = v_k , v_{k+1} , \dots , v_{k+n} = w_2 $ the $n+1$ domain variables inside $[w_1, w_2]$; because of step \ref{alg1:action2:case3} of the algorithm, $n\! \geq \! 2$, i.e., there is at least one domain variable $v$ such that $w_1 \! \! < \!  \! v \! < w_2$. These variables divide $[w_1 , w_2]$ into $n$ intervals $ [v_k , v_{k+1}],\ [v_{k+1}, v_{k+2}],$ $ \dots , [v_{k+n-1} , v_{k+n}]$. Let us first consider the $n-2$ intervals $[v_{k+1} ,v_{k+2}], \dots , [v_{k+n-2} , v_{k+n-1}]$ (if $n=2$ there are no such intervals). Since these intervals are closed and bounded, we can apply the same strategy used before. It remains to be proved that $f$ exceeds $g$ in the two extremal intervals $] v_k , v_{k+1}] = ]w_1 , v_{k+1}]$ and $[v_{k+n-1} , v_{k+n}[ = [v_{k+n-1}, w_2 [$.

Our treatment is limited to the interval $]v_k, v_{k+1}]= ]w_1 , v_{k+1}]$, the other case being closely analogous. We want to prove that, when $\al \! \in \! I$ is sufficiently small, $(Mf)_{\al} (\eta) \! > \! (Mg)_{\al} (\eta)$ holds for all $ \eta \in ]{\eta}_k , {\eta}_{k+1}]$; in the following, it will be necessary to make the value $\al$ explicit. We begin by proving the following two propositions.

\begin{proposition}\label{prop:2.3} Given a model $M$ as before,  for every $\bar{\eta} \!  \in  ]{\eta}_k ,{\eta}_{k+1}]$ there exists an $\bar{\al} \! \in \! \I$ such that for every $\eta \! \in ]{\bar{\eta}} , {\eta}_{k+1}]$ we have $(Mf)_{\bar{\al}}(\eta) \! > \! (Mg)_{\bar{\al}}(\eta)$. Moreover, this holds for every $\al \!  \in \! \I$ such that $\al < \bar{\al}$.
\end{proposition}

\dimostraz Because of the steps \ref{itemAlgoOne}.\ref{alg1:actionOne}) and \ref{itemAlgoFour}) of the algorithm, we have ${\bar{y}}_k^f = {\bar{y}}_k^g$. Therefore, for every $\eta \in ]{\eta}_k , {\eta}_{k+1}]$ and for every $\al \in \I$, we have:
\[
\begin{aligned}
&(Mf)_{\al} (\eta) > (Mg)_{\al} (\eta)\\
&\qquad \qquad \Longleftrightarrow~~ (Mf)_{\al} (\eta) - (Mg)_{\al} (\eta) >0 \\
&\qquad \qquad \Longleftrightarrow~~ s_k^f (p_k (\eta)) + c_k^f (p_k (\eta)) -s_k^g (p_k (\eta)) -c_k^g (p_k (\eta)) >0 \\
&\textstyle \qquad \qquad \Longleftrightarrow~~ {\bar{y}}_k^f + ({\bar{y}}_{k+1}^f - {\bar{y}}_k^f) \x \frac{\eta - {\eta}_k }{{\eta}_{k+1} - {\eta}_k} + c_k^f (p_k (\eta )) - {\bar{y}}_k^g - ( {\bar{y}}_{k+1}^g - {\bar{y}}_k^g ) \x \frac{\eta - {\eta}_k}{{\eta}_{k+1} - {\eta}_k} - c_k^g (p_k (\eta ))>0 \\
&\textstyle \qquad \qquad \Longleftrightarrow~~\ \ -c_k^f (p_k (\eta )) + c_k^g (p_k (\eta )) < \left( \frac{{\bar{y}}_{k+1}^f - {\bar{y}}_{k+1}^g}{{\eta}_{k+1} - {\eta}_k} \right) \x ( \eta - {\eta}_k )~~~~~~~~~~(\mbox{remembering that}\ {\bar{y}}_k^f = {\bar{y}}_k^g ).
\end{aligned}
\]
Because of step \ref{itemAlgoFour}.\ref{alg4:actC}) and \Lem{lem:1.7} we have ${\bar{y}}_{k+1}^f - {\bar{y}}_{k+1}^g >0$ and also:
\[
-c_k^f (p_k (\eta )) + c_k^g (p_k (\eta )) \leq | -c_k^f (p_k (\eta )) + c_k^g (p_k (\eta ))| \leq | c_k^f (p_k (\eta )) | + | c_k^g (p_k (\eta )) | \leq 2\al .
\]
It is sufficient to take $\al < \left( \frac{{\bar{y}}_{k+1}^f - {\bar{y}}_{k+1}^g}{{\eta}_{k+1} - {\eta}_k} \right) \x \frac {( \bar{\eta} - {\eta}_k )}{2}$. \qed 

\begin{proposition}\label{prop:2.4} Given a model $M$ as before, for every $\al \! \in \! \I$ sufficiently small, there exists an $\bar{\eta} \! \in ] {\eta}_k , {\eta}_{k+1}]$ such that $(Mf)_{\al} (\eta) > (Mg)_{\al}(\eta )$ for all $\eta \in ]{\eta}_k , \bar{\eta}]$.
\end{proposition}

\dimostraz Because of step \ref{itemAlgoFour}.\ref{alg4:actC}) of the algorithm, we have ${\bar{t}}_k^f \geq {\bar{t}}_k^g$, from this (by \Lem{lem:2.2}) it follows that for every $\al \! \in \! \I$ we have $(Mf)_{\al}'({\eta}_k) \geq (Mg)_{\al}'({\eta})$. We now separately consider the two cases ${\bar{t}}_k^f > {\bar{t}}_k^g$ and ${\bar{t}}_k^f = {\bar{t}}_k^g$.

If ${\bar{t}}_k^f > {\bar{t}}_k^g$, for every $\al \in \I$ we have:
\[
\begin{aligned}
(Mf)_{\al}'({\eta}_k) > (Mg)_{\al}'({\eta}_k) &~~\Longleftrightarrow~~   (Mf)_{\al}'({\eta}_k) - (Mg)_{\al}'({\eta}_k) > 0\\
&~~\Longleftrightarrow~~ ((Mf)_{\al}-(Mg)_{\al})'({\eta}_k)>0.
\end{aligned}
\]
Thus, $(Mf)_{\al} - (Mg)_{\al}$ is an increasing function in ${\eta}_k$. Moreover, since ${\bar{y}}_k^f = {\bar{y}}_k^g$ we have:
\[
(Mf)_{\al}({\eta}_k) = (Mg)_{\al}({\eta}_k),
\]
and therefore 
\[
(Mf)_{\al}({\eta}_k) - (Mg)_{\al}({\eta}_k)=0
\]
and there exists an $\bar{\eta}$ such that
\[
(Mf)_{\al}({\eta}) - (Mg)_{\al}({\eta})>0\ \ \ \forall \eta \in ]{\eta}_k ,\bar{\eta}],
\]
and hence
\[
(Mf)_{\al}({\eta}) > (Mg)_{\al}({\eta}) \ \ \ \forall \eta \in ]{\eta}_k , \bar{\eta} ].
\]

If ${\bar{t}}_k^f = {\bar{t}}_k^g $, the following equivalences hold:
\[
\begin{aligned}
{\bar{t}}_k^f = {\bar{t}}_k^g &~\Longleftrightarrow~ (Mf)_{\al}'({\eta}_k) = (Mg)_{\al}'({\eta}_k) \\
&~\Longleftrightarrow~  (s_k^f (p_k))'({\eta}_k) + (c_k^f (p_k))'({\eta}_k) = (s_k^g (p_k))'({\eta}_k) + (c_k^g (p_k))'({\eta}_k) ~  (\mbox{\small opening the two derivatives})~\\
&~\Longleftrightarrow~
\textstyle \frac{{\bar{y}}_{k+1}^f-{\bar{y}}_k^f}{{\eta}_{k+1} - {\eta}_k} + (c_k^f (p_k))'({\eta}_k) = \frac{{\bar{y}}_{k+1}^g - {\bar{y}}_k^g}{{\eta}_{k+1} - {\eta}_k} + (c_k^g (p_k))'({\eta}_k).
\end{aligned}
\]
We observe that
\[
\textstyle (c_k^f (p_k))'({\eta}_k) = {\theta}_k^f ({\bar{t}}_k^f) \x \frac{1}{{\eta}_{k+1} - {\eta}_k}\ \ \ \mbox{and} \ \ \ (c_k^g (p_k))'({\eta}_k)={\theta}_k^g ({\bar{t}}_k^g) \x \frac{1}{{\eta}_{k+1} - {\eta}_k}.
\]
This implies that $(c_k^f (p_k))'({\eta}_k)$ and $(c_k^g (p_k))'({\eta}_k)$ do not depend on $\al$. Moreover, it must hold that
\[
(c_k^f (p_k))'({\eta}_k) < (c_k^g (p_k))'({\eta}_k);
\]
else, if
\[
(c_k^f (p_k))'({\eta}_k) \geq (c_k^g (p_k))'({\eta}_k),
\]
then, we would obtain
\[
\frac{{\bar{y}}_{k+1}^f - {\bar{y}}_k^f}{{\eta}_{k+1} - {\eta}_k} \leq \frac{{\bar{y}}_{k+1}^g - {\bar{y}}_k^g }{{\eta}_{k+1} - {\eta}_k}.
\]
Applying step \ref{itemAlgoOne}.\ref{alg1:actionOne}), this would imply
\[
{\bar{y}}_{k+1}^f \leq {\bar{y}}_{k+1}^g.
\]
However, by step \ref{itemAlgoFour}.\ref{alg4:actC}), we know that
\[
{\bar{y}}_{k+1}^f > {\bar{y}}_{k+1}^g,
\]
which leads to a contradiction.

In case ${\bar{t}}_k^f = {\bar{t}}_k^g$ we have four subcases, namely $(c_k^f (p_k))'({\eta}_k) >0$ and $(c_k^g (p_k))'({\eta}_k) >0$; $(c_k^f (p_k))'({\eta}_k) <0$ and $(c_k^g (p_k))'({\eta}_k) <0$; $(c_k^f (p_k))'({\eta}_k) < 0$ and $(c_k^g (p_k))'({\eta}_k) \geq 0$; and finally, $(c_k^f (p_k))'({\eta}_k) \leq 0$ and $(c_k^g (p_k))'({\eta}_k) >0$. We consider the first one, whereas the others have a similar proof.

If $(c_k^f (p_k))'({\eta}_k) >0$ and $(c_k^g (p_k))'({\eta}_k) >0$, then
\[
{\theta}_k^f ({\bar{t}}_k^f) = ({\eta}_{k+1} - {\eta}_k ) \x ( c_k^f (p_k))'({\eta}_k) > 0
\]
and
\[
{\theta}_k^g ({\bar{t}}_k^g) = ({\eta}_{k+1} - {\eta}_k ) \x ( c_k^g (p_k))'({\eta}_k) > 0,
\]
besides, because of $(c_k^f (p_k))'({\eta}_k) < (c_k^g (p_k))'({\eta}_k)$, we have ${\theta}_k^f ({\bar{t}}_k^f) < {\theta}_k^g ({\bar{t}}_k^g)$. From definitions and properties of \Subsec{subsec:tools} we can compute the following second derivative:
\[
(c_k^f)''(0)= \left\{
\begin{array}{ll}
-({\theta}_k^f ({\bar{t}}_k^f))^2 \x \frac{1}{\al} + 8\al & \mbox{if}\ c_k^f\ \mbox{is single},\\[1pt]
-({\theta}_k^f ({\bar{t}}_k^f))^2 \x \frac{1}{\al} + 32\al & \mbox{if}\ c_k^f\ \mbox{is double or special},
\end{array}
\right.
\]
and
\[
(c_k^g)''(0)= \left\{
\begin{array}{ll}
-({\theta}_k^g ({\bar{t}}_k^g))^2 \x \frac{1}{\al} + 8\al & \mbox{if}\ c_k^g\ \mbox{is single},\\[1pt]
-({\theta}_k^g ({\bar{t}}_k^g))^2 \x \frac{1}{\al} + 32\al & \mbox{if}\ c_k^g\ \mbox{is double or special}.
\end{array}
\right.
\]
Then, for $\al $ sufficiently small, we have:
\begin{align*}
&(s_k^f (p_k) +c_k^f (p_k))''({\eta}_k) - (s_k^g (p_k) + c_k^g (p_k) )''({\eta}_k) \\
& \quad = \textstyle {\left( \frac{1}{{\eta}_{k+1} -{\eta}_k} \right)}^2 \x (c_k^f)''(p_k({\eta}_k)) - {\left( \frac{1}{{\eta}_{k+1} - {\eta}_k} \right) }^2 \x (c_k^g)''(p_k({\eta}_k))\\
& \quad = \textstyle {\left( \frac{1}{{\eta}_{k+1} - {\eta}_k } \right) }^2 \x (c_k^f)''(0) - {\left( \frac{1}{{\eta}_{k+1} - {\eta}_k} \right) }^2 \x (c_k^g)''(0) >0
\end{align*}
thus, by \Lem{lem:1.13}, for $\al$ sufficiently small it exists $\bar{\eta} \! \in ]{\eta}_k, {\eta}_{k+1}]$ such that:
\[
s_k^f (p_k(\eta )) + c_k^f (p_k(\eta )) > s_k^g (p_k(\eta )) + c_k^g (p_k(\eta ))\ \ \ \forall \eta \in ]{\eta}_k , \bar{\eta} ]
\]
and 
\[
(Mf)_{\al}(\eta ) > (Mg)_{\al} (\eta ) \ \ \ \forall \eta \in ]{\eta}_k , \bar{\eta} ].  \tag*{\qed}
\]

We can summarize the two preceding propositions as the following properties:
\[
\begin{array}{ll}
\mbox{\Prop{prop:2.3}} & \forall \bar{\eta} \! \in ]{\eta}_k , {\eta}_{k+1}]\ \exists {\al}_{\bar{\eta}} : \forall \al \in \I , \al < {\al}_{\bar{\eta}} \imp (Mf)_{\al} > (Mg)_{\al}\ \mbox{in} \ ] \bar{\eta} , {\eta}_{k+1}]; \\[0.1cm]
\mbox{\Prop{prop:2.4}} & \forall \tilde{\al} \! \in \! \I \ \exists {\eta}_{\tilde{\al}} : (Mf)_{\tilde{\al}} > (Mg)_{\tilde{\al}} \ \mbox{in} \ ]{\eta}_k , {\eta}_{\tilde{\al}}].
\end{array}
\] 
Let us go back to our problem and let $M$ be a model with ${\al}^* \! \in \! \I$. By \Prop{prop:2.4} we have:
\[
\exists {\eta}_{{\al}^*}={\eta}^* \in ]{\eta}_k , {\eta}_{k+1}]\ \mbox{such that}\ (Mf)_{{\al}^*} > (Mg)_{{\al}^*}\ \mbox{in}\ ]{\eta}_k , {\eta}^* ];
\]
now, ${\eta}^* \in ] {\eta}_k , {\eta}_{k+1}]$, thus, by \Prop{prop:2.3} we have:
\[
\exists {\al}_{{\eta}^*} = \hat{\al} \ \mbox{such that}\ \forall \al \in \I , \al < \hat{\al} \imp (Mf)_{\al} > (Mg)_{\al} \ \mbox{in}\ ]{\eta}^* , {\eta}_{k+1}], 
\]
if ${\al}^* \! < \! \hat{\al}$ then we have $(Mf)_{{\al}^*} > (Mg)_{{\al}^*}$ in $]{\eta}_k , {\eta}_{k+1}]$ and we are done.

It remains the possibility that ${\al}^* \geq \hat{\al}$.

In this case, let $\bar{\eta} \in ]{\eta}_k , {\eta}^*]$ with $\bar{\eta} < {\eta}_k + \frac{{\eta}_{k+1} - {\eta}_k}{4}$, then by \Prop{prop:2.3}:
\[
\exists {\al}_{\bar{\eta}} = \bar{\al}\ \mbox{such that} \ \forall \al \in \I , \al < \bar{\al} \imp (Mf)_{\al} > (Mg)_{\al} \ \mbox{in} \ ]\bar{\eta} , {\eta}_{k+1}].
\]
Let now $\tilde{\al} \in \I $ with $\tilde{\al} < \bar{\al}$ and $\tilde{\al} < {\al}^*$, then by \Prop{prop:2.4} we have:
\[
\exists {\eta}_{\tilde{\al}}= \tilde{\eta} \in ]{\eta}_k ,{\eta}_{k+1}]\ \mbox{such that}\ (Mf)_{\tilde{\al}} > (Mg)_{\tilde{\al}} \ \mbox{in}\ ]{\eta}_k , \tilde{\eta}],
\]
and, because $\tilde{\al} < \bar{\al}$, we also have $(Mf)_{\tilde{\al}} > (Mg)_{\tilde{\al}}$ in $[\bar{\eta} , {\eta}_{k+1}]$; thus:
\[
(Mf)_{\tilde{\al}} > (Mg)_{\tilde{\al}}\ \mbox{in}\ ]{\eta}_k , \tilde{\eta}] \cup [\bar{\eta} , {\eta}_{k+1}],
\]
it remains to prove the property inside the interval $[\tilde{\eta} , \bar{\eta}]$.

Let us consider the functions $S^h(\eta) \putAs s^h_k(p_k(\eta))$ and $C^h_{\al}(\eta) \putAs c^h_{k,\al}(p_k(\eta))$ where $c^h_{k,\al}$ is the $[\al, \theta^h_k(\bar{t}^h_k),\theta^h_k(\bar{t}^h_{k+1})]$-defined function, with $\al \! \in  ]0,\al^*]$ and $h \! \in \! \{f,g\}$.

Then, we may decompose the functions $(Mf)_{{\al}^*}$ and $(Mg)_{{\al}^*}$ as:
\[
(Mf)_{{\al}^*} = S^f + C_{{\al}^*}^f \ \ \mbox{and}\ \ (Mg)_{{\al}^*} = S^g + C_{{\al}^*}^g\ \ \mbox{in}\ \ [{\eta}_k , {\eta}_{k+1}]
\]
where $S^f$ and $S^g$ are the interpolating linear functions between the points $({\eta}_k , {\bar{y}}_k^f), ({\eta}_{k+1} ,{\bar{y}}_{k+1}^f)$ and $({\eta}_k , {\bar{y}}_k^g), ({\eta}_{k+1}, {\bar{y}}_{k+1}^g)$ (with ${\bar{y}}_k^f = {\bar{y}}_f^g$ and ${\bar{y}}_{k+1}^f > {\bar{y}}_{k+1}^g$), while $C_{{\al}^*}^f$ and $C_{{\al}^*}^g$ are $[\al , {\theta}_1 , {\theta}_2]$-defined functions rescaled from $[0,1]$ to $[{\eta}_k , {\eta}_{k+1}]$, with $\theta$s defined by the derivative conditions at the endpoints.

Since $\bar{\eta} \in ]{\eta}_k , {\eta}^* ]$ and ${\bar{y}}_{k+1}^f > {\bar{y}}_{k+1}^g$, we have:
\[
(Mf)_{{\al}^*} > (Mg)_{{\al}^*}\ \ \mbox{and} \ \ S^f > S^g \ \ \mbox{in}\ \ [\tilde{\eta} , \bar{\eta}].
\] 
We now consider $(Mf)_{\tilde{\al}}$ and $(Mg)_{\tilde{\al}}$, they have the following form:
\[
(Mf)_{\tilde{\al}} = S^f + C_{\tilde{\al}}^f\ \ \mbox{and} \ \ (Mg)_{\tilde{\al}} = S^g + C_{\tilde{\al}}^g \ \ \mbox{in}\ \ [\eta_k , \eta_{k+1}],
\]
let us denote ${\theta}_1^f \putAs(C_{{\al}^*}^f)'({\eta}_k)=(C_{\tilde{\al}}^f)'({\eta}_k)$ and ${\theta}_1^g \putAs(C_{{\al}^*}^g)'({\eta}_k)=(C_{\tilde{\al}}^g)'({\eta}_k)$. Since $\tilde{\al} \! < \! {\al}^*$, by \Lem{lem:1.9} we also have $|C_{\tilde{\al}}^f| \! < \! |C_{{\al}^*}^f|$ and $|C_{\tilde{\al}}^g| \! < \! |C_{{\al}^*}^g|$.

\smallskip

Let $\eta_0 \in [\tilde{\eta} , \bar{\eta}]$, we have five possibilities, depending on ${\theta}_1^f$ and ${\theta}_1^g$. The first three of these five sub-cases are rather straightforward; whereas the last two require more attention.

1) ${\theta}_1^f=0$ or ${\theta}_1^g=0$, in this case $(Mf)_{{\al}^*}(\eta_0)=S^f(\eta_0)$ or $(Mg)_{{\al}^*}(\eta_0)=S^g(\eta_0)$ in $[\tilde{\eta},\bar{\eta}]$, for the first one we have:
\[
 (Mg)_{\tilde{\al}}(\eta_0) \leq \max\{(Mg)_{{\al}^*}(\eta_0), S^g(\eta_0)\} < (Mf)_{{\al}^*}(\eta_0)=S^f(\eta_0)=(Mf)_{\tilde{\al}}(\eta_0)
\]
thus $ (Mg)_{\tilde{\al}}(\eta_0)<(Mf)_{\tilde{\al}}(\eta_0)$; the other case is similar.

2) ${\theta}_1^f<0$ and ${\theta}_1^g>0$, here $(Mf)_{{\al}^*}(\eta_0)<S^f(\eta_0)$ and $(Mg)_{{\al}^*}(\eta_0)>S^g(\eta_0)$ in $[\tilde{\eta},\bar{\eta}]$, so we have
\[
S^g(\eta_0)<(Mg)_{\tilde{\al}}(\eta_0)<(Mg)_{{\al}^*}(\eta_0)<(Mf)_{{\al}^*}(\eta_0)<(Mf)_{\tilde{\al}}(\eta_0)<S^f(\eta_0)
\]
thus $ (Mg)_{\tilde{\al}}(\eta_0)<(Mf)_{\tilde{\al}}(\eta_0)$. 

3) ${\theta}_1^f>0$ and ${\theta}_1^g<0$, in this case we obtain $(Mf)_{{\al}^*}(\eta_0)>S^f(\eta_0)$ and $(Mg)_{{\al}^*}(\eta_0)<S^g(\eta_0)$ in $[\tilde{\eta},\bar{\eta}]$, therefore:
\[
(Mg)_{{\al}^*}(\eta_0)<(Mg)_{\tilde{\al}}(\eta_0)<S^g(\eta_0)<S^f(\eta_0)<(Mf)_{\tilde{\al}}(\eta_0)<(Mf)_{{\al}^*}(\eta_0)
\]
and thus $ (Mg)_{\tilde{\al}}(\eta_0)<(Mf)_{\tilde{\al}}(\eta_0)$.

4-5) Of the two remaining cases, namely ${\theta}_1^f>0,\ {\theta}_1^g>0$ and ${\theta}_1^f<0,\ {\theta}_1^g<0$, we consider only the former, as the treatment of the other case is strictly analogous.

Since the only interval under consideration is $[\eta_k,\eta_{k+1}]$, in the following we omit the index $k$ for functions, e.g. $s^f$ instead of $s^f_k$; on the other hand, given its relevance inside the proof, we keep the appropriate index $\al$ for the $[\al,\theta_1,\theta_2]$-defined functions involved, e.g. $c^f_{\tilde{\al}}$. Thus, by definition, we have:
\[
(Mf)_{\tilde{\al}}(\eta) = s^f(p(\eta))+c_{\tilde{\al}}^f(p(\eta))\ \ \mbox{and}\ \ (Mg)_{\tilde{\al}}(\eta) = s^g(p(\eta))+c_{\tilde{\al}}^g(p(\eta))\ \ \mbox{in}\ \ [{\eta}_k,{\eta}_{k+1}],
\]
where
\[
 p \colon [{\eta}_k,{\eta}_{k+1}] \frec [0,1]\ \ \mbox{with}\ \  p(\eta) = \frac{\eta -{\eta}_k}{{\eta}_{k+1}-{\eta}_k},
\]
and
\[
s^h\colon [0,1]\frec \mR\ \ \mbox{with}\ \ s^h(x)={\bar{y}}_k^h+({\bar{y}}_{k+1}^h-{\bar{y}}_k^h)\x x,\ \ h\! \in \{ f,g \}.
\]
Then
\begin{align*}
(Mf)_{\tilde{\al}}(\eta) >(Mg)_{\tilde{\al}}(\eta) \ \mbox{in}\  [\tilde{\eta}, \bar{\eta}] & 
~~\Longleftrightarrow~~ s^f(p(\eta))+c_{\tilde{\al}}^f(p(\eta)) > s^g(p(\eta))+c_{\tilde{\al}}^g(p(\eta)) \ \mbox{in} \ [\tilde{\eta}, \bar{\eta}] \\
& ~~\Longleftrightarrow~~ s^f(x)+c_{\tilde{\al}}^f(x)>s^g(x)+c_{\tilde{\al}}^g(x) \ \mbox{in}\  [\tilde{x}, \bar{x}] \\ 
& ~~\Longleftrightarrow~~ c_{\tilde{\al}}^f(x)+L(x)>c_{\tilde{\al}}^g(x) \ \mbox{in}\  [\tilde{x}, \bar{x}],
\end{align*}
where $\tilde{x}$ and $\bar{x}$ are the projection through $p$ of $\tilde{\eta}$ and $\bar{\eta}$, and $L(x)=k\x x=s^f(x)-s^g(x)$ with $k>0$ a real number; in particular, $L(x) \geq 0$ in $[0,1]$ and $0 <\bar{x}<\frac{1}{4}$.

Moreover, because
\[
(Mf)_{\tilde{\al}}(\eta)>(Mg)_{\tilde{\al}}(\eta)\ \mbox{in}\ ]{\eta}_k,\tilde{\eta}]\cup [\bar{\eta},{\eta}_{k+1}]
\]
we have
\[
(Mf)_{\tilde{\al}}(\tilde{\eta})>(Mg)_{\tilde{\al}}(\tilde{\eta})\ \mbox{and}\ (Mf)_{\tilde{\al}}(\bar{\eta})>(Mg)_{\tilde{\al}}(\bar{\eta})\
\]
thus, with a slight abuse of notation,
\[
(Mf)_{\tilde{\al}}(\tilde{x})>(Mg)_{\tilde{\al}}(\tilde{x})\ \mbox{and}\ (Mf)_{\tilde{\al}}(\bar{x})>(Mg)_{\tilde{\al}}(\bar{x}).\
\]
We now consider three cases, depending on the type (single, double or special) of functions $c^f_{\tilde{\al}}$ and $c^g_{\tilde{\al}}$.

Since in the following the focus shifts from the functions $C^f_{\tilde{\al}}$ and $C^g_{\tilde{\al}}$ to $c^f_{\tilde{\al}}$ and $c^g_{\tilde{\al}}$, let us define $\theta^h \putAs \left( c^h_{\tilde{\al}}\right)'(0)\ \mbox{with} \ h \in \{ f,g \}$, observing that $\theta^h_1 = \left( C^h_{\tilde{\al}}\right)'(\eta_k)=\left( c^h_{\tilde{\al}}\right(p))'(\eta_k)=\left( c^h_{\tilde{\al}}\right)'(p(\eta_k)) \x p'(\eta_k)=\theta^h \x \frac{1}{\eta_{k+1}-\eta_k}$; in particular, $\theta^h > 0 \Leftrightarrow \theta^h_1 > 0$ and $\theta^f \bowtie \theta^g \Leftrightarrow \theta^f_1 \bowtie \theta^g_1$ with ${\bowtie} \in \{ <,\leq,=,\geq,>\}$.

\bigskip

\centerline{\fbox{\texttt{First case: $c^f_{\tilde{\al}}$ and $c^g_{\tilde{\al}}$ of the same type}}}

We assume that both $c_{\tilde{\al}}^f$ and $c_{\tilde{\al}}^g$ are single-defined. By definition, and because $[\tilde{x}, \bar{x}] \subset ]0, \frac{1}{4}[$, we have:
\begin{align*}
 c_{\tilde{\al}}^f(x)+L(x)> c_{\tilde{\al}}^g(x)  & ~~\Longleftrightarrow~~
\tilde{\al} [ 1-(2x-1)^2 e^{-2 \left( \frac{{\theta}^f}{2\tilde{\al}}-2 \right) x} ] +k\x x  > \tilde{\al}[ 1-(2x-1)^2 e^{-2\left( \frac{{\theta}^g}{2\tilde{\al}}-2\right)x} ] \\
& ~~\Longleftrightarrow~~ \tilde{\al} [ 1-(2x-1)^2 e^{- \frac{{\theta}^f}{\tilde{\al}}x+4x} - 1 +(2x-1)^2 e^{- \frac{{\theta}^g}{\tilde{\al}}x+4x} ] > -k\x x \\
& ~~\Longleftrightarrow~~ (2x-1)^2 e^{4x} [ -e^{- \frac{{\theta}^f}{\tilde{\al}}x}+e^{- \frac{{\theta}^g}{\tilde{\al}}x} ]>  -\Kappa \x x \\
& \textstyle ~~\Longleftrightarrow~~ e^{-Gx}-e^{-Fx}> -\frac{\Kappa\,x}{e^{4x} (2x-1)^2}\\
& \textstyle ~~\Longleftrightarrow~~ \frac{\Kappa\,x}{e^{4x} (2x-1)^2} > e^{-Fx}-e^{-Gx},     
\end{align*}
where $\Kappa=\frac{k}{\tilde{\al}}>0$, $F=\frac{{\theta}^f}{\tilde{\al}}>0$, and $G=\frac{{\theta}^g}{\tilde{\al}}>0$.

We consider the functions:
\[
FG(x) \putAs e^{-Fx}-e^{-Gx}\ \ \mbox{and}\ \ K(x) \putAs \frac{\Kappa\,x}{e^{4x} (2x-1)^2}\ \mbox{in}\ \left[ 0,\frac{1}{4} \right],
\]
for which we have:
\[
FG(0)=K(0)=0\ \ \mbox{and}\ \ K(x)=\frac{\Kappa\,x}{e^{4x} (2x-1)^2}>0\ \mbox{in}\ \left] 0, \frac{1}{4} \right].
\]

We consider two sub-cases:

\noindent {\bf\boldmath($ {\theta}^g \leq {\theta}^f $):}\ If $0<{\theta}^g \leq {\theta}^f$ then $0<G\leq F$, thus $e^{-Gx}\geq e^{-Fx}$ in $[0,\frac{1}{4}]$, therefore $e^{-Fx} - e^{-Gx}\leq 0< K(x)$, whence the sought conclusion $(Mf)_{\tilde{\al}}(\eta)>(Mg)_{\tilde{\al}}(\eta)$  follows for all $\eta\in [\tilde{\eta},\bar{\eta}]$.

\noindent {\bf\boldmath($ {\theta}^f < {\theta}^g $):}\ Let us study $K(x)$. We have:
\begin{align*}
K'(x) & =  \Kappa\frac{e^{4x}(2x-1)^2-x[4e^{4x}(2x-1)^2+4(2x-1)e^{4x}]}{[e^{4x}(2x-1)^2]^2}\\
      & =  \Kappa\frac{2x-1-8x^2}{e^{4x} (2x-1)^3}
\end{align*}
thus, $K'(0)=\Kappa$ and $K'(x)>0$ in $[0,\frac{1}{4}]$.
\begin{align*}
K''(x) & =  \Kappa\frac{(2-16x)e^{4x}(2x-1)^3-(2x-1-8x^2)[4e^{4x}(2x-1)^3+6e^{4x}(2x-1)^2]}{[e^{4x}(2x-1)^3]^2]} \\
       & =  \Kappa\frac{(2-16x)(2x-1)-(2x-1-8x^2)[4(2x-1)+6]}{e^{4x}(2x-1)^4} \\
       & =  8\Kappa\frac{8x^3+3x-4x^2}{e^{4x}(2x-1)^4}\ =\ 8\Kappa\frac{x(8x^2-4x+3)}{e^{4x}(2x-1)^4}
\end{align*}
thus, $K''(0)=0$ and $K''(x)>0$ in $]0,\frac{1}{4}]$, therefore in $[0,\frac{1}{4}]\ K(x)$ is an  \ita{increasing} and \ita{convex} function.

Now we study $FG(x)$. $0 \! < \! {\theta}^f \! < \! {\theta}^g$ implies that $0 \! < \! F \! < \! G$, thus $e^{-Fx}>e^{-Gx}$ and therefore $FG(x)>0$ in $[0,\frac{1}{4}]$; moreover:
\begin{align*}
    FG'(x) & =Ge^{-Gx}-Fe^{-Fx},~ \text{with } FG'(0)=G-F>0,
    \intertext{and}
    FG''(x) &=F^2 e^{-Fx}-G^2 e^{-Gx},~ \text{with } FG''(0)=F^2-G^2<0.
\end{align*}

In particular, after some trivial computation we obtain:
\begin{align*}
    FG'(x) > 0 & \Longleftrightarrow  e^{\frac{x}{\tilde{\al}}(\theta^g- \theta^f)} < \frac{\theta^g}{\theta^f},
    \intertext{and}
    FG''(x) > 0 & \Longleftrightarrow  e^{\frac{x}{\tilde{\al}}(\theta^g- \theta^f)} > \left(\frac{\theta^g}{\theta^f}\right)^{\! 2}.
\end{align*}

Since $\left(\frac{\theta^g}{\theta^f}\right)^{\! 2} > \frac{\theta^g}{\theta^f}$ (because $\theta^g > \theta^f$ and thus $\frac{\theta^g}{\theta^f} >1$), if $FG(x)$ is increasing, then it is concave; and since $FG(0)>0$ there exists $\be >0$ such that $FG(x)$ is increasing, and concave, in $[0,\be]$. There are three possibilities $\be \in ]0,\tilde{x}]$, $\be \in ]\tilde{x}, \bar{x}[$ or $\be \geq \bar{x}$; we consider only the second one, being the other two treatable in an analogous way. 

We know that $K(x) \! > \! FG(x)$ in $]0,\tilde{x}] \cup [\bar{x},1]$, with $0 < \tilde{x} < \be < \bar{x} < \frac{1}{4}$.
\begin{enumerate}

\item let us consider the point $\tilde{X}=(\tilde{x},K(\tilde{x}))$, with $K(\tilde{x})>FG(\tilde{x})$, and the line $r$ passing through the origin $(0,0)$ and the point $\tilde{X}$. In $[0,\be]$ functions $\Kappa$ and $FG$ are both increasing and, respectively, convex and concave; thus, in the interval $[\tilde{x},\be]$ the function $\Kappa$ will be above the line $r$, while $FG$ will be below it, and so $K(x) > FG(x)$ in $[\tilde{x},\be]$.

\item from the previous point we have $\Kappa(\be) > FG(\be)$; since in $[\be, \bar{x}]$ $\Kappa$ is increasing whereas $FG$ is decreasing, we obtain $\Kappa(x) > FG(x)$ also in $[\be,\bar{x}]$.

\end{enumerate}

Therefore, we have:
\[
\forall x\! \in \! [\tilde{x},\bar{x}]\ K(x)>FG(x) \imp (Mf)_{\tilde{\al}}(x) \! > \! (Mg)_{\tilde{\al}}(x)\imp (Mf)_{\tilde{\al}}(\eta) > (Mg)_{\tilde{\al}}(\eta)\ \mbox{in}\ [\tilde{\eta},\bar{\eta}].
\]
If both $c_{\tilde{\al}}^f$ and $c_{\tilde{\al}}^g$ are double- or special-defined the proof is completely analogous.

\bigskip

\centerline{\fbox{\texttt{Second case: $c^f_{\tilde{\al}}$ double- or special-defined and $c^g_{\tilde{\al}}$ single-defined}}}
For this case, in $[0,\frac{1}{4}]$ we have:
\begin{align*}
 c_{\tilde{\al}}^f(x)+L(x)> c_{\tilde{\al}}^g(x) &~~\Longleftrightarrow~~ \tilde{\al} [ 1-(4x-1)^2 e^{-2 \left( \frac{{\theta}^f}{4\tilde{\al}}-2 \right) 2x} ] +k\x x  > \tilde{\al}[ 1-(2x-1)^2 e^{-2\left( \frac{{\theta}^g}{2\tilde{\al}}-2\right)x} ] \\
&~~\Longleftrightarrow~~ \tilde{\al} [ 1-(4x-1)^2 e^{- \frac{{\theta}^f}{\tilde{\al}}+8x} - 1 +(2x-1)^2 e^{- \frac{{\theta}^f}{2\tilde{\al}}+4x} ] > -k\x x \\
&\textstyle ~~\Longleftrightarrow~~ (2x-1)^2 e^{4x} [ -\left(\frac{4x-1}{2x-1}\right)^2 e^{- \frac{{\theta}^f}{\tilde{\al}}x+4x}+e^{- \frac{{\theta}^g}{\tilde{\al}}x} ]>  -\Kappa \x x \\
&\textstyle ~~\Longleftrightarrow~~ e^{-Gx}-\left(\frac{4x-1}{2x-1}\right)^2 e^{-Fx}> -\frac{\Kappa\,x}{e^{4x} (2x-1)^2}\\
&\textstyle ~~\Longleftrightarrow~~ \frac{\Kappa\,x}{e^{4x} (2x-1)^2} > \left(\frac{4x-1}{2x-1}\right)^2 e^{-Fx}-e^{-Gx},
\end{align*}
where $\Kappa=\frac{k}{\tilde{\al}}>0$, $F=\frac{{\theta}^f}{\tilde{\al}}-4>0$, and $G=\frac{{\theta}^g}{\tilde{\al}}>0$.\\
Thus, we have in $[0,\frac{1}{4}]$:
\[
\textstyle \frac{4x-1}{2x-1} \! \leq \! 1 \! \imp  \! {\left( \frac{4x-1}{2x-1}\right)}^2 \! \! \! \leq \! 1 \! \imp \! \! \left( \frac{4x-1}{2x-1} \right)^2 \! \! e^{-Fx} \! \leq \! e^{-Fx} \imp \! \left( \frac{4x-1}{2x-1} \right)^2 \! \! e^{-Fx}-e^{-Gx} \! \leq \! e^{-Fx}-e^{-Gx}
\]
therefore:
\[
\textstyle \frac{\Kappa\,x}{e^{4x} (2x-1)^2} > e^{-Fx}-e^{-Gx} \geq \left( \frac{4x-1}{2x-1}\right)^2 e^{-Fx}-e^{-Gx} \ \mbox{in}\ [\tilde{x}, \bar{x}],
\]
where the first inequality comes from the single-single case treated before.

\bigskip

\centerline{\fbox{\texttt{Third case: $c^f_{\tilde{\al}}$ single-defined and $c^g_{\tilde{\al}}$ double- or special-defined}}}

As before, we consider the interval $[0,\frac{1}{4}]$ and we have:
\begin{align*}
 c_{\tilde{\al}}^f(x)+L(x)> c_{\tilde{\al}}^g(x)& ~~\Longleftrightarrow~~ \tilde{\al} [ 1-(2x-1)^2 e^{-2 \left( \frac{{\theta}^f}{2\tilde{\al}}-2 \right) x} ] +k\x x  > \tilde{\al}[ 1-(4x-1)^2 e^{-2\left( \frac{{\theta}^g}{4\tilde{\al}}-2\right) 2x} ] \\
& ~~\Longleftrightarrow~~ \tilde{\al} [ 1-(2x-1)^2 e^{- \frac{{\theta}^f}{\tilde{\al}}x+4x} - 1 +(2x-1)^2 e^{- \frac{{\theta}^g}{\tilde{\al}}x+8x} ] > -k\x x \\
&\textstyle ~~\Longleftrightarrow~~ (2x-1)^2 e^{4x} [ -e^{- \frac{{\theta}^f}{\tilde{\al}}x}+ \left(\frac{4x-1}{2x-1}\right)^2e^{- \frac{{\theta}^g}{\tilde{\al}}x+4x} ]>  -\Kappa \x x \\
&\textstyle ~~\Longleftrightarrow~~ \left(\frac{4x-1}{2x-1}\right)^2e^{-Gx}-e^{-Fx}> -\frac{\Kappa\,x}{e^{4x} (2x-1)^2}\\
&\textstyle ~~\Longleftrightarrow~~ \Kappa\frac{x}{e^{4x} (2x-1)^2} >  e^{-Fx}-(\frac{4x-1}{2x-1})^2e^{-Gx} ,
\end{align*}
with $\Kappa=\frac{k}{\tilde{\al}}>0$, $F=\frac{{\theta}^f}{\tilde{\al}}>0$, and $G=\frac{{\theta}^g}{\tilde{\al}}-4>0$.

We consider:
\[
K(x) \putAs\Kappa\frac{x}{e^{4x} (2x-1)^2}\ \ \mbox{and}\ \ FG(x)\putAs e^{-Fx}-\left( \frac{4x-1}{2x-1}\right)^{\! 2} e^{-Gx}.
\]
As before, $K(x)$ is increasing and convex in $[0,\frac{1}{4}]$.

For $FG(x)$, we have:
\begin{align*}
FG'(x) & \textstyle =  -Fe^{-Fx}-\left[ -Ge^{-Gx} \x \left( \frac{4x-1}{2x-1} \right)^{\! 2} + e^{-Gx} \x 2 \x \left( \frac{4x-1}{2x-1}\right) \x \frac{4(2x-1)-2(4x-1)}{(2x-1)^2} \right] \\[0.25cm]
 & \textstyle =  -Fe^{-Fx}-\left[ -Ge^{-Gx}\left( \frac{4x-1}{2x-1} \right)^{\! 2} - e^{-Gx} \left( \frac{4x-1}{2x-1} \right) \frac{4}{(2x-1)^2} \right] \\
 & \textstyle =  -Fe^{-Fx}+e^{-Gx}\left[ \left( \frac{4x-1}{2x-1} \right)^{\! 2} G +4\frac{4x-1}{(2x-1)^3} \right]
\end{align*}
and, in particular,
\[
\textstyle FG'(0)=-F+G+4=-\frac{{\theta}^f}{\tilde{\al}}+\frac{{\theta}^g}{\tilde{\al}}-4+4=\frac{{\theta}^g-{\theta}^f}{\tilde{\al}}\ \ \mbox{and}\ \ FG'\left (\textstyle{\frac{1}{4}} \right) <0.
\]
As for the second derivative, we have:
\begin{align*}
FG''(x) & = \textstyle  F^2 e^{-Fx} + \left\{  -Ge^{-Gx}  \x   \left[  \left( \frac{4x-1}{2x-1} \right)^2  G + 4\frac{4x-1}{(2x-1)^3}  \right] + \right.  \\
 & \textstyle  \hspace{2.6cm}  + \, e^{-Gx} \left[ 2G  \left( \frac{4x-1}{2x-1} \right)  \x  \frac{-2}{(2x-1)^2}   + 4  \frac{8x-4-6(4x-1)}{(2x-1)^4} \right]  \bigg\} \\
 & =  \textstyle  F^2e^{-Fx}+\left\{-Ge^{-Gx}  \x  \left[ \! \left( \frac{4x-1}{2x-1} \right)^2 \! G\! +\! 4\frac{4x-1}{(2x-1)^3} \! \right] \! +e^{-Gx} \left[-4G\frac{4x-1}{(2x-1)^3}+8\frac{1-8x}{(2x-1)^4} \right] \right\} \\
 & = \textstyle  F^2e^{-Fx}+e^{-Gx} \left[ -G^2\left( \frac{4x-1}{2x-1}\right)^2 -4G\frac{4x-1}{(2x-1)^3}-4G\frac{4x-1}{(2x-1)^3}+8\frac{1-8x}{(2x-1)^4} \right]\\[0.25cm]
 & = \textstyle  F^2e^{-Fx}-e^{-Gx}\left[ G^2 \left( \frac{4x-1}{2x-1}\right)^2 + 8G\frac{4x-1}{(2x-1)^3}-8\frac{1-8x}{(2x-1)^4} \right]
\end{align*}
and, in particular,
\[
\textstyle FG''(0)=F^2-[G^2+8G-8]=F^2-[(G+4)^2-24]= \! \left(\frac{{\theta}^f}{\tilde{\al}}\right)^2 \! -\left(\frac{{\theta}^g}{\tilde{\al}}\right)^2 \! +24=\frac{({\theta}^f)^2-({\theta}^g)^2}{{\tilde{\al}}^2} +24.
\]
The previous computations can be extended to the generic case with $0 < \al \leq \al$. More precisely, we can consider the parameters $\Kappa_{\al}=\frac{k}{\al}, F_{\al}=\frac{\theta^f}{\al},G_{\al}=\frac{\theta^g}{\al}-4$, and the functions
\[
K_{\al}(x) \putAs\Kappa_{\al}\frac{x}{e^{4x} (2x-1)^2}\ \ \mbox{and}\ \ FG_{\al}(x)\putAs e^{-F_{\al}x}-\left( \frac{4x-1}{2x-1}\right)^{\! 2} e^{-G_{\al}x}. 
\]
We also have the corresponding results for what concerns derivatives, namely:
\[
K_{\al}'(0)=\Kappa_{\al},\ FG_{\al}'(0)=\frac{\theta^g-\theta^f}{\al},\ FG'_{\al}\! \left(\frac{1}{4}\right) <0\ \mbox{and}\ FG''_{\al}(0)=\frac{({\theta}^f)^2-({\theta}^g)^2}{{\al}^2} +24
\]
Regarding notation, we keep the unlabelled case for $\al=\tilde{\al}$, e.g. $K(x)=K_{\tilde{\al}}(x)$.

As in the previous cases, our goal is to prove that, at least for $0  < \al \leq  \tilde{\al}$ sufficiently small, if
\[
K_{\al}(x)>FG_{\al}(x)\ \ \mbox{in}\ ]0,\tilde{x}] \cup \left[ \bar{x}, \textstyle{\frac{1}{4}} \right],
\]
then
\begin{equation} \tag{*}
\label{eqn:thesis}
K_{\al}(x)>FG_{\al}(x)\ \ \mbox{in}\ [\tilde{x}, \bar{x} ] .
\end{equation}
To this end, we distinguish the following three subcases, depending on the comparison between ${\theta}^g$ and ${\theta}^f$: 
\[
{\theta}^g = {\theta}^f, \qquad {\theta}^g < {\theta}^f, \qquad {\theta}^g > {\theta}^f.
\]

\bigskip

\centerline{\underline{\texttt{\boldmath$ {\theta}^g = {\theta}^f$}}}

Given the subcase hypothesis, previous expressions can be simplified, namely:
\begin{align*}
FG_{\al}(x) & \textstyle = e^{-\frac{{\theta}^f}{\al}x}-\left( \frac{4x-1}{2x-1}\right)^2 e^{-(\frac{{\theta}^f}{\al}-4)x} \\[0.1 cm]
 & \textstyle =  e^{-\frac{{\theta}^f}{\al}x}-\left( \frac{4x-1}{2x-1}\right)^2 e^{-\frac{{\theta}^f}{\al}x}\x e^{4x} \\[0.1 cm]
 & \textstyle =  e^{-\frac{\theta^f}{\al}x}\left[1-\left( \frac{4x-1}{2x-1}\right)^{\! 2}e^{4x} \right];
\end{align*}
in particular, we have $FG_{\al}(x)>0$ in $]0,\frac{1}{4}]$, $FG_{\al}'(0)=0$ and $FG_{\al}''(0) \geq 0$.

Let $x_0 \in ]0, \bar{x}]$ and $0 < \be < \al \leq \tilde{\al}$, from the expression of $FG_{\al}$, and the one for $K_{\al}(x)$, we can immediately derive the following properties:
\begin{itemize}

\item $\lim_{\al \frec 0^+} FG_{\al}(x_0)=0$,

\item $K_{\al}(x_0) < K_{\be}(x_0)$,

\item $FG_{\be}(x_0) < FG_{\al}(x_0)$.

\end{itemize}

Given these properties, for $\al$ sufficiently small we obtain:
\[
FG_{\al}(x) \leq \max_{\tau \in [\tilde{x}, \bar{x}]}FG_{\al}(\tau) < K_{\al}(\tilde{x}) \leq K_{\al}(x) \ \ \forall x \in [\tilde{x}, \bar{x}].
\]

\bigskip

\centerline{\underline{\texttt{\boldmath${\theta}^g < {\theta}^f$}}}

The treatment proceeds similarly as follows.

Since $FG'(0)<0$ and $FG(0) \! = \! 0$, there exists an $\be \! \in ]0,\frac{1}{4}]$ such that $FG(x) \! < \! 0$ in $]0,\be]$. If $\be > \bar{x}$ we are done; otherwise, we prove that $FG_{\al}(x)<0$ for all $\al \leq \tilde{\al}$ and all $x \in \, ]0,\be]$. This suffices to obtain $K_{\al} (x)> 0 > FG_{\al}(x)$ in $]0,\be]$; for what concerns $[\be, \bar{x}]$, we can rely on the same strategy used in the previous subcase $\theta^f = \theta^g$.

Given $x_0 \! \in ]0,\be]$, we consider $t \! = \! \frac{1}{\al}$, $\fhi \! = \! {\theta}^g x_0$, $\psi \! = \! {\theta}^f x_0$ and $A_{x_0}=\left(\frac{4x_0-1}{2x_0-1}\right)^2 e^{4x_0}$, with $\psi >\fhi $. 

From $FG_{\al}(x_0)=e^{-F_{\al}x_0}-\left( \frac{4x_0-1}{2x_0-1}\right)^{\! 2} e^{-G_{\al}x_0}$, we obtain by substitution the following function in $t$:
\[
E_{x_0}(t)=e^{-\psi t}-A_{x_0}e^{-\fhi t},
\]
and we consider its derivative
\[
E'_{x_0}(t)=-\psi e^{-\psi t}+\fhi A_{x_0} e^{-\fhi t}.
\]
In particular:
\[
\begin{array}{rclrcl}
E'_{x_0}(t) > 0   
& \Longleftrightarrow & -\psi e^{-\psi t} + \fhi A_{x_0} e^{-\fhi t} > 0 
& \Longleftrightarrow & \fhi A_{x_0} e^{-\fhi t}  > \psi e^{-\psi t} \\[1ex]
& \Longleftrightarrow &  e^{-\fhi t + \psi t}  > \dfrac{\psi}{\fhi} \cdot \dfrac{1}{A_{x_0}}
& \Longleftrightarrow & e^{(\psi - \fhi)t}  > \dfrac{\psi}{\fhi} \cdot \dfrac{1}{A_{x_0}}\/,
\end{array}
\]
with $\psi -\fhi > 0$.
Moreover, the following holds:
\[
\lim_{\al \to 0^+}FG_{\al}(x)=\lim_{t \to +\infty } E_{x}(t)=0\ \ \ \ \forall x \! \in \left[ 0,\frac{1}{4} \right] .
\]
Thus, given $x_0 \in ]0,\be]$ there are two possibilities:

\begin{enumerate}

\item the derivative $E'_{x_0}(t)$ is negative, hence $FG_{\al}(x_0)$ is decreasing, with respect to $\al \frec 0^+$, and still negative,
\item the derivative $E'_{x_0}(t)$ is positive, hence $FG_{\al}(x_0)$ is increasing with respect to $\al \frec 0^+$; but, since from that point the derivative will always be positive (while $t$ grows and $\al$ decreases) and $\lim_{t \to +\infty } E_{x_0}(t)=0$, $E_{x_0}(t)=FG_{\al}(x_0)$ will always be negative.

\end{enumerate}

Therefore, $K(x) \! > \! FG(x)$ holds in $]0,\be]$ for every $\al \! \leq \! \tilde{\al}$. As before, since $\lim_{\al \to 0^+ } FG_{\al}(x)=0$ also holds in $[\be, \bar{x}]$, for $\al$ sufficiently small we have:
\[
\textstyle K(x)>FG(x)\ \ \mbox{in}\ \ \left] 0,\frac{1}{4} \right],
\]
whence \eqref{eqn:thesis} follows.

\bigskip

\centerline{\underline{\texttt{\boldmath${\theta}^g > {\theta}^f$}}}

For the last case, we apply the same approach used in the case single-single. In particular, we prove that, at least for $\al$ sufficiently small, the two following properties hold:

\begin{enumerate}
\item $FG_{\al}'(0)>0$, $FG_{\al}'(\frac{1}{4})<0$, $FG_{\al}''(0)<0$ \mbox{and} $FG_{\al}''(\frac{1}{4})>0$,
\item $FG_{\al}$ has only one stationary tangent point, which is a maximum, and only one inflection point.
\end{enumerate}

If they hold, there exists a real value $\be \in ]0,\frac{1}{4}[$ such that $FG$ is increasing and concave in an interval $[0,\be]$ and decreasing in $[\be ,\frac{1}{4} ]$; then the proof is analogous to the single-single case. We now prove the two  properties.

Concerning the former, we have:
\[
\textstyle FG_{\al}'(0)= \frac{{\theta}^g -{\theta}^f}{\tilde{\al}} >0,\ \ \ FG_{\al}' \left( \frac{1}{4} \right)= -F_{\al}e^{-\frac{F_{\al}}{4}} < 0,
\]
and, for $\al$ sufficiently small,
\[
\textstyle FG_{\al}''(0)= \frac{({\theta}^f)^2 - ({\theta}^g)^2}{{\tilde{\al}}^2} + 24 <0,\ \ \ FG_{\al}'' \left( \frac{1}{4} \right) =  \frac{({\theta}^f)^2}{(\tilde{\al})^2}e^{-\frac{{\theta}^f}{4\tilde{\al}}} - 128e^{-\frac{{\theta}^g}{4\tilde{\al}} -1} > 0.
\]
As for the latter:
\begin{align*}
FG_{\al}'(x)=0 &~~ \textstyle \Longleftrightarrow ~~ F_{\al}e^{-F_{\al}x}=e^{-G_{\al}x}\left[ \left( \frac{4x-1}{2x-1}\right)^2 G_{\al}+4\frac{4x-1}{(2x-1)^3} \right] \\
&~~ \textstyle \Longleftrightarrow ~~ e^{(G_{\al}-F_{\al})x}= \frac{1}{F}\left[ \left( \frac{4x-1}{2x-1}\right)^2 G_{\al}+4\frac{4x-1}{(2x-1)^3} \right],
\end{align*}
thus, we consider the functions:
\[
\textstyle e^{(G_{\al}-F_{\al})x}\ \  \ \mbox{and}\  \ \ G_{F_{\al}} (x)=\frac{1}{F_{\al}}\left[ \left( \frac{4x-1}{2x-1} \right)^2 G_{\al}+4\frac{4x-1}{(2x-1)^3} \right] \ \mbox{in}\ \left[ 0,\frac{1}{4} \right].
\]
For $\al$ sufficiently small, the following holds:
\[
\textstyle G_{\al}-F_{\al}=\frac{{\theta}^g}{\al}-4-\frac{{\theta}^f}{\al}=\frac{{\theta}^g-{\theta}^f}{\al}-4>0 \imp e^{(G_{\al}-F_{\al})x}\ \mbox{increasing}\ \mbox{in}\ \left[ 0,\frac{1}{4} \right]
\]
moreover:
\[
\textstyle G_{F_{\al}} (0)=\frac{{\theta}^g}{{\theta}^f}>1\ \ \mbox{and}\ \ G_{F_{\al}} \left (\frac{1}{4} \right) =0
\]
for the existence and uniqueness of the stationary point, we prove that $G_{F_{\al}}$ is decreasing in $\left [0,\frac{1}{4} \right]$.
\[
\textstyle G'_{F_{\al}} (x)= \frac{1}{F_{\al}}\left[-4G_{\al}\frac{4x-1}{(2x-1)^3}+8\frac{1-8x}{(2x-1)^4}\right]
\]
thus:
\begin{align*}
G'_{F_{\al}}(x)<0\ \quad \mbox{in}\ \left[ 0,\textstyle{\frac{1}{4}} \right] &~~ \textstyle \Longleftrightarrow ~~ \frac{1}{F_{\al}}\left[-4G_{\al}\frac{4x-1}{(2x-1)^3}+8\frac{1-8x}{(2x-1)^4}\right] <0\ && \textstyle \mbox{in}\ \left[ 0,\textstyle{\frac{1}{4}} \right] \\
&\textstyle ~~ \Longleftrightarrow ~~ 4G_{\al} \frac{4x-1}{(2x-1)^3} > 8\frac{1-8x}{2x-1}\ && \mbox{in}\ \left[ 0,\textstyle{\frac{1}{4}} \right] \\
&\textstyle ~~ \Longleftrightarrow ~~ G_{\al}(4x-1) > 2\frac{1-8x}{2x-1}\ && \mbox{in}\ \left[ 0,\textstyle{\frac{1}{4}} \right] \\
&~~ \Longleftrightarrow ~~ 8G_{\al}x^2 - 6G_{\al}x +G_{\al} -2 +16x >0 && \mbox{in}\ \left[ 0,\textstyle{\frac{1}{4}} \right].
\end{align*}
Given:
\[
Par(x)=8G_{\al}x^2 - 6G_{\al}x +G_{\al} -2 +16x,
\]
we have:
\[
\textstyle Par(0)=G_{\al}-2>2\ \mbox{for}\ \al \ \mbox{sufficiently small and}\ Par\left( \frac{1}{4} \right) =2
\]
and
\[
\textstyle Par'(x)=16G_{\al}x+16-6G_{\al}<0\ \mbox{in}\ \left[ 0,\frac{1}{4} \right]\ \mbox{for}\ \al \ \mbox{sufficiently small}.
\]
Therefore, $FG_{\al}$ has only one stationary point, which is a maximum. For the second derivative and the inflection point, the proof is similar.

As stated above, the case ${\theta}^f<0\ \et \ {\theta}^g <0$ is proven in a completely analogous way.

\smallskip

\paragraph{Subcase \boldmath $(f\mathord{>})$, unbounded and closed intervals:}

Lastly, we consider the case of unbounded intervals; more precisely, we deal with atoms of the form $(f>g)_{[v_r,+\infy[}$ with $v_r$ the greatest domain variable, being the opposite case $(f>g)_{]-\infty,v_1]}$ treatable in an analogous way. Observe that due to Step \ref{itemAlgoOne}.\ref{alg1:actionOne}), there are no more literals of the form $(f>g)_{]-\infty,v_1[}$ or $(f>g)_{]v_r,+\infty[}$. In this final subcase, we exploit the numeric variables $k^f_r$, and the related literals that have been introduced in Steps \ref{alg4:actionB:case2} and \ref{itemAlgoFour}.\ref{alg4:actH}) of the algorithm; since the only interval we explicitly treat is $[v_r, \infty[$, variables of type $y^f_r, t^f_r, \ga^f_r, k^f_r$ will be simply denoted by $y_f, t_f, \ga_f, k_f$. Furthermore, slightly deviating from previous notation, in this paragraph we adopt a tilde for the real value assigned to the variable, e.g. $\tilde{t}_f$ denotes the real value of the variable $t_f$ in the numeric model under consideration. (We briefly recall that $y_f$ and $t_f$ represent, respectively, the value and the derivative of (the function interpreting) $f$ in the last point $v_r$ of the domain variables.)

Our goal is to assign to any function variable $f$ which occurs in an atom of the form $(f>g)_{[v_r,+\infy[}$ a real parameter $\phi_f$ such that the following expression
\begin{equation}\label{eq:inf}\tag{**}
(Mf)({\eta}) \putAs \tilde{y}_f + (\tilde{t}_f -\phi_f)(1- e^{{\eta}_r -\eta }) + \phi_f (\eta - {\eta}_r)
\end{equation}
defines, over $[\eta_r, +\infty[$ , a function $(Mf)$ that complies simultaneously with order constraints, e.g $(f>g)_{[v_r,+\infy[}$, and derivative constraints, e.g. $(D[f]>y)_{[v_r,+\infy[}$; these latter literals will be treated in a later case.

With respect to \eqref{eq:inf}, the following computations hold:
\[
(Mf)(\eta_r)=\tilde{y}_f, \ (Mf)'(\eta_r)=\tilde{t}_f, \ \lim_{\eta \frec +\infty} (Mf)'(\eta )={\phi}_f
\]
in particular, $(Mf)$ has an oblique asymptote given by the straight line:
\[
r(\eta)={\phi}_f \eta + \tilde{y}_f +\tilde{t}_f - {\phi}_f \left( {\eta}_r +1 \right),\ \ \mbox{with}\ \ r ({\eta}_r)=\tilde{y}_f + \tilde{t}_f - {\phi}_f
\]
Let us define the following sets:\footnote{Observe that, differently from $\sK, \sF$ and $\Gamma$, the last set $\sM$ is a set of real numbers not variables.}
\begin{align*}
\sK & \putAs  \{ \mbox{set of all} \ k_f \ \mbox{occurring in}\ \phi_4 \} ,\\ 
\sF & \putAs  \{ \mbox{set of all function variables} \ f \ \mbox{such that} \ k_f \! \in \! \sK \} \\
\Gamma & \putAs  \{ \mbox{set of all} \ \ga_f \ \mbox{occurring in}\ \phi_4 \ \mbox{such that} \ f\! \in \! \sF  \} , \\ 
\sM & \putAs  \{ \tilde{\mu} \, | \, \ga_f \bowtie \mu \ \mbox{appears in}\ \phi_4 \ \mbox{with} \ {\bowtie}  \in \! \{ <, \leq, =, \geq , > \} \ \mbox{and} \ \ga_f \in \Gamma \} ,
\end{align*}
fixing also:
\[
m \putAs \min \sM \ , \quad M \putAs \max \sM \ \quad \mbox{and} \ \quad m=M=0 \ \mbox{if} \ \sM=\emptyset.
\]
To each $f \! \in \! \sF$, we associate a bounded interval $I_f$ as follows: the ends points $\al_f$ and $\be_f$ of $I_f$ are defined as $\al_f \putAs \max \left(A_f \cup \{m-1\} \right)$ with $A_f \putAs \{ \tilde{\mu} \, | \, \ga_f \trianglerighteq \mu \ \mbox{appears in}\ \phi_4 \ \mbox{with} \ \trianglerighteq  \in \! \{ =, \geq , > \} \}$ and $\be_f \putAs \min \left( B_f \cup \{ M+1 \} \right)$ with $B_f \putAs \{ \tilde{\mu} \, | \, \ga_f \trianglelefteq \mu \ \mbox{appears in}\ \phi_4 \ \mbox{with} \ \trianglelefteq  \in \! \{ <, \leq, =,  \} \}$; moreover, if $\al_f = \tilde{\mu}_f$ with $\tilde{\mu}_f = \max A_f$ and $\ga_f > \mu_f$ appears in $\phi_4$, then $\al_f \notin I_f$, similarly for $\be_f$.

The interval $I_f$ represents the admissible values for $\phi_f$; for example, if $\ga_f \geq \mu_1$ and $\ga_f < \mu_2$ are the only literals involving $\ga_f$ that occur in $\phi_4$ with $\tilde{\mu}_1 = 2$ and $\tilde{\mu}_2 = 5$, then $\al_f =2$, $\be_f=5$ and $I_f = [2 , 5[$.

Finally, let $\sI$ be the set of all previously defined intervals $I_f$ with $f \! \in \! \sF$.

By partially ordering $\sI$ with $I_f \geq I_g$ iff $k_f \geq k_g$, we can apply \Lem{lem:intervals} to obtain a family $\left(\phi_f\right)_{f\in \sF}$ of real numbers such that
\[
\phi_f \in I_f \ \quad \mbox{and} \ \quad k_f \geq k_g \imp \phi_f \geq \phi_g.
\]
Unfortunately, this choice for the parameters $\left(\phi_f\right)_{f\in \sF}$ alone does not suffice to ensure the satisfaction of the constraints $(f > g)_{[v_r, +\infty[}$, it could be the case that $g>f$ in a right neighborhood of $\eta_r$; an example of such a phenomenon is given by the following formula
\[
\phi_4 \ \equiv \ y_f = 1 \, ~\et~ \, t_f=0 \, ~\et~ \, 0 \! \leq \! {\ga}_f \! < \! 3 \, ~\et~ \, y_g=0 \, ~\et~ \, t_g=3 \, ~\et~ \, 2 \! < \! {\ga}_g \! \leq \! 5 \, ~\et~ \,  k_f \! \geq k_g \, ~\et~ \, v_r=0,  
\]
applying \eqref{eq:inf} and \Lem{lem:intervals} to these numeric constraints leads to the following functions over $[0, +\infty[$
\begin{align*}
(Mf)(\eta) & =  \textstyle 1  -\frac{11}{4}(1- e^{ -\eta }) + \frac{11}{4} \eta,  \\
(Mg)(\eta) & =  \textstyle (3 -\frac{5}{2})(1- e^{ -\eta }) + \frac{5}{2} \eta,
\end{align*}
for which we have $(Mf)(1)=1 + \frac{11}{4}e^{-1} < 3 - \frac{1}{2}e^{-1}= (Mg)(1)$.

 To overcome this problem, we proceed as follows. If $\phi_f = \tilde{t}_f$, then from \eqref{eq:inf} $(Mf)$ is a straight line and we are done; let us consider the case $\phi_f < \tilde{t}_f$ (if $\phi_f > \tilde{t}_f$ the approach is completely analogous). First, a new interpolation point $\eta_{\infty} \putAs \eta_r +1$ is added. Secondly, for every function variable $f$ in $\sF$, let us denote by $\sF_{<f}$ and $\sF_{>f}$ the sets of function variables which are smaller, resp. greater, than $f$ with respect to the partial order given by variables $k$, for example $\sF_{<f}=\{ g \, | \, g \! \in \! \sF \ \mbox{and}\ k_f \geq k_g \}$. Thirdly, we fix $y^f_{inf} \putAs \max \{ \tilde{y}_g \, | \, g \! \in \! \sF_{<f} \}$ and $y^f_{sup} \putAs \min \{ \tilde{y}_g \, | \, g \! \in \! \sF_{>f} \}$ with $y^f_{inf} \putAs \tilde{y}_f -1$ and $y^f_{sup} \putAs \tilde{y}_f +1$ if $\sF_{<f}$, resp. $\sF_{>f}$, is empty; setting also $\underline{y_f} \putAs \frac{\tilde{y}_f + y^f_{inf}}{2}$ and $\overline{y_f} \putAs \frac{\tilde{y}_f + y^f_{sup}}{2}$. 

We now consider the straight lines $\overline{r_f}$, $r_f$ and $\underline{r_f}$ with angular coefficient $\phi_f$ passing, respectively, through the points $(\eta_r,\overline{y_f})$, $(\eta_r,\tilde{y}_f)$ and $(\eta_r,\underline{y_f})$; these lines define the new points $\overline{y^{\infty}_f} \putAs \overline{r_f}(\eta_{\infty})$, $y^{\infty}_f \putAs r_f(\eta_{\infty})$ and $\underline{y^{\infty}_f} \putAs \underline{r_f}(\eta_{\infty})$. Since we are considering the subcase $\phi_f < \tilde{t}_f$, and thus we search for a concave function in $[\eta_{\infty}, +\infty[$, let us fix $\widehat{y_f} \putAs \frac{y_f^{\infty} + \overline{y_f^{\infty}}}{2}$ and $ \widehat{t_f} \putAs \frac{\widehat{y_f} + \overline{y_f^{\infty}}}{2} + \phi_f - \widehat{y_f}$. Finally, over $[\eta_{\infty}, +\infty[$ we define the following function:
\[
(Mf)({\eta}) = \widehat{y_f} + (\widehat{t_f} -{\phi}_f )(1- e^{{\eta}_{\infty} -\eta }) + {\phi}_f (\eta - {\eta}_{\infty}),
\]
given our choice for the parameters $\widehat{y_f}$, $\widehat{t_f}$ and $\phi_f$ this function passes through the point $(\eta_{\infy},\widehat{y_f})$ and, after $\eta_{\infy}$, lies strictly between $\overline{r_f}$ and $\underline{r_f}$ (in this case actually between $\overline{r_f}$ and $r_f$). The following picture summarizes this method.

\begin{figure}[!htb]\begin{center}
\resizebox{14cm}{!}{
  \includegraphics{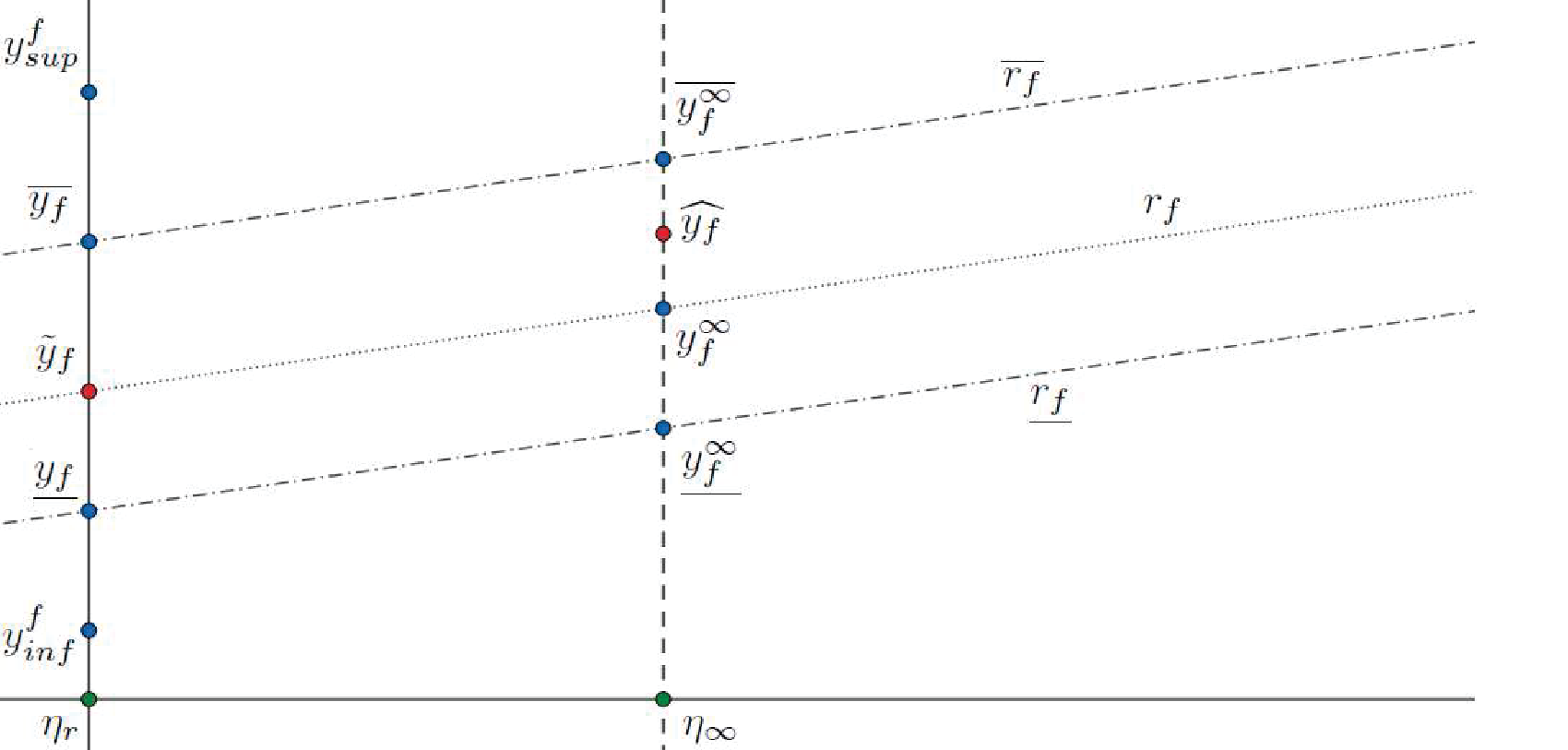}
}
\end{center}
\caption{\label{fig:inf}\Nv\footnotesize The green dots denote the domain values $\eta_r$ and $\eta_{\infty}$. The blue dots represent the constructed points and the two red ones the points through which function $f$ passes.}
\end{figure}

Applying this procedure to each function variable in $\sF$, we obtain a family of functions such that if $k_f \geq k_g$, then $(Mf)(\eta) > (Mg)(\eta)$ for all $\eta \geq \eta_{\infty}$.

Concerning the closed bounded interval $[\eta_r, \eta_{\infy}]$, this can be handled as outlined before.

\paragraph{Case \boldmath$(D\mathord{\geq})$:} As before, we start with the closed case. Let $(D[f]\geq y)_{[z_1 ,z_1]}$ be a literal occurring in ${\fhi}_3$ and let $Mz_1 \leq Mz_2$, otherwise the literal is vacuously true. In every single interval $[{\eta}_j ,{\eta}_{j+1}]$ with $j\in \{ \ind(z_1), \dots , \ind(z_2)-1 \}$, we have to verify that:
\[
(Mf)'(\eta ) \geq \bar{y},\ \ \forall \eta \in [{\eta}_j , {\eta}_{j+1}].
\]
For each $\eta \! \in \! [{\eta}_j , {\eta}_{j+1}]$, we have:
\begin{align*}
(Mf)'(\eta) ~=~  (s_j^f (p_j))'(\eta ) + (c_j^f (p_j))'(\eta)   ~=~   (p_j)(\eta ) \x {\left[ (s_j^f)'(x) + (c_j^f)'(x) \right] }_{x=p_j(\eta )} . 
\end{align*}
Moreover, from the clauses $t_j^f \geq y$ introduced in Step~\ref{itemAlgoFour}.\ref{alg4:actC}), we have:
\[
\textstyle {\bar{t}}_j^f = (Mf)'({\eta}_j)=\left[ ({\bar{y}}_{j+1}^f -{\bar{y}}_j^f)+ (c_j^f)'(0) \right] \x \frac{1}{{\eta}_{j+1}-{\eta}_j} \geq \bar{y} .
\]
Thus,
\[
(c_j^f)'(0) \geq ({\eta}_{j+1}-{\eta}_j )\x \bar{y} - ({\bar{y}}_{j+1}^f - {\bar{y}}_j^f )= {\theta}_j^f (\bar{y}),
\]
and, analogously, we have $(c_j^f)'(1) \geq {\theta}_j^f (\bar{y})$.\\
At this point, let us distinguish the following two cases. If $c_j^f$ is a single-$[\al ,{\theta}_j^f ({\bar{t}}_j^f), {\theta}_j^f ({\bar{t}}_{j+1}^f) ]$-defined function, then by \Cor{cor:1.2}, we have:
\[
(c_j^f)'(x) \geq {\theta}_j^f (\bar{y})\ \ \ \forall x \! \in \! [0,1].
\] 
Otherwise, if $c_j^f$ is either double- or special-$[\al , {\theta}_j^f ({\bar{t}}_j^f) , {\theta}_j^f ({\bar{t}}_{j+1}^f)]$-defined, then let $\theta = \pm 4\al$ be its middle slope. Using clauses $\frac{y_{j+1}^f-y_j^f}{v_{j+1}^f-v_j^f} \geq y$ introduced in step \ref{itemAlgoFour}.\ref{alg4:actC}), ${\theta}_j^f (\bar{y}) \leq 0$ holds. In particular, it must be ${\theta}_j^f (\bar{y}) \! < \! 0$; indeed, if ${\theta}_j^f \! = \! 0$ then ${\theta}_j^f ({\bar{t}}_j^f)= {\theta}_j^f ({\bar{t}}_{j+1}^f)=0$ because of the satisfiability of formula $(t_j^f =y \et t_{j+1}^f =y)$ introduced in step \ref{itemAlgoFour}.\ref{alg4:actC}), and this implies that $c_j^f$ is a single-$[\al , 0, 0]$-defined function, contradicting the hypothesis.

In order to verify that $\pm 4\al > {\theta}_j^f (\bar{y})$, we proceed as follows. Trivially, ${\theta}_j^f (\bar{y}) \! < \! 0 \! < \! \al $ holds. Furthermore, because ${\theta}_j^f (\bar{y}) \! < \! 0$ we can use minimality of $\al$ with respect to $A$, and we get $4\al \! < \! |{\theta}_j^f (\bar{y})|$, that is $-4\al > {\theta}_j^f (\bar{y})$. Now, \Lem{lem:1.4} can be applied to get $(c_j^f)'(x) \geq {\theta}_j^f (\bar{y})$ for each $x\in [0,1]$. Finally, in both cases, we have:
\[
\textstyle (Mf)'(\eta) \geq \left[  ({\bar{y}}_{j+1}^f -{\bar{y}}_j^f) + {\theta}_j^f (\bar{y}) \right] \x \frac{1}{{\eta}_{j+1} -{\eta}_j} = \bar{y}
\]
for each $\eta \in [{\eta}_j ,{\eta}_{j+1}]$.

Moreover, if $z_2 =+\infty$ (resp. $z_1=-\infty$), the inequality holds in the interval $[\eta_r, +\infty[$ (resp. $] -\infty, \eta_1 ])$. In fact, for each $\eta \in [\eta_r, +\infty[$, we have:\footnote{See the unbounded case of $(f>)$ for the definition of $\phi_f$.}
\[\begin{array}{rcl}
(Mf)'(\eta) & = & \phi_f + ( {\bar{t}}_r^f -\phi_f )e^{\eta -{\eta}_1} \geq \phi_f + \underset{\eta \in [\eta_r,+\infty[}{\mbox{inf}} [({\bar{t}}_r^f -\phi_f) e^{{\eta}_r - \eta} ],
\end{array}\]
depending on the sign of $({\bar{t}}_r^f -\phi_f)$, the last sum may assume the following values:
\[\begin{array}{rcl}
\phi_f + \underset{\eta \in [\eta_r,+\infty[}{\mbox{inf}} [({\bar{t}}_r^f -\phi_f) e^{{\eta}_r - \eta} ] &=& \begin{cases}
     {\bar{t}}_1^f & \mbox{if}\ {\bar{t}}_1^f < {\bar{\ga}}_0^f ,\\
     {\bar{\ga}}_0^f & \mbox{if}\ {\bar{t}}_1^f \geq {\bar{\ga}}_0^f,
     \end{cases}
\end{array}\] 
in both cases $ \phi_f + \underset{\eta \in [\eta_r,+\infty[}{\mbox{inf}} [({\bar{t}}_r^f -\phi_f) e^{{\eta}_r - \eta} ] \geq \bar{y}$ due to the satisfiability of the formulas introduced in step \ref{itemAlgoFour}.\ref{alg4:actC}).

Let us now consider the case of open or semi-open intervals; in particular, consider the case $(D[f]>y)_{]w_1 ,w_2 [}$ with $w_1 <w_2$, otherwise the literal is vacuously true. Let $w_1 =v_k , v_{k+1}, \dots ,v_{k+n} =w_2$ be the $n+1$ domain variables inside $[w_1 ,w_2 ]$; because of point \ref{alg1:action4:case3} of the algorithm $n \! \geq \! 2$, i.e there is at least a domain variable $v$ such that $w_1 \! < \! v \! < \! w_2 $. These variables split $[w_1 , w_2 ]$ into $n$ intervals $[v_k , v_{k+1} ], [v_{k+1} ,v_{k+2}], \dots, [v_{k+n-1}, v_{k+n}]$. First, we consider the $n-2$ intervals $[v_{k+1} ,v_{k+2}], [v_{k+2} ,v_{k+3}], \dots, [v_{k+n-2}, v_{k+n-1}]$ (for $n=2$ there are no such intervals). Since these intervals are closed and bounded, we can use the same strategy applied before, and we get $D[(Mf)]>y$ inside them. It remains to prove the property for the two endmost intervals $]v_k, v_{k+1}] = ]w_1 ,v_{k+1}]$ and $[v_{k+n-1}, v_{k+n}[ = [v_{k+n-1}, w_2[$.

Let $]v_k ,v_{k+1} ]= ] w_1 ,v_{k+1}]$. Due to steps \ref{itemAlgoFour}.\ref{alg4:actD}) and \ref{itemAlgoThree}) of the algorithm, we have $ \bar{y} = {\bar{t}}_k^f $; moreover, for the step \ref{itemAlgoFour}.\ref{alg4:actD}) we also have:
\[
\textstyle \dfrac{{\bar{y}}_{k+1}^f - {\bar{y}}_k^f}{{\eta}_{k+1} -{\eta}_k} > \bar{y}
\]
so ${\theta}_k^f ({\bar{t}}_k^f) ={\theta}_k^f (\bar{y})<0$. Since
\[
(Mf)(\eta) = s_k^f (p_k (\eta )) + c_k^f (p_k (\eta ))
\]
for every $\eta \in [{\eta}_k ,{\eta}_{k+1}]$, where $c_k^f \colon [0,1] \frec \mR $ is the function $[\al, {\theta}_k^f ({\bar{t}}_k^f ), {\theta}_k^f ({\bar{t}}_{k+1}^f ) ]$-defined, we have:
\[
(c_k^f)' (0)= {\theta}_k^f ({\bar{t}}_k^f)<0
\]
We now separately consider the three cases in which the function $c_k^f$ is \ita{single}, \ita{double}, or \ita{special}-defined.

If $c_k^f$ is a \ita{single}-defined function, from $(c_k^f)'(0)<0$, because of \Cor{cor:1.1}, it follows that $c_k^f$ is strictly convex; thus its derivative increases monotonically. Thus, for $c_k^f$ single-defined, we have
\[
(c_k^f)'(0) < (c_k^f)'(t)\ \ \forall t \in\ ]0,1].
\]
If $ c_k^f $ is a \ita{double}-defined function, from $(c_k^f)'(0)<0$, because of \Cor{cor:1.1} and \Def{def:1.4}, it follows that $c_k^f$ is strictly convex in $[0,\frac{1}{2}]$ and strictly concave in $[\frac{1}{2}, 1]$.  Thus, its derivative increases monotonically in $[0,\frac{1}{2}]$ and decreases monotonically in $[\frac{1}{2}, 1]$. Therefore
\begin{align*}
(c_k^f)'(0) &< (c_k^f)'(t)\ \ \forall t \in \left] 0, \textstyle{\frac{1}{2}}  \right] 
\intertext{and}
(c_k^f)'(1) &< (c_k^f)'(t)\ \ \forall t \in \left[ \textstyle{\frac{1}{2}}, 1 \right[.
\end{align*}
Furthermore, because of steps \ref{itemAlgoOne}.\ref{alg1:actionThree}), \ref{itemAlgoThree}) and \ref{itemAlgoFour}.\ref{alg4:actD}) of the algorithm, we have
\[
\bar{y}={\bar{t}}_k^f\ \ \mbox{and}\ \ {\bar{t}}_{k+1}^f > \bar{y},
\]
then
\[
{\theta}_k^f ({\bar{t}}_k^f) < {\theta}_k^f ({\bar{t}}_{k+1}^f),
\]
and this means
\[
(c_k^f)'(0)<(c_k^f)'(1).
\]
Thus, if $c_k^f$ is a double-defined function we have:
\[
(c_k^f)'(0)<(c_k^f)'(t)\ \ \forall t \in ]0,1].
\]

If $c_k^f$ is a \ita{special}-defined function, from $(c_k^f)'(0)<0$, because of \Cor{cor:1.1} and \Def{def:1.5}, it follows that $(c_k^f)'(1)=0$, $c_k^f$ is strictly convex in $[0, \frac{1}{2}]$ and strictly concave in $[\frac{1}{2}, 1]$; then its derivative is monotonically increasing in $[0,\frac{1}{2}]$ and monotonically decreasing in $[\frac{1}{2}, 1]$. Thus
\begin{align*}
(c_k^f)'(0) &<(c_k^f)'(t)\ \ \forall t \in \left] 0,\textstyle{\frac{1}{2}} \right]
\intertext{and}
(c_k^f)'(1) &\leq (c_k^f)'(t)\ \ \forall t \in \left[ \textstyle{\frac{1}{2}} ,1 \right[ .
\end{align*}
Moreover
\[
(c_k^f)'(0) < 0 = (c_k^f)'(1).
\]
Thus, if $c_k^f$ is a \ita{special} function, we have:
\[
(c_k^f)'(0) < (c_k^f)'(t)\ \ \forall t \in ]0, 1].
\]
Then, for each of the three cases (\ita{single}, \ita{double}, \ita{special}), for the function $c_k^f$ the following property holds:
\[
(c_k^f)'(0) < (c_k^f)'(t)\ \ \forall t \in ]0,1].
\] Let us now consider the derivative of the function $(Mf)$, in the interval $[{\eta}_k ,{\eta}_{k+1}]$ it has the form:
\[
\textstyle  ((Mf))'(\eta )=(({\bar{y}}_{k+1}^f -{\bar{y}}_k^f )+ (c_k^f)'(p_k (\eta ))) \x \frac{1}{{\eta}_{k+1}-{\eta}_k} \ \ \ \forall \eta \! \in \! [{\eta}_k , {\eta}_{k+1} ].
\]
Moreover, from
\[
(c_k^f)'(0) < (c_k^f)'(t)\ \ \forall t \in\ ]0,1],
\]
it follows
\[
(c_k^f)'(0) = (c_k^f)'(p_k ({\eta}_k )) < (c_k^f)'(p_k (\eta ))\ \ \forall \eta \! \in ]{\eta}_k , {\eta}_{k+1}]
\]
and
\[
((Mf) )'({\eta}_k ) < ((Mf))'(\eta )\ \ \forall \eta \in ]{\eta}_k , {\eta}_{k+1}].
\]
Moreover, by the steps \ref{itemAlgoOne}.\ref{alg1:actionThree}) and \ref{itemAlgoThree}) of the algorithm and \Lem{lem:2.2}, we have:
\[
\bar{y} = {\bar{t}}_k^f = ((Mf))'({\eta}_k).
\]
Thus, we finally obtain
\[
\bar{y} < ((Mf))'(\eta )\ \ \forall \eta \in ]{\eta}_k , {\eta}_{k+1}].
\]

Satisfiability of the other clauses, $(D[f]\bowtie y)_{[z_1 ,z_2]}$ with ${\bowtie} \in \{ =, >,<, \leq \}$ or the other extremal interval $[v_{k+n-1}, v_{k+n}[ = [v_{k+n-1} , w_2 [$ can be proved in a completely analogous manner.

\paragraph{Case \textbf{(Mon):}} 
Let Strict$\_$Up$(f)_{[z_1 , z_2]}$ be a literal appearing in $\fhi_3$, and suppose $Mz_1 \! < \! Mz_2 $; otherwise, the literal holds trivially. The function $(Mf)$ is strictly increasing over the interval $[Mz_1,Mz_2]$. To verify this, it is sufficient to check that its first-order derivative remains positive throughout the interval. The formal proof is omitted, as it would simply repeat the reasoning given in the previous step. A similar approach applies to analogous literals, e.g. Strict$\_$Down$(f)_{[z_1, z_2]}$.

\paragraph{Case \textbf{(Con):}}
Let Convex$(f)_{[z_1,z_2]}$ be a literal occurring in ${\fhi}_3$ and let $Mz_1 \! < \! Mz_2$, otherwise the literal is vacuously true. In every single interval $]{\eta}_j , {\eta}_{j+1}[$ with $j \in \{\mathit{ind}(z_1), \dots , \ind(z_2) - 1\}$, the function $(Mf)$ is convex. It is enough to verify that its second-order derivative is not negative:
\[
(Mf)''(\eta) =(c_j^f)''(p_j(\eta )) \x ((p_j)'(\eta) )^2 \geq 0,
\]
for each $\eta \in {]}{\eta}_j ,\frac{{\eta}_j +{\eta}_{j+1}}{2}{[} \cup {]} \frac{{\eta}_j +{\eta}_{j+1}}{2},{\eta}_{j+1}{[}$, as follows.

From satisfiability of formulas introduced in step \ref{itemAlgoFour}.\ref{alg4:actF}), we have ${\theta}_j^f ({\bar{t}}_j^f)\leq 0$ and ${\theta}_j^f ({\bar{t}}_{j+1}^f) \geq 0$ for each $j \! \in \! \{ \ind(z_1), \dots, \ind(z_2)-1\}$. Moreover, observe that ${\theta}_j^f ({\bar{t}}_j^f ) \! = \! 0$ implies ${\theta}_j^f ({\bar{t}}_{j+1}^f) \! = \! 0$ and conversely, by virtue of satisfiability of the formula $t_j^f = t_{j+1}^f$ introduced in step \ref{itemAlgoFour}.\ref{alg4:actF}). This means, by \Lem{lem:1.1}, that $c_j^f$ is necessarily a single-$[\al , {\theta}_j^f ({\bar{t}}_j^f), {\theta}_j^f ({\bar{t}}_{j+1}^f)]$-defined function; and by \Cor{cor:1.1}, we have $(c_j^f)''(x) \geq 0$ for each $x\in {]}0, \frac{1}{2}{[} \cup {]}\frac{1}{2}, 1{[}$.

Moreover, if $z_1 =-\infty $ (resp. $z_2 =+\infty $), the function $(Mf)$ is convex in the interval $]-\infty ,{\eta}_1{[}$ (resp. ${]}{\eta}_r , +\infty {[}$); indeed we have:
\[
(Mf)''(\eta )= ({\bar{t}}_1^f - {\bar{\ga}}_0^f ) e^{\eta -{\eta}_1 } \geq 0\ \ \forall \eta \in ] -\infty, {\eta}_1 [.
\]
where ${\bar{t}}_1^f \geq {\bar{\ga}}_0^f $ because of the clauses introduced in step \ref{itemAlgoFour}.\ref{alg4:actF}).

The convexity of the function $(Mf)$ in the whole interval $[Mz_1,Mz_2 ]$ easily follows from its differentiability in that interval. Satisfiability of clauses Concave$(f)_{[z_1,z_2]}$, Strict\_Concave$(f)_{[z_1,z_2]}$, Strict\_Convex$(f)_{[z_1,z_2]}$ can be verified in a completely analogous manner.

Finally, all purely arithmetic clauses occurring in ${\fhi}_3$ also occur in ${\fhi}_4$ and they are hence automatically satisfied by $M$.

\begin{center}
    ------------
\end{center}

\noindent {\boldmath\large${\fhi}_3 \imp {\fhi}_4 )$}\ We need to demonstrate that if a model for ${\fhi}_3$ exists, then it is possible to construct a model that satisfies all the formulas introduced in ${\fhi}_4$.

Let $M$ be a concrete model for ${\fhi}_3$. Note that only the numerical variables $\ga$ and $k$, which were added during the transformation process, are not interpreted by $M$; moreover, observe that the majority of the clauses in $\phi_4$, where no numerical variable $\ga$ or $k$ appears, are direct consequences of $\phi_3$, and thus are already satisfied by $M$.

Now, let $ {\eta}_1 < {\eta}_2 < \dots < {\eta}_r $ denote the domain-variable interpretations in the model $M$.

\begin{definition}[Variables $\ga$] Fix two points ${\ep}_0 \in {]-\infty} , {\eta}_1 [$ and ${\ep}_r \in {]{\eta}_r} , +\infty [$, and, for each function variable $f \! \in \! F$, consider the real values
\[
\textstyle c_0^f \! \putAs \frac{{\bar{y}}_1^f - (Mf)({\ep}_0)}{{\eta}_1 - {\ep}_0}\ \ \mbox{and}  \ \ c_r^f \! \putAs \frac{(Mf)({\ep}_r) - {\bar{y}}_r^f}{{\ep}_r - {\eta}_r}.
\]
The model $M$ gets extended with the following interpretations of $\ga$s, where, for each function variables $f\! \in \! F$, we set
\[
{\bar{\ga}}_i^f = M({\ga}_i^f) = c_i^f,\ \ \ i\in \{ 0, r \} . \tag*{\eod}
\]
\end{definition}
\begin{definition}[Variables $k$] For each function variable $f\! \in \! F$, consider the following real values:
\[
 l_0^f \! \putAs
  \left\{
     \begin{array}{lll}
      1 & \mbox{if} & \liminf_{x \frec -\infty}(Mf)(x)=+\infty, \\
      \frac{2\:arctg(\tau)}{\pi} & \mbox{if} & \liminf_{x \frec -\infty}(Mf)(x)= \tau, \\
      -1 & \mbox{if} & \liminf_{x \frec -\infty}(Mf)(x)=-\infty;
     \end{array}
   \right. 
 \ \ 
l_r^f \! \putAs
  \left\{
     \begin{array}{lll}
      1 & \mbox{if} & \liminf_{x \frec +\infty}(Mf)(x)=+\infty, \\
      \frac{2\:arctg(\tau)}{\pi} & \mbox{if} & \liminf_{x \frec +\infty}(Mf)(x)= \tau, \\
      -1 & \mbox{if} & \liminf_{x \frec +\infty}(Mf)(x)=-\infty.
     \end{array}
   \right.   
\]
The model $M$  gets extended with the following interpretation of $k$'s, where, for each function variable $f \! \in \! F$, we set:
\[
{\bar{k}}_i^f = M({k}_i^f) = l_i^f,\ \ \ i\in \{ 0, r \} .\tag*{\eod}
\]
\end{definition}

To complete the proof, we need to confirm that all clauses in $\fhi_4$, introduced in step \ref{itemAlgoFour}), are satisfied, particularly those that involve the numerical variables $\ga$ and $k$.

\paragraph{Case \boldmath{$(f\mathord{=})$:}} 
Let $(f=g)_{[z_1 , z_2 ]}$ be a literal occurring in ${\fhi}_3$. The formulas introduced in step \ref{itemAlgoFour}.\ref{alg4:actA}) are trivially satisfied; in particular, if $z_1 = - \infty $, we have ${\bar{\ga}}_0^f = {\bar{\ga}}_0^g $, because $c_0^f = c_0^g $. The case $z_2 = +\infty$ is similar.

\paragraph{Case \boldmath{$(f\mathord{>})$:}} 
As previously, we begin by considering the closed case. Let $(f>g)_{[w_1 ,w_2 ]}$ represent a literal occurring in ${\fhi}_3$. It is straightforward to confirm that the corresponding formulas introduced in step \ref{itemAlgoFour}.\ref{alg4:actB}) are satisfied.

Let now $(f>g)_{]w_1 ,w_2 [}$ be a literal occurring in ${\fhi}_3$, we have:
\[
(Mf)(\eta) > (Mg)(\eta )\ \ \ \forall \eta \! \in ]{\eta}_{\mathit{ind}(w_1)} , {\eta}_{\mathit{ind}(w_2)}[.
\]
Thus, by the literal added in step \ref{itemAlgoThree}), we have:
\[
{\bar{y}}_i^f = (Mf)({\eta}_i) > (Mg)({\eta}_i) = {\bar{y}}_i^g
\]
for all $i$ such that $\ind(w_1)+1 \leq i \leq \ind(w_2)-1$. If $w_1 < w_2 $ as domain ordered variables, because of step \ref{itemAlgoOne}.\ref{alg1:actionOne}) of the algorithm we have:
\[
\begin{array}{c}
(Mf)({\eta}_{\mathit{ind}(w_1)})=(Mg)({\eta}_{\mathit{ind}(w_1)}),\\
(Mf)({\eta}_{\mathit{ind}(w_2)})=(Mg)({\eta}_{\mathit{ind}(w_2)}).
\end{array}
\]
Then, because of step \ref{itemAlgoOne}.\ref{alg1:actionThree}) and \Lem{lem:1.14}, we have:
\[
\begin{array}{c}
{\bar{t}}_{\mathit{ind}(w_1)}^f = (Mf)'({\eta}_{\mathit{ind}(w_1)}) \geq (Mg)'({\eta}_{\mathit{ind}(w_1)}) = {\bar{t}}_{\mathit{ind}(w_1)}^g ,\\
{\bar{t}}_{\mathit{ind}(w_2)}^f = (Mf)'({\eta}_{\mathit{ind}(w_2)}) \leq (Mg)'({\eta}_{\mathit{ind}(w_2)}) = {\bar{t}}_{\mathit{ind}(w_2)}^g .
\end{array}
\]
For literals of the types $(f>g)_{] w_1, w_2]}$ and $(f>g)_{[w_1 ,w_2 [}$, the proof is similar.

Finally, for unbounded literals, such as $(f \! > \! g)_{[w_1, +\infty[} $, if $(Mf)\! > \! (Mg)$ in $[w_1, +\infty[$, then $\liminf_{x \frec +\infty}(Mf)(x) \geq \liminf_{x \frec +\infty}(Mg)(x)$. Thus, $l_r^f \! \geq \! l_r^f$, and so $\bar{k}_r^f \! \geq \! \bar{k}_r^g $. Similarly, one can proceed in the case of literals of the types $(f>g)_{] -\infty, w_1]}$ and $(f>g)_{]-\infty , +\infty[}$.

\paragraph{Case \boldmath{$(D\mathord{\geq})$:}} 
Let $(D[f]\geq y )_{[z_1 ,z_2]}$ be a literal occurring in ${\fhi}_3$. The formulas introduced in step \ref{itemAlgoFour}.\ref{alg4:actC}) are satisfied. Trivially, we have $(Mf)'({\eta}_i)={\bar{t}}_i^f \geq \bar{y}$, for each $i \in \{ \ind(z_1), \dots , \ind(z_2)-1 \}$; while satisfiability of formulas $\frac{y_{j+1}^f -y_j^f}{v_{j+1} -v_j} \geq y$, for each $j\in \{ \ind(z_1), \dots , \ind(z_2)-1 \}$, can easily be checked as follows. In each interval $[{\eta}_j , {\eta}_{j+1} ] \subset [Mz_1, Mz_2 ]$, the function $(Mf)$ is continuous and differentiable and we can apply the mean value theorem; that is, $\exists \ep \in ]{\eta}_j , {\eta}_{j+1}[$ such that
\[
\textstyle \dfrac{(Mf)({\eta}_{j+1})-(Mf)({\eta}_j)}{{\eta}_{j+1} -{\eta}_j}=(Mf)'(\ep ).
\]
The satisfiability of formulas follows from the hypothesis $(Mf)'(x)\geq \bar{y}$ for each $x \in [Mz_1, Mz_2]$.\\
Moreover, if $\frac{{\bar{y}}_{j+1}^f -{\bar{y}}_j^f}{{\eta}_{j+1} - {\eta}_j}= {\bar{t}}_{j+1}^f = \bar{y}$ we have $(Mf)'({\eta}_j)=(Mf)'({\eta}_{j+1}) = \bar{y}$ by \Lem{lem:1.11}, that is, $ {\bar{t}}_j^f = {\bar{t}}_{j+1}^f = \bar{y}$.

If $z_1= -\infty $, we have ${\bar{\ga}}_0^f = c_0^f \geq \bar{y}$ because of the mean value theorem; the case $z_2= +\infty$ can be treated analogously.

Let $(D[f]>y)_{]w_1 w_2[}$ be a literal occurring in ${\fhi}_3$, we have:
\[
(Mf)'(\eta )> \bar{y} \ \ \ \forall \eta \in ]{\eta}_{\mathit{ind}(w_1)}, {\eta}_{\mathit{ind}(w_2)}[.
\]
Thus, because of the formulas introduced in step \ref{itemAlgoThree}), we have:
\[
{\bar{t}}_i^f =(Mf)'(\eta ) > \bar{y}
\]
for all $i$ such that $ \ind(w_1) \! < \! i \! < \! \ind(w_2) $. Moreover, if ${\eta}_h$ and ${\eta}_{h+1}$ are two consecutive domain variables, because of the mean value theorem it exists $\ep \in ]{\eta}_h , {\eta}_{h+1}[$ such that:
\[
\textstyle \dfrac{(Mf)({\eta}_{h+1}) -(Mf)({\eta}_h)}{{\eta}_{h+1} - {\eta}_h }= (Mf)'(\ep ).
\]
Then, because of step \ref{itemAlgoThree}) of the algorithm, we have:
\[
\frac{{\bar{y}}_{j+1}^f - {\bar{y}}_j^f}{{\eta}_{j+1}-{\eta}_j} =\frac{(Mf)({\eta}_{j+1}) - (Mf)({\eta}_j)}{{\eta}_{j+1} - {\eta}_j } > \bar{y}
\]
for all $j$ such that $\ind(w_1) \! \leq \! j \! \leq \! \ind(w_2)-1$.

The satisfiability of the other literals, $(D[f]\bowtie y)_{[z_1, z_2]}$, $(D[f]\bowtie y)_{]w_1, w_2]}$ and $(D[f]\bowtie y)_{[w_1, w_2[}$ with ${\bowtie} \in \{=, >, <,\\\leq \}$, is again verified in a similar manner.

\paragraph{Case \textbf{(Mon):}} 
Let Strict$\_$Up$(f)_{[z_1 ,z_2]}$ (resp. Strict$\_$Down$(f)_{[z_1 ,z_2]}$) be a literal occurring in ${\fhi}_3$, it is immediate to verify that the corresponding formulas introduced in step \ref{itemAlgoFour}.\ref{alg4:actE}) are satisfied.

\paragraph{Case \textbf{(Con):}} 
Let Convex$(f)_{[z_1 ,z_2 ]}$ be a literal occurring in ${\fhi}_3$. Due to the convexity of the function $(Mf)$ in $[Mz_1 , Mz_2]$, the formulas introduced in step \ref{itemAlgoFour}.\ref{alg4:actF}) are satisfied, because in that interval $(Mf)$ does not lie below the straight line tangent to $(Mf)$ in any point $(x, (Mf)(x))$ for each $x\in [Mz_1 , Mz_2]$; e.g., for each $i \in \{ \ind(z_1), \dots, \ind(z_2)-1 \}$, the point $({\eta}_{i+1}, {\bar{y}}_{i+1}^f)$ does not lie below the straight line tangent to $(Mf)$ in $({\eta}_i , {\bar{y}}_{i+1}^f)$.

Moreover, if $\frac{{\bar{y}}_{i+1}^f - {\bar{y}}_i^f}{{\eta}_{i+1} - {\eta}_i }$ equals ${\bar{t}}_i^f$ or ${\bar{t}}_{i+1}^f$, we have $(Mf)'({\eta}_i)=(Mf)'({\eta}_{i+1})$ by \Lem{lem:1.12}, i.e., ${\bar{t}}_i^f = {\bar{t}}_{i+1}^f$.

Finally, in a completely analogous way, if $z_1 = -\infty$ (resp. $z_2 = +\infty$) we have ${\bar{\ga}}_0^f \leq {\bar{t}}_1^f$ (resp. ${\bar{\ga}}_r \geq {\bar{t}}_r^f$).

Likewise, one verifies that all literals of the forms Strict$\_$Concave$(f)_{[z_1 ,z_2]}$, Concave$(f)_{[z_1 ,z_2]}$, Strict$\_$Convex$(f)_{[z_1, z_2]}$ appearing in ${\fhi}_3$ are satisfied.

We can now restate and prove our goal, namely:
\begin{theorem} The $RDF^+$ theory is decidable.
\end{theorem}
\dimostraz We have built up and proved the following chain of equisatisfiability
results:
\[\begin{array}{rcccl}
\fhi_{i-1} & \mbox{and} & \fhi_{i} & \mbox{are equisatisfiable,} &
 \mbox{for\ $i=1,2,3,4$}\:. \end{array}
\] 
Formula ${\fhi}_4$, which contains only numerical variables, the numerical operators $+,\x\, $, and the comparators $=,<$ between reals, is decidable using Tarski's decision method. \qed

\section{Related Work and Conclusions}\label{sec:final}

We now briefly consider several aspects of the topics discussed so far, with particular attention to complexity-related results and potential future developments.
 
\subsection{Related literature and complexity issues} \label{subsec:relatedWork} 

The decidability of the \RDFp theory, as outlined above, builds upon a series of earlier results concerning the \ita{RMCF}, \ita{$RMCF^+$}, and \ita{RDF} theories \cite{DC06,DC07,DC87,GC00}. A comprehensive overview of these findings is available in \cite{DC12}, which also discusses additional decidability results in real analysis, notably those related to the Friedman–Seress theory \cite{FS89,FS90}.  

Since the decidability of \RDFp is established via an explicit algorithm, assessing its computational complexity is both natural and necessary. Our procedure incorporates Tarski’s decision method and therefore inherits its complexity as a lower bound. The first substantial improvement over Tarski's original approach was introduced by Collins \cite{GC 75}, whose cylindric algebraic decomposition (CAD) method has doubly exponential complexity in the number of variables (or exponential, if the number of variables is fixed). 
This technique also found applications in problems such as the \ita{Piano Movers} problem \cite{SS 83}. 

Grigoriev further refined Collins' method by devising an algorithm tailored for sentences in prenex normal form, exhibiting doubly exponential complexity with respect to the number of quantifier alternations~\cite{DG 88}. In the specific context of the \ita{existential} theory of reals, known decision procedures generally have exponential complexity in the number of variables~\cite{BasuPR96}. However, if the number of variables is fixed, the complexity drops to polynomial-time~\cite{JR 88}.

Lastly, while Tarski proved the undecidability of the elementary algebra \EAR of the real numbers when extended with certain transcendental functions such as $\sin x$~\cite{AT 51}, Richardson later showed that even the existential theory becomes undecidable when it includes the constants $\log 2$, $\pi$, and the functions $e^x$ and $\sin x$~\cite{DR 68}.

\subsection{Conclusions and follow-up work} \label{subsec:Conclusion}

This article has presented a decision algorithm for a fragment of real analysis, called \RDFp, which extends \EAR with variables designating real-valued functions possessing continuous derivatives. We began by introducing the syntax and semantics of \RDFp, illustrating how to define derived relators and express common analytic properties and classical theorems within the theory. We then described in full detail the reduction algorithm, which transforms a generic \RDFp formula $\fhi$ into an equisatisfiable \EAR formula $\widehat{\fhi}$ through four successive steps that eliminate function variables. A concrete example followed to demonstrate the algorithm in practice.

The correctness of the decision procedure---specifically, the equisatisfiability of $\fhi$ and $\widehat{\fhi}$---relies on the construction of suitable functional models. These models are obtained from a particular class of functions, the so-called $[\al, {\theta}_1, {\theta}_2]$-defined functions, introduced to meet the needs of the completeness proof.

The present article reorganizes and fills in gaps left by previous published work \cite{BCCOS20,DC07} as well as by earlier unpublished notes. The framework developed here has since been extended in a series of follow-up papers. In particular, \cite{BCCOS23} introduces a new theory, \RDFs, which enriches \RDFp with two functional term operators: function addition and scalar multiplication. Specifically, if $f,g$ are function terms, then $f+g$ denotes their pointwise sum, and if $s$ is a numerical term, then $s \x f$ denotes the pointwise product of $s$ and $f$. These operators permit more expressive atoms---e.g., $(f+g = h)_A$ and $(D[s \x f] > t)_A$---necessitating a refinement of the decision algorithm and a new class of interpolating functions for the completeness proof.

A further extension is proposed in \cite{BCCOS24}, where a family of theories \RDFn (with $n \! \in \! \mN$) is introduced. For each fixed $n$, the syntax of \RDFn extends that of \RDFs by adding derivative operators $D^{\al}[\cdot]$ for each $1 \leq \al \leq n$. These operators enable reasoning about functions of class $C^n$, thus broadening the expressivity beyond merely continuously differentiable ($C^1$) functions.

Future directions include further expansions of the syntax---for example, the addition of function product operators or the generalization of the theory to multivariate functions. In the latter case, the standard semantics would shift from unary functions $\mR \to \mR$ to multivariate functions $\mR^n \to \mR^m$, a change that seems well-supported by the current structure of the algorithm. Lastly, although Tarski established the undecidability of \EAR when extended with certain transcendental functions, a deeper investigation into the boundaries of undecidability for such extensions remains an open and relevant research avenue. A preliminary exploration in this direction is presented in \cite{BCCOS24}.

\section*{Acknowledgments}
This work was partially supported by the research program \textit{PIAno di inCEntivi per la Ricerca di Ateneo 2024/2026 – Linea di Intervento I “Progetti di ricerca collaborativa”}, Università di Catania, under the project \textit{“Semantic Web of EveryThing through Ontological Protocols” (SWETOP)}.
The first author is a member of the ``Gruppo Nazionale per le Strutture Algebriche, Geometriche e le loro Applicazioni'' (GNSAGA) of the ``Istituto Nazionale di Alta Matematica'' (INdAM), while the second and fourth authors are members of the ``Gruppo Nazionale per il Calcolo Scientifico" (GNCS) of INdAM.
The second author is also affiliated with the National Centre for HPC, Big Data and Quantum Computing (Project CN00000013, Spoke 10), co-funded by the European Union – NextGenerationEU.


\end{document}